\documentclass[12pt]{article}

\usepackage{amstext,amsmath,amssymb,amsfonts}
\usepackage[latin1]{inputenc}
\usepackage{epsfig}
\usepackage{hyperref}
\usepackage{amsthm}
\usepackage{subfigure}
\usepackage{color}
\usepackage{multirow}

\newcommand{\bea}{\begin{eqnarray}}
\newcommand{\eea}{\end{eqnarray}}

\newcommand{\So}{S_{1D}}
\newcommand{\Lo}{{\mathcal{L}_{1D}}}
\newcommand{\Sosca}{S_{1D}^{\text{scale}}}
\newcommand{\Losca}{{\mathcal{L}^{\text{scale}}_{1D}}}

\newcommand{\Stri}{S_{3D}}
\newcommand{\Ltri}{\mathcal{L}_{3D}}
\newcommand{\Skin}{S_{\text{kin},\,3D}}
\newcommand{\Sint}{S_{\text{int},\,3D}}
\newcommand{\Lkin}{\mathcal{L}_{\text{kin},\,3D}}
\newcommand{\Lint}{\mathcal{L}_{\text{int},\,3D}}

\newcommand{\Strisca}{S^{\text{scale}}_{3D}}
\newcommand{\Ltrisca}{\mathcal{L}^{\text{scale}}_{3D}}
\newcommand{\Skinsca}{S^{\text{scale}}_{\text{kin},\,3D}}
\newcommand{\Lkinsca}{\mathcal{L}^{\text{scale}}_{\text{kin},\,3D}}

\newcommand{\Scol}{S^{\text{color}} }
\newcommand{\Lcol}{\mathcal{L}^{\text{color}} }
\newcommand{\Lkincol}{\mathcal{L}^{\text{color}}_{\text{kin}}}

\newcommand{\Solsca}{S^{\text{color},\text{scale}} }
\newcommand{\Lcolsca}{\mathcal{L}^{\text{color},\text{scale}} }

\newcommand{\fifib}{(\phi \leftrightarrow \bar\phi) }

\newcommand{\buss}{ (\bullet)_{\check{s}} }

\makeatletter

\begin{document}

\begin{titlepage}
\begin{flushright}
pi-qg-229\\
ICMPA-MPA/007/2011
\end{flushright}

\vspace{20pt}

\begin{center}

{\Large\bf Classical Group Field Theory}
\vspace{15pt}

Joseph Ben Geloun

\vspace{15pt}

$^{a}${\sl Perimeter Institute for Theoretical Physics}\\
{\sl 31 Caroline St. N., ON, N2L 2Y5, Waterloo, Canada}\\
\vspace{10pt}
$^{c}${\sl International Chair in Mathematical Physics and Applications\\ (ICMPA-UNESCO Chair),
University of Abomey-Calavi,\\
072B.P.50, Cotonou, Rep. of Benin}\\
\vspace{5pt}
E-mail:  {\em jbengeloun@perimeterinstitute.ca}

\vspace{10pt}

\begin{abstract}
The ordinary formalism for classical field theory
is applied to dynamical group field theories. 
Focusing first on a local group field theory over
one copy of $SU(2)$ and, then, on more involved nonlocal
theories (colored and non colored) defined over a tensor product of
the same group, 
we address the issue of translation
and dilatation symmetries and the corresponding Noether theorem. 
The energy momentum tensor and dilatation current are  derived and
their properties identified for each case.
\end{abstract}
\end{center}

\setcounter{footnote}{0}

\noindent  Pacs numbers:  03.50.-z, 11.30.-j, 11.40.-q, 04.60.-m
\\

\setcounter{footnote}{0}

\end{titlepage}

\section{Introduction}
\label{Intro}

Group field theories (GFTs)  are usually defined 
as tensor field theories over a group manifold.  Introduced
in the beginning of the 90's \cite{boul},  they rapidly become
pertinent candidates for  quantum gravity \cite{Freidel,oriti,oriti2}. 
In a nutshell, GFTs provide a framework for addressing the problem 
of the emergence of the topology and the metric properties of
spacetime \cite{oriti2}.

Since the inception of GFT, 
several  studies have been led using the path integral approach \cite{vincentrev}.
Interesting facts pertaining to the renormalization 
program such as power
counting theorems \cite{fgo}-\cite{Geloun:2010vj}, 
an emergent locality principle \cite{Krajewski:2010yq,Geloun}, 
Ward-Takahashi identities for unitary symmetries \cite{BenGeloun:2011xu} have been highlighted. Furthermore, 
a large $1/N$ topological/combinatorial expansion \cite{Gurau:2011xq}\cite{Gurau:2011tj} for colored GFT models \cite{Gurau:2009tw, Gurau:2010nd,Gurau:2011xp} portends new and fertile contacts with models in statistical mechanics
\cite{Gurau:2011kk}-\cite{Benedetti:2011nn}. 

Classical aspects of GFT have been also examined.
For instance, the equations of motion for GFT models
possessing a trivial kinetic term\footnote{ The kinetic term
of the most basic GFT is trivial: it is simply composed by 
a mass term. 
A typical nontrivial dynamics is governed, for instance, by a Laplacian 
on the group manifold. } 
have been solved, thus providing an explicit class of instantons for these
theories \cite{Livine}.
These solutions were used to compute a class of effective actions. 
We can also emphasize that, using a  group Fourier transform 
(for seminal works on this topic and more applications, 
\cite{Oriti:2011ac} affords a recent review), 
GFTs can be seen as noncommutative field models with a
set of diffeomorphisms which turns out to be related to a
deformed Poincar\'e group \cite{Girelli:2010ct, Baratin:2011tg}.
Interestingly, one notices that 
the group manifold initially associated with
the background space on which the fields are defined, becomes finally
a curved momentum space via this group Fourier
transform \cite{Freidel:2005me}. The direct space 
generated after inverse transform is flat 
whereas the algebra of fields is traded for a noncommutative algebra endowed with a $\star$-product.  
Nevertheless, other interesting properties 
 at the classical level have been yet investigated.
Another contribution emphasizes to what extent the dynamics 
matters in the renormalization program for GFT \cite{Geloun}.  
Implementing a dynamical part for GFTs
will certainly affect the Noether analysis for symmetries
for these models whereas Noether currents in the absence
of dynamics are trivial, in general.  
Thus the present contribution addresses, in a more traditional 
field theory spirit and by reconsidering the group 
as a base manifold, the issue of the classical formalism for GFTs.

We should emphasize that some anterior investigations
have been carried out on field theories on
a Lie group regarding both classical symmetries and 
also some of their quantum implications. For example, 
a $\phi^4$ quantum field theory on the affine group has
been studied in \cite{Altaisky:2000gq}.
Furthermore, diffeomorphisms 
and Weyl transformations in a curved $\phi^4$ theory  and their
implications have a long history (see for instance \cite{Ford:1997hb}\cite{diffeos}). 
We will not use, in the present contribution, the same
route but will focus either on the gauge invariance of fields
or on the nonlocal feature of GFT. 
 Indeed, these two features of GFTs, and among these
the nonlocal character of GFTs which should be 
regarded as the most peculiar, 
will have drastic consequences on the regular properties of classical 
symmetries that one is accustomed in ordinary field theory.
It can be anticipated that the core notion of local conservation of currents will be drastically 
affected as it is the case in other well known nonlocal field
theories. This is indeed the case of the so called 
field theories on a  noncommutative
 spacetime \cite{Douglas:2001ba}. For instance, in any Moyal type noncommutative field theories, the action possesses Poincare currents with an explicit breaking for the local conservation
property \cite{Gerhold:2000ik}-\cite{Hounkonnou:2009qt}.

In this paper, we study the classical dynamical GFT over 
$D=1$ and then $D=3$ copies (that we shortly call dimensions
or rank) of $SU(2)$. The case $D=2$ turns out to be equivalent 
to the situation $D=1$ due to the gauge invariance condition
on fields.
Our results can be extended without ambiguity in any dimension.
As it will stand the action simply describes a gauge invariant 
tensor scalar field theory over $D$ copies of the sphere $S^3$
with a local (for $D=1$) or nonlocal (for $D=3$) interaction. 
The main issue in the present
study is to show that the procedure for solving
equations of motion or for studying the symmetries of the action 
finds an extension in a curved, tensored and nonlocal theory.
The translation symmetry and the corresponding energy momentum
tensor (EMT) have been worked out. We find that
the EMT appears to be symmetric in a certain sense but not locally conserved for $D>1$ in ordinary GFT. 
However, we surprisingly find that the colored
GFT possesses a covariantly conserved quantity 
obtained by integrating some EMT components.
We then address another interesting question
 which is the implementation 
of dilatation symmetry or scale transformation at the GFT level. 
Requiring the invariance under dilatations, in the
way we perform it here, yields a radically 
different action from the translation invariant action.
We compute and characterize the current tensor associated with this transformation. 

The content of this paper is the following: Section 2 is
devoted to the model presentation and the first steps
of the classical study: we solve the equation of motion
for free fields for the dynamical Boulatov model in $D=3$.
 Section 3  thoroughly undertakes the Noether
theorem for translation and dilatation symmetries for
a $1D$ GFT as a guiding example for more
complicated higher rank GFTs. 
Section 4 discusses the same symmetries for 
general GFTs. Section 5 deals with 
colored models and their particular characteristics.
Section 6 summarizes our results and also provides outlooks of
this work. Finally, 
a detailed appendix collects the proofs of our claims 
and useful identities invoked along the text.

\section{The dynamical Boulatov-Ooguri model}
\label{sect1}

The prominent properties of the dynamical three dimensional 
GFT over $G = SU(2)$ are quickly introduced (for more details on  the general formulation see \cite{Freidel,oriti,oriti2}). 
This will be followed by  the resolution of the field equation of motion without coupling constant. 
The Noether analysis of symmetries will be differed to the next sections. 
This section admits a straightforward extension in any GFT 
dimension $D>1$.

\noindent{\bf The model -} The fields belong to the Hilbert space of square integrable 
and gauge invariant functions on $G^{3}$ which  satisfy
\begin{equation}
\phi(g_1h,g_2h,g_3h)=\phi(g_1,g_2,g_3)\;, \quad \forall h \in 
G\;. 
\label{inv}
\end{equation}
The shorthand notation $\phi(g_1,g_2,g_3) = \phi_{1,2,3}$ 
will be used henceforth.

The action $\Stri$ is formed by a kinetic term 
and interacting part. The kinetic term has the form 
\begin{equation}
\Skin[\phi]:= \int [\prod_{\ell=1}^{3}dg_\ell]\ \left[ 
\frac{1}{2} \sum_{s=1}^3\mathbf{g}_s^{ij}\nabla_{(s)\;i} \;\phi_{1,2,3}
\nabla_{(s)\;j}\; \phi_{1,2,3}
+\frac{1}{2}m^2  \phi_{1,2,3}\;\phi_{1,2,3} \right],
\end{equation}
where $dg_i$ denotes the Haar measure on $G=SU(2)$, the operator
$\nabla_{(s)}^{i}$ represents the covariant derivative (acting here merely as a partial derivative
on above fields) defined with the Levi-Civita connection on $S^3\simeq SU(2)$. 
The index $(s)$ will always refer to the tensor structure and so 
to the particular group element $g_s$ with respect to which one  derivates.\footnote{It will be also 
called strand index in the following, $s=1,2,\dots, D$.}
The labels $i,j$ refer to the local coordinates and, therefore,
are lowered or raised by the $S^3$ metric $\mathbf{g}_{ij}$.
Note that the Haar measure of $SU(2)$,  $dg$ can be written 
in a more standard fashion with respect to a theory on a curved background 
as $dg=(2\pi^2)^{-1} \sqrt{|\det\mathbf{g}|} d\theta  d\varphi_1d\varphi_2$ 
with $\mathbf{g} = d\theta^2 +
\sin^2\theta (d\varphi^2_1 + \sin^2\varphi_1 d \varphi^2_2)$.

The interaction in $D$ dimensional GFTs is nonlocal and dually associated with a $D$-simplex. 
For $D=3$ dimensions, the interaction is
\begin{equation}
\Sint[\phi]:=\frac{\lambda}{4} \int [\prod_{\ell=1}^{6}dg_\ell]\ \phi_{1,2,3}\,
\phi_{3,4,5}\,\phi_{5,2,6}\,\,\phi_{6,4,1}\;,
\label{intera}
\end{equation}
with a particular pairing of the six variables 
according to the pattern of the edges of a tetrahedron.

By reducing the kinetic part to a pure massive term and considering 
the interaction term (\ref{intera}), one gets a model belonging 
to the class Boulatov-Ooguri models \cite{boul}. We will refer to models 
including Laplacian dynamics as dynamical GFT models. 

Formally, a Lagrangian density can be defined as
\bea
\Ltri = \frac{1}{2} \sum_{s}\nabla_{(s)}^{i} \;\phi_{1,2,3}
\nabla_{(s)\;i}\; \phi_{1,2,3}
+\frac{1}{2}m^2  \phi_{1,2,3}\;\phi_{1,2,3} + 
\frac{\lambda}{4} \; \phi_{1,2,3}\phi_{3,4,5} \,\phi_ {5,2,6}\, \phi_{6,4,1} 
\;.
\label{lagr}
\eea
The density (\ref{lagr}) should be integrated 
here over six copies of the group
(one copy for each $g_{i}$, $i=1,2,\dots,6$), 
this is the base manifold of the present GFT. 
Remark that, the kinetic part
does not involve $g_{j}$, $j=4,5,6$, 
but the integration of these three variables is 
without any effect since the Haar measure is normalized. 

Defining the quantity (\ref{lagr}) as the Lagrangian
density will not affect the remaining developments. 
Indeed, one should keep in mind that, in the Noether procedure,
the field variations are taken with respect to the action and 
provide equations of motion for fields. The latter are, 
together with the data of infinitesimal field variations 
under a given transformation and boundary conditions,
 the main ingredients in order to apply the Noether theorem. 
Current calculations have to be performed,
as it should be in ordinary field theory, by varying all quantities
invoked in the action, up to a surface term. 
It turns out that, for the field symmetries treated hereafter, what we have called formal Lagrangian density appears
as a natural object which is a part of that surface term. 
This is in exact agreement with the appearance
of any ordinary Lagrangian in the computation of a Noether current,
for instance the EMT.

The equation of motion for the field results from the action variation:
\bea
0= \frac{\delta \Skin}{\delta \phi_{1,2,3}} + \frac{\Sint }{\delta \phi_{1,2,3}}
 =- \sum_{s} \Delta_{(s)} \phi_{1,2,3} + m^2 \phi_{1,2,3}+ 
\lambda \int [\prod_{\ell=4}^{6} dg_i] \; \phi_{3,4,5} \,\phi_ {5,2,6}\, \phi_{6,4,1} 
\;,
\label{eqmotion}
\eea
with $\Delta_{(s)}$ being the Laplace operator on the group.
Remark that in order to get (\ref{eqmotion}), we implicitly used an integration by parts
and the fact that the sphere does not have a boundary. 
Furthermore, one should also rename cyclically the group arguments in the interaction 
in order to vary properly this nonlocal term. 

\noindent{\bf Colored GFT -} Colored GFT models 
\cite{Gurau:2009tw}\cite{Gurau:2010nd}
have mainly the same definition as above with the crucial attribute
that fields possess an extra ``color'' index $\phi^a$. We will choose 
them to be complex valued functions. The number
of colors being the number of fields in the interaction. For the Boulatov
colored model, we have $a=1,2,3,4$. More generally for a $D$ dimensional GFT,  the color indices are  $a=1,2,,\dots,D+1$. 
All field properties remain the same as previously and a 
Lagrange density for the $3D$ theory can be given as
\bea
\Lcol &=&  \sum_{a=1}^4 \left[
\sum_{s} \nabla_{(s)}^{i} \;\bar\phi^a_{1,2,3} \nabla_{(s)\;i}\; \phi^a_{1,2,3}
+ m^2  \bar\phi^a_{1,2,3}\;\phi^a_{1,2,3} \right] 
\crcr
&&
+  \lambda \;\phi^1_{1,2,3}\phi^2_{3,4,5} \,\phi^3_ {5,2,6}\, \phi^4_{6,4,1}  +  \bar\lambda \;\bar\phi^1_{1,2,3}\bar\phi^2_{3,4,5} \,\bar\phi^3_ {5,2,6}\, \bar\phi^4_{6,4,1} \;.
\label{colorlag}
\eea
Important quantum topological aspects lie in the ``coloring''
of GFT \cite{Gurau:2010nd}\cite{caravel}. For the present work,
we will indeed see that even at the level of classical analysis,  implementing this extra color index to field might lead 
to an improvement of the features of the Noether currents
for a given symmetry.

\noindent{\bf Solving the tensor Klein-Gordon equation -} 
In \cite{Livine}, treating the Boulatov model, a class of solutions for the 
equation of motion has been found. 
In the present situation, another issue due to the dynamics arises. However
the equivalent of Klein-Gordon equation can be again worked out. 
We have to solve
\bea
-\sum_s \Delta_{(s)} \phi_{1,2,3} + m^2 \phi_{1,2,3} =0\;,
\eea
for gauge invariant fields. 
Using Peter-Weyl decomposition
(see Appendix \ref{app:gaug} for a summary of following notations), the above
equation is equivalent to 
\bea
\sum_{j_a,m_a,n_a} \phi^{j_a}_{m_a,n_a}\left(-\sum_{a=1}^3 C(j_a) + m^2\right) \prod_{a=1}^3 \sqrt{d_{j_a}} D^{j_a}_{m_a,n_a} (g_a) =0  \;,
\eea
with $C(j_a)=j_a(j_a+1)$ denoting the Casimir or eigenvalue of $\Delta_{(a)}$. 
A  solution of the Klein-Gordon equation for $D=3$ GFT is therefore
\bea
\phi_{1,2,3} = 
\sum_{j_a,m_a,n_a} \phi^{j_a}_{m_a,n_a} 
\delta_{\sum_{a=1}^3 C(j_a)-m^2,0} \prod_{a=1}^3 \sqrt{d_{j_a}} D^{j_a}_{m_a,n_a} (g_a) \;,
\label{solu}
\eea
where $\phi^{j_a}_{m_a,n_a}$ is assumed to satisfy also (\ref{boujoum}). 
For large spin $j_a$,  the solutions (\ref{solu}) are such that only
modes $\phi^{j_a}$ with $j_1^2 + j_2^2 + j^2_3=m^2$ 
remain in the field expansion.

\section{Translations and dilatations: 1D GFT}
\label{sect1dNother}

In this section, as a preliminary and essential study to the full picture for any GFT dimension,  we first start by Noether theorem for translations
and dilatations for GFT in one dimension. The latter theory is local  and the analysis in this local framework will be 
 compared to the analogous for a GFT in any dimension 
$D\geq 3$ which is nonlocal. 

In $1D$ GFT,  the gauge invariant condition (\ref{inv}) for fields 
should be abandoned as it is equivalent to the requirement 
of constant fields. The bottom line is the data of an 
action over one copy of $G$ of the form
\bea
\So [\phi]= \int dg\; \Lo (\phi,\nabla\phi)\;,\qquad 
\mathcal L_{1D} = 
\frac{1}{2} \mathbf{g}^{ij} 
\nabla_i \phi(g) \nabla_j \phi(g) + \frac12 m^2 \phi^2(g) 
+ \frac{\lambda}{4} \phi^{4}(g) \;.
\label{act1d}
\eea

\subsection{Translations and EMT}
\label{subsect11:ogft}

A right translation\footnote{Left translations can be carried out in a similar
manner.} symmetry by an element $h$ is simply 
the right group multiplication $g \mapsto gh$. Under this 
symmetry, a field transforms as
\bea
\phi(g) \mapsto \phi(gh)\;.
\eea
At the infinitesimal level, given a local coordinate system, the variation of any field is given by 
\bea
\delta_{X} \,\phi =  X \cdot \partial \; \phi = 
\sum_{i=1}^3 X^i \partial_{i}\,\phi \;.
\label{varia1d}
\eea
The  operator\footnote{In the following, the normalization $1/(2\pi^2)$ of the Haar measure will be dropped.}
\bea
W(X) (\cdot)= \int  d\theta d \varphi^1 d \varphi^2\;
\left(\delta_{X} \mathbf{g}^{ij} \, \frac{\delta }{\delta \mathbf{g}^{ij}}(\cdot) 
+  \delta_{X} \phi \, \frac{\delta }{\delta \phi}(\cdot) \right)
\label{wiopinit}
\eea
acting on the action $\So$ (\ref{act1d}) allows one to define the Noether theorem for
a given symmetry with parameter $X$ for which an infinitesimal field 
variation  $\delta_{X}\phi$ is given.
Operators of the kind (\ref{wiopinit}) prove to be useful tool either 
in the situation of a gauge symmetry (and are indeed related
to Ward identity operators when acting on a partition function), 
or when one deals with nonlocal interaction
as appear in noncommutative geometry or matrix models
\cite{Gerhold:2000ik}-\cite{Hounkonnou:2009qt}.
In the following and according to the context, this operator
will take different forms and will enable us to treat the nonlocal
interaction properly.

Considering  (\ref{varia1d}), one obtains after some algebra 
(Appendix \ref{app:trans:w1D} provides details of the derivations)
\bea
\frac{\partial}{\partial X^i} W(X) \So  =
 - \sum_{k} \int d\theta d\varphi^1d\varphi^2\;\;
\partial_{k} ( \sqrt{|\det\mathbf{g}|}\; \mathbf{g}^{kj} T_{ij})\;,
\eea
where $T_{ij}$ is the EMT given in a covariant form as 
\bea
T_{ij} = \nabla_{i}  \phi \;\nabla_{j} \phi
 - \mathbf{g}_{ij}  \mathcal{L}_{1D} \;.
\label{emtrig1d}
\eea
The properties of the EMT are quite straightforward: 
$T_{ij}$ is symmetric and covariantly conserved. 
Using the equation of motion,
it can be proved (see Appendix \ref{app:trans:w1D}) that
\bea
\nabla^i T_{ij} = 0\;. 
\label{consemt}
\eea
Nevertheless, the sense of conserved charges 
remains unclear at this level. Indeed, there is {\it a priori}
no preferred coordinate embodying the ordinary 
role of time and no obvious partial integration on the remaining variables
for which a correct conserved quantity could be generated from
(\ref{consemt}). 

For a massless theory, the trace of the
EMT (\ref{emtrig1d}) is not vanishing. 
Note that also the usual EMT in a massless $\phi^4_4$ theory is not traceless. 
A traceless EMT can be only built by adding a correction to the original EMT. Here, the naive improvement procedure by adding 
an extra term to the EMT such that
\bea
\hat T_{ij} = T_{ij;\; m=0} + \frac{1}{\beta} (\mathbf{g}_{ij} \nabla^k \nabla_k - \nabla_i \nabla_j) \phi^2 
\eea
 yields still a symmetric tensor but it is neither traceless nor  covariantly conserved (the obstruction of that local conservation can be 
expressed in terms of the Ricci tensor associated with the connection). 
Insisting on the traceless improvement procedure for the EMT (\ref{emtrig1d}), one can perform the following modification:
\bea
\hat T'_{ij} = T_{ij;\; m=0} 
+ \frac{1}{\beta} \mathbf{g}_{ij}  \phi \nabla^k \nabla_k \phi 
+ \frac{1}{\beta'} \nabla_i \phi\nabla_j\phi \;,
\eea
such that $\text{Tr}\; \hat T' =0$ is recovered for
for $\beta' = 2$ and  $\beta=4$. Note that, 
$\hat T'$, even though symmetric, is not covariantly conserved.

\subsection{Dilatations and current vector}
\label{subsect12:dilat}

\noindent{\bf Group dilatations -}
Unlike in the flat and noncompact space case, the notion of dilatation
symmetry on a compact manifold like the sphere is not an obvious concept. We use here an idea familiar 
to wavelet analysis on the two-sphere \cite{antoine}
for discussing the concept of dilatations on the sphere $S^3$.
We will show that these dilatations can be implemented  
for particular GFT models. 

Let $a$ be a real strictly positive number.  Given a group element
$g=g(\theta, \vec n) \in G \simeq S^3$, characterized by the class angle $\theta$ and the unit vector $\vec n \in S^2$,  one defines the map
$d_a: G \to G$ such that $g\mapsto g_a$ with 
\bea
g_a= g_a(\theta_a, \vec n)\;,\qquad \tan \frac{\theta_a}{2} = a \tan\frac{\theta}{2} \;. 
\label{scale}
\eea
More intuitively, the group element $g_a$ can be viewed as follows:
given $g\in S^3$, project $g$ on the tangent $3D$ hyperplane at the North pole 
by a stereographic projection from the South pole; 
apply an usual Euclidean dilatation by $a$ to the projected element in the flat
space and then project back the result onto the sphere $S^3$ by the inverse stereographic
projection. Remark that the stereographic projection is not well
defined at the South pole and this will also  have consequences
in the formulation with some undefined ratios.

Under this mapping, the $\theta$ dependence of the
 Haar measure undergoes
(Appendix \ref{app:dilat:sph} provides justifications of the following
results)
\bea
d\theta (\sin\theta)^2  \mapsto d\theta_a  (\sin\theta_a)^2  = (\mu(a,\theta))^{3} d\theta  (\sin\theta)^2
\;, \qquad
\mu(a,\theta) = \frac{ 2a}{(1-a^2)\cos \theta+ 1+ a^2}
\;.
\label{deter}
\eea
In fact, restricted to the two-sphere,
dilatations of this kind together with translations 
belong to a subgroup of the Lorentz group $SO_0(3,1)$, 
the component of $SO(3,1)$ connected to the identity,
which acts conformally on $S^2$ \cite{antoine}. 
For our situation of the three-sphere,
we foresee that dilatations and translations will reasonably
belong to a subgroup of conformal group acting of $S^3$ \cite{Okuyama:2002zn}.

Discussing infinitesimal variations, the angle $\theta$ transforms as
\bea
\delta_{\epsilon}\theta =  
 2 \arctan  [ (1+\epsilon) \tan\frac{\theta}{2} ]  - \theta
= 2 \epsilon\cos \frac{\theta}{2}\sin \frac{\theta}{2}
= \epsilon \sin \theta\;.
\eea

\noindent{\bf Dilatations and current vector -}
Scale invariance for fields for $1D$ GFT corresponds
to the requirement
\bea
\phi(g) \mapsto \widetilde{\phi}(g_a) = \mu(a,\theta)^{-1} \phi(g)\;. 
\label{dilatation1d}
\eea
Infinitesimally, the above transformation finds the variation
(see (\ref{inftfield1d}) in Appendix \ref{app:dilat:inftsym})
\bea
\delta_\epsilon \phi(g) =-\epsilon [\cos\theta + \sin\theta \partial_\theta]\phi(g)\;.
\label{infini1d}
\eea
One notices that the group field dilatations (\ref{dilatation1d}) might be different from the so-called canonical Weyl transformations considered in \cite{diffeos}.

Considering the infinitesimal generator associated with this transformation (acting on fields)
\bea
{\mathcal D} = \cos\theta + \sin\theta \partial_\theta
\eea
together with translation generators $\partial_j$, we have
\bea
[\partial_j , {\mathcal D} ] = 
\delta_{j\theta} {\mathcal D}' \;,\qquad 
[{\mathcal D}',\partial_j] = \delta_{j\theta}  {\mathcal D} 
 \;,\qquad 
[{\mathcal D} ,{\mathcal D}' ]
= - \partial_\theta \;,\qquad
{\mathcal D}' := (-\sin\theta + \cos\theta \partial_\theta) \;.
\eea
The generator ${\mathcal D}'$ can be seen as a rotation
of ${\mathcal D}$ by an angle of $\pi/2$ and so defines
a generator of a distinct dilatation seen from another pole
(West, up to a sign). 
Hence on the algebra of fields, the translation generator 
$\partial_\theta$, ${\mathcal D}$ and ${\mathcal D}'$ associated each with a different dilatation,
form a closed $\mathfrak{so}(2,1)$ Lie algebra of vector fields. 
This can be seen by first multiplying each generator 
by the complex $i$ and then rename $K_0 = i \partial_\theta$,
$K_1 =i{\mathcal D}$ and $K_2 =i{\mathcal D}'$.
Note also that other translations generators $\partial_{j'}$, $j'\neq \theta$, just
span a central extension to be added to this Lie algebra.

Since the Haar measure transforms according to (\ref{deter}),
a scale invariant action is of the form
\bea
\Sosca[\phi]&=& \int dg
\Big{[} 
 (\sin\theta)^{-1}\frac{\mathbf{g}^{kl}}{2}  
(\partial_{k} \;(\sin\theta\phi))
(\partial_{l}\; (\sin\theta\phi))  + \frac{\lambda}{4} \sin\theta \phi^4\Big{]} \crcr
&=&\int dg \Big{[}  \; 
(\sin\theta)^{-1}\frac{\mathbf{g}^{kl}}{2}
\left\{   \delta_{k,\theta}\delta_{l,\theta}[\cos\theta\phi]^2 
+ 2\delta_{l,\theta} \cos\theta\sin\theta\phi \partial_{k}\phi
 + (\sin\theta)^{2} \partial_{k}\phi \partial_{l}\phi \right\} \crcr
&&\qquad +\frac{\lambda}{4} \sin\theta \phi^4
\Big{]}.
\label{actscale1d}
\eea
It is worth emphasizing that, due to the explicit appearance of 
the coordinate $\theta$ in the Lagrangian, we expect a breaking of the ordinary notion of local conservation of current in this theory. 
Note also that a  mass term could be also included but, 
for simplicity, we will not consider it.

The field equation of motion reads
\bea
&&
\frac{\delta \Sosca }{\delta \phi} =0
=  
(\bullet)\frac{ (\cos\theta)^2}{\sin\theta} \phi
+ (\bullet)
 \cos\theta \partial_{\theta} \phi 
-  \partial_{\theta} [  (\bullet)\cos\theta \phi ] 
 - \widetilde{\Delta} \phi 
+ (\bullet)\lambda \sin\theta\phi^3\;,
\label{eqscale1d}\\ 
&&(\bullet) := \sqrt{|\det \mathbf{g}|} \;,\quad
\widetilde{\Delta}\phi:= \partial_{k}\left\{(\bullet)  
\sin\theta \mathbf{g}^{kl}\partial_{l}\phi\right\}  ,
\nonumber
\eea
where $\widetilde{\Delta}$ is a modified Laplacian. 
The metric variation will be not considered this time and
rather consider the functional operator for solely field dilatations given by
\bea
W(\epsilon) (\cdot)= \int  d\theta d\varphi^1d\varphi^2 \; 
\;
 \delta_{\epsilon} \phi \, \frac{\delta }{\delta \phi}(\cdot) \;.
\label{wiopdilat1d}
\eea
We have to evaluate the variation of the action under  (\ref{wiopdilat1d})
\bea
&&
\frac{\partial }{\partial \epsilon}
W(\epsilon) \Sosca = \frac{\partial }{\partial \epsilon_i}
\int  d\theta d\varphi^1d\varphi^2
\Big{\{ } \left(  -\epsilon  \mathcal{D} \phi  \right) \times \crcr
&& \Big{[} 
(\bullet)\frac{ (\cos\theta)^2}{\sin\theta} \phi
+ (\bullet)
 \cos\theta \partial_{\theta} \phi 
-  \partial_{\theta} [  (\bullet)\cos\theta \phi ] 
 - \widetilde{\Delta} \phi 
+ (\bullet)\lambda \sin\theta\phi^3 \Big{]}\Big{\}} 
\eea
and will prove that this can be computed as a surface
term.
A direct, though lengthy, calculation (see Appendix \ref{app:dilat:1dgft}) yields the current
\begin{equation}
D_j = \sin\theta\,\left[ \cos \theta  + \sin \theta\, \nabla_{\theta} 
  \right] \phi\;  \nabla_{j} \phi     
+\mathbf{g}_{j\theta} \cos\theta \phi \left[ \cos\theta +\sin\theta \nabla_\theta   \right] \phi
-\mathbf{g}_{j\theta} 
\sin\theta \Losca\,, 
\end{equation}
that we put in another form
\bea
D_j = 
\nabla_{\theta} (\sin\theta \phi)\nabla_{j} (\sin\theta\phi)     
-\mathbf{g}_{j\theta} 
\sin\theta \Losca\;.
\label{dilatationcurrent1d}
\eea
Concerning the local conservation property,  as expected, we find that 
the current is not covariantly conserved 
 (a proof of this can be found in Appendix \ref{app:dilat:1dgft}).
The breaking term for the covariant conservation to hold can be written
as
\bea
\nabla^j D_j &=&  \cos \theta    \sin\theta \;
\Big[ -(\cot\theta)^2 \phi^2  
 + 
 \nabla_{\theta} \phi\;  \nabla_{\theta} \phi 
+\frac{\lambda}{2} \phi^4 
\Big] \crcr
&=&
  2\cos \theta  \;
\Big[ -\frac12 \frac{(\cos\theta)^2 }{\sin\theta}\phi^2  
 + 
    \frac12\sin\theta\nabla_{\theta} \phi\;  \nabla_{\theta} \phi 
+\frac{\lambda}{4} \sin\theta\phi^4 
\Big].
\label{break}
\eea
The breaking expression comes mainly from partial derivative
in $\theta$ of factors containing an explicit $\theta$
dependence in the initial Lagrangian $\Losca$. A non-trivial task,
going beyond the scope of this paper, is to understand the breaking (\ref{break})
in terms of coordinate dependent regular Lagrangian systems \cite{carinena}. 

As a remark, from the expression (\ref{dilatationcurrent1d}),
one could be tempted to argue 
that the current $D_j$ should be related to an EMT $\widetilde{T}_{ij}$ by just a field  redefinition
$
\phi \to \widetilde{ \phi } = \sin\theta \phi .
$
Then one ought to check that $\sin\theta \Losca$
is the correct Lagrangian of the form 
$
\widetilde{ \Lo}  = 
(1/2)\mathbf{g}^{ij} 
\nabla_i  \widetilde{ \phi }  \nabla_j  \widetilde{ \phi } 
+ (\lambda/4)  \widetilde{ \phi } ^{4}
$
so that the EMT in this theory can be related to
the current by  $\widetilde{T}_{\theta j} = D_j$
and thereby $\nabla^j \widetilde{T}_{\theta j} =\nabla^j  D_j
 = 0$ should reasonably hold. We then compute
\bea
\sin\theta \Losca 
 = \frac{\mathbf{g}^{kl}}{2}
 \partial_{k}\widetilde{ \phi } \partial_{l}\widetilde{ \phi }
+\frac{\lambda}{4} (\sin\theta)^{-2}\widetilde{ \phi }^4 
\eea
and clearly find that $\sin\theta \Losca  \neq \widetilde{ \Lo} $. 
The new Lagrangian $\sin\theta \Losca$ as a function 
of $\widetilde{\phi}$ and $\theta$ is not translation invariant
even up to a surface term.
This means that dilatation invariance
cannot be reduced, at least in that naive way, to translation invariance.

Finally, in the ordinary $\phi^4$ theory, the dilatation
current $d_j$ can be related to the improved local and traceless
energy momentum tensor by $d_\mu = x^\nu \hat{T}_{\nu \mu}$,
where $\hat{T}_{\nu\mu} = T_{\nu\mu} + (1/6)(\eta_{\mu\nu}
\partial^\kappa \partial_k - \partial_\nu \partial_\mu)\phi^2$, 
$T_{\nu\mu}$ being the ordinary EMT and  $\eta$ 
the Minkowski metric. Here trying to obtain an analogous
relation, one may write at best
\bea
D_j =
 \sin\theta\, T^{\text{scale}}_{\theta j} + 
\sin\theta\cos \theta \phi\;  \nabla_{j} \phi     
+\mathbf{g}_{j\theta} \cos\theta \phi \left( \cos\theta +\sin\theta \nabla_\theta   \right) \phi\,,
\eea
and,  remarkably, $\sin\theta$ plays the old role of the coordinate position in flat spacetime.

\section{Translations and dilatations: 3D case}
\label{sect2}

\subsection{Translations and EMT}
\label{subsect21:ogft}

Right translations conserve the sense of Section \ref{subsect11:ogft}.
In a higher rank tensor GFT, each group argument can be translated by a fixed quantity, but fields will transform according to a particular rule. 
It has been highlighted recently that diffeomorphisms 
can be implemented for GFT models through quantum deformed
symmetries. The deformed Poincare group 
acts on the base manifold defined by the Lie
algebra dual to the group \cite{Girelli:2010ct}. 
Translations in Lie algebra or ``metric'' variables have been scrutinized
by Baratin {\it et al.} still in the form of a quantum symmetry
realization \cite{Baratin:2011tg}. For colored GFT models in $3D$ and $4D$, quantum translation invariance corresponds to the invariance under translations of the vertices of the Euclidean tetrahedron representing the GFT interaction. 
The same symmetry reflects, in the group formulation,
the fact that the boundary connection associated with 
field variables is flat. Even though the fields used in this
section are non colored and so the geometric interpretation 
of translations will certainly differ, the way of implementing 
group translations here (by acting one some particular field
arguments) is somehow similar to the anterior formalisms.

First, we need to introduce complex valued fields
 and consider the new Lagrangian density $\Ltri=\Lkin+\Lint$
with
\bea
\Lkin&=& 
\sum_{s=1}^3\mathbf{g}_s^{ij}\nabla_{(s)\;i} \;\bar \phi_{1,2,3}
\nabla_{(s)\;j}\; \phi_{1,2,3}
+m^2  \bar\phi_{1,2,3}\;\phi_{1,2,3}\;,\\
\Lint&=&
\frac{\lambda}{2}\phi_{1,2,3} \bar\phi_{5,4,3} \phi_{5,2,6}\bar\phi_{1,4,6} \;.
\label{denslagint}
\eea
The action defined by 
\begin{equation}
\Skin[\phi]:= \int [\prod_{\ell=1}^{3}dg_\ell]\Lkin \;,\qquad
\Sint[\phi] = \int 
[\prod_{\ell=1}^{6}dg_\ell]\;\Lint \;,
\label{noncolaction}
\end{equation}
is real and may be written $\Stri =\Skin[\phi] +\Sint[\phi]  := \int 
[\prod_{\ell=1}^{6}dg_\ell]\; \Ltri$.
The equation of motion for fields $\phi_{1,2,3}$ can be inferred from (\ref{eqmotion})
after renaming  properly some variables  in the interaction. 

Under right translations by $h_\ell$, $\ell=1,2,3$,  
the fields transform according to 
\bea
&&
\phi(g_1,g_2,g_3) \mapsto \phi(g_1h_1, g_2h_2, g_3h_3)\;, \crcr
&&
\bar\phi(g_5,g_4,g_3) \mapsto \bar\phi(g_5h_1, g_4h_2, g_3h_3) \;,\crcr
&&
\phi(g_5,g_2,g_6) \mapsto \phi(g_5h_1, g_2h_2, g_6h_3) \;,\crcr
&&
\bar\phi(g_1,g_4,g_6) \mapsto \bar\phi(g_1h_1, g_4h_2, g_6h_3) \;.
\label{rtrans4}
\eea
Note that the group arguments with labels $(1,5)$, $(2,4)$ 
and $(3,6)$ are shifted by the same amount.  This defines a correct
field symmetry.  Actually, there is a simpler field transformation
which can be extracted from the above mappings 
by just shifting either the first, the second or the last field argument. 
Thus an ordinary $D$ dimensional GFT will have
$D$ such basic translations.  The results and conclusions 
obtained with the ``3-translation'' (\ref{rtrans4}) are in some sense more general and will again hold for any of these simpler symmetries.

Infinitesimal variations of a tensor field can be inferred from the variations 
of a field defined over a single copy of the group.
Three translations defined by the group elements $h_\ell$, $\ell=1,2,3,$
yield the infinitesimal variations
\begin{equation}
\delta_{X_{(1)},X_{(2)},X_{(3)}} \,\phi_{1,2,3} = \sum_{s} \; X_{(s)} \cdot \partial_{(s)} \; \phi_{1,2,3} = \sum_{s,i}
 X_{(s)}^i \partial_{(s)\; i}\,\phi_{1,2,3} \;.
\label{variatensor}
\end{equation}
Henceforth, $\delta_{X_{(1)},X_{(2)},X_{(3)}}$ will be simply denoted $\delta_{X}$. 
The following operator
\bea
W(X) (\cdot)= \int [\prod_{\ell=1}^6 d\theta_\ell d \varphi_\ell^1 d \varphi_\ell^2] \;
\left[ \sum_{s=1}^6 \delta_{X} \mathbf{g}_s^{ij} \, \frac{\delta }{\delta \mathbf{g}_s^{ij}}(\cdot) 
+  \delta_{X} \bar\phi_{1,2,3} \, \frac{\delta }{\delta \bar\phi_{1,2,3}}(\cdot) +  \delta_{X} \phi_{1,2,3} \, \frac{\delta }{\delta \phi_{1,2,3}}(\cdot) \right]
\label{wiop}
\eea
generalizing (\ref{wiopinit}) will act
again on the action $\Stri$ in order to find the Noether current 
for translation symmetry with parameters $X_{(s)}$.

Considering  (\ref{variatensor}), some calculations yield
(see  Appendix \ref{app:trans:calc3D})
\begin{equation}
\frac{\partial}{\partial X_{(s)}^i} W(X) \Stri  =
 - \sum_{s',k} \int [\prod_{\ell=1}^6 d\theta_\ell d\varphi_\ell^1d\varphi_\ell^2] \;\;
\partial_{(s')\;k} (  [\prod_{\ell=1}^6 \sqrt{|\det\mathbf{g}_\ell|}]\; \mathbf{g}^{kj}_{s'} T_{(s,s');\, (i,j)})\;,
\end{equation}
where $T_{(s,s');\, (i,j)}$  is the ``stranded'' EMT given by
\begin{equation}
T_{(s,s');(i,j)} =\partial_{(s)\;i}  \phi_{1,2,3} \,
\partial_{(s')\;j} \bar\phi_{1,2,3} +
\partial_{(s)\;i}  \bar\phi_{1,2,3} \,
\partial_{(s')\;j} \phi_{1,2,3}  
-\delta_{s,s'}\mathbf{g}_{s'\;ij}  \Ltri
- \delta_{s+[\alpha_{s}],s'}\mathbf{g}_{s'\;ij} \Lint\,,
\label{emtrig}
\end{equation}
with indices such that $s=1,2,3$, $s'=1,2,\dots,6$ and $([\alpha_1],[\alpha_2], [\alpha_3])=(4,2,3)$.
In any dimension $D$, the EMT will hold this form, the strand indices
will become $s=1,2,\dots,D$ and $s'=1,2,\dots,D(D+1)/2$ whereas
the indices $[\alpha_s]$ have to be combinatorially well chosen. 
For a basic field transformation acting, for example, 
only on the strands $(1,5)$, the corresponding EMT is nothing 
but the component $T_{(1,s');(i,j)} $.

First, the EMT (\ref{emtrig}) is symmetric under the permutation $(s,i) \leftrightarrow(s',j)$, 
if $s'=1,2,3$. For $s'=4,5,6$, the EMT breaks down to the interaction Lagrangian and therefore this component remains also symmetric. However, the EMT (\ref{emtrig}) turns out to be not covariantly 
conserved (see Appendix \ref{app:trans:calc3D}).
This is mainly due to the fact that the nonlocal interaction
clashes with the specific way that the field symmetry is imposed
in (\ref{rtrans4}) (we will come back to this fact in depth in the next
paragraph). Such an oddity prevents to form proper equations of motion for fields thanks to which, usually, the local conservation
could be guaranteed.  
 It could be asked if removing the dynamical
part in the definition of the Lagrange density will not help 
to recover the conservation of the EMT. 
It can be shown, in that particular instance, 
that the EMT reduces to the Lagrangian itself,
and in the non colored case, this quantity is still not locally conserved. The local conservation breaking of the EMT and, in fact, of all other currents as we will find in the subsequent analysis, 
has a deeper reason to hold. 

Let us digress a little from our present purpose
and understand what is the main reason why
nonlocal field theories under a group 
of transformations will generally fail to obey
the Noether theorem.  
The local conservation of currents by Noether theorem 
assumes different features of the initial theory. 
For the sake of clarity, let us consider a theory with local fields 
described by an action $S[\phi,\partial \phi]$ 
which computes up to surface term $\partial_{\mu} T^{\mu}$
under a group of field symmetry $Q$
\begin{equation}
Q \triangleright S[\phi,\partial \phi]  = \int_{\Omega} \partial_{\mu} T^{\mu} =0\;.
\label{loc}
\end{equation}
The last equation is valid on-shell and  the integration domain $\Omega$ should be any subspace of the domain $M$ of fields $\phi: \Omega \subset M \to \mathbb{C}$. From the arbitrariness of $\Omega$, one concludes to
the local continuity equation: 
\begin{equation}
\partial_{\mu} T^{\mu} =0\;.
\end{equation}
However, there may exist nonlocal theories having an action
$S^{\text{nonlocal}}[\phi,\partial \phi]$
invariant under a group of transformations, the surface term 
can be still computed but now the equation (\ref{loc}) may be only valid
on the entire domain of the fields, namely
\begin{equation}
Q \triangleright S[\phi,\partial \phi]  = \int_{M} \partial_{\mu} T^{\mu} =0\;,
\end{equation}
then nothing else can be said about the local conservation
properties of the current $T^{\mu}$.
In general, the corresponding current is not locally conserved. 
Nevertheless, depending on the structure of the field theory, 
the current could appear by itself a locally conserved quantity
or one may use different tools in order to regularize or to improve the properties of that current. 
This last case includes, again,  the so-called noncommutative
field theories defined over the algebra of fields equipped
with a Moyal star products \cite{Douglas:2001ba}: the noncommutative
tensor $i\theta^{\mu\nu}=[x^\mu,x^{\nu}]_{\star}$ can be chosen 
in a specific way to extract from the initial non locally
conserved EMT of the scalar $\phi^{4}_{\star}$ theory,
 a conserved four-momentum. Obviously, the success 
of getting the relevant local properties of currents
in these kinds of theories is strongly dependent on 
the type of framework one is dealing with. 

In the present setting of nonlocal GFTs, we are in a similar situation. 
Indeed, let us scrutinize the translations given in (\ref{rtrans4})
by reducing them to the minimal translation, for instance $(1,5)$.
This transformation acts on the same field on two distinct
points of the manifold: 
it can be called a ``nonlocal translation''. 
The  associated surface term is computed using the
structure of, at least, two Haar integrals over, at least, two group copies 
in order to vary properly the nonlocal interaction (involving two different 
group variables $g_1$ and $g_5$; in fact, the computations 
involve all the six copies of the group).  
As a simple consequence, an ordinary local conservation which 
occurs on a local group copy of the EMT will be explicitly broken. An improvement procedure, if exists,
would be necessary in order to recover that local property.
Note that, as shown in Appendix (\ref{app:trans:calc3D}), 
integrating the local conservation breaking term
on the full manifold vanishes again. 
By this, one is ensured
of the validity of a vanishing divergence on the entire
manifold as the calculation of the surface term claims
this fact. Nevertheless, we will see that for colored GFTs, 
the EMT appear locally conserved because, in comparison
to the above nonlocal translation, in colored GFT
one can implement ``local translation'' (the meaning
of these will become clear in the following).
This confirms the point of view that 
the local conservation of the currents in nonlocal field
theories becomes actually model dependent. 

Let us investigate alternatively the meaning of group 
translations for fields invariant under permutation of their 
arguments and understand if this extra feature could not
improve the currents. 
Hence, for any permutation $\sigma \in \Sigma_3$,
we assume
$
\phi_{1,2,3} = \phi_{\sigma(1),\sigma(2),\sigma(3)}.
$
We can think about the consequences of having such an invariance
on translations in a combinatoric way. 
We focus
on a translation with respect to one field argument, say
$g_1\to g_1 h$. $g_1$ appears twice in the interaction,
 in  $\phi_{1,2,3}$ and in $\bar \phi_{1,4,6}$, and will be 
 translated in both of these fields.
The same translation remains to be defined on two
 fields: $\bar\phi_{5,4,3}$ and $\phi_{5,2,6}$.
The transformation could only affect $g_{5,4}$ in the former
and $g_{5,6}$ in the latter field (otherwise, $g_2$ or $g_3$
will be assumed to be transformed and then the situation become
more nonlocal with the, by now, expected features). Then, two transformations
could be consistently implemented: one of which is the minimal translation 
involving $(1,5)$ as performed earlier and a second one defined by
\bea
&&\phi(g_1,g_2,g_3) \to \phi(g_1h,g_2,g_3)\;, \qquad
\bar \phi(g_1,g_4,g_6) \to \bar\phi(g_1h,g_4,g_6)\;,\crcr
&&\bar \phi(g_5,g_4,g_3) \to \bar\phi(g_5,g_4h,g_3)\;, \qquad
\phi(g_5,g_2,g_6)  \to  \phi(g_5,g_2,g_6 h) \;.
\label{tord}
\eea
As peculiar as it appears, 
 considering (\ref{tord}) as a valid field translation,
 one agrees that it is even more nonlocal than the minimal $(1,5)$
translation. The transformation acts on the same field 
on three distinct points. It becomes obvious that all previous broken currents will remain broken under this transformation. 
Otherwise, if one disputes the fact that $g_4$ and $g_6$
have been translated and then should be translated 
in the remaining terms, a rapid checking shows that 
the said translation is equivalent to a translation 
of all field arguments by a unique element.  This transformation
is nothing but the identity from  
the gauge invariance of the fields. Thus, considering 
fields invariant  under permutation of their arguments leads to the same
conclusions as given in the above analysis or to triviality. 
Nevertheless, we can comment that, through
 any new meaning that one could give to 
``nonlocal translations,'' that we did not investigated 
in depth in this paper focusing on the most nontrivial  
situations, it is not excluded that properties of currents
can be improved using new types of transformations.

Next, for a massless theory, let us evaluate the trace of the
EMT (\ref{emtrig})  in the following sense:
\begin{equation}
\text{Tr}\; T_{m=0} =
\sum_{s=1}^3 \left[
\mathbf{g}_{s}^{ij} T_{(s,s);(i,j);\; m=0}  +
\mathbf{g}_{s+[\alpha_{s}]}^{ij} T_{(s,s+[\alpha_{s}]);(i,j);\; m=0} 
\right], 
\label{surtrace}
\end{equation}
where $T_{(s,s');(i,j);\; m=0} $ is defined from $\mathcal{L}_{3D,m=0}$.
A trace of this form is justified by the fact that a contribution
for each strand represented in the Lagrangian is needed. 
The calculations of this trace yield (in covariant notations)
\bea
\text{Tr}\; T_{m=0} 
=\sum_{s=1}^3 
[\nabla_{(s)}^{i}  \bar\phi_{1,2,3} \;
\nabla_{(s)\;i} \phi_{1,2,3} ] -9(\mathcal{L}_{3D,m=0} +\Lint) 
\eea
which is not a vanishing quantity. 
A traceless EMT can be built by considering instead
\bea
\hat T_{(s,s');(i,j)}  = T_{(s,s');(i,j);\; m=0} + 
\frac{1}{\beta}
\delta_{s,s'}\mathbf{g}_{s'\;ij} \sum_{s''=1}^3 \phi_{1,2,3} \Delta_{(s'')} \bar\phi_{1,2,3}
+ \frac{1}{\beta'}\nabla_{(s)\;i} \bar\phi_{1,2,3} \,\nabla_{(s')\;j} 
\phi_{1,2,3}\;.
\label{emttrace}
\eea
The trace of this tensor is 
\bea
\text{Tr}\; \hat{T} =
\frac{-8\beta' +1}{\beta'}
\sum_{s=1}^3
\nabla_{(s)}^{i} \bar\phi_{1,2,3}  \nabla_{(s)\;i} \phi_{1,2,3} 
 - 9\lambda \phi_{123}\bar\phi_{543}\phi_{526}\bar\phi_{146} 
+\frac{9 }{\beta}\sum_{s=1}^3  \phi_{1,2,3} \Delta_{(s)} \bar\phi_{1,2,3}.
\label{traceless}
\eea
The improved EMT $\hat{T}$ is traceless if $\beta' = 1/8$ and  $\beta=1$ after integration of the variables coined by $4,5,6$.
We should emphasize that 
$\hat T$ is not covariantly conserved. 

In general, a traceless property of a locally conserved 
EMT hints at a scale invariant theory. 
Scale or dilatation symmetry is an important
feature of any field theory because it preludes,
in certain cases, to a larger conformal symmetry.
The latter has a cortege of physical implications
among which rests, for example, the existence
of universality classes having for fixed point
a particular conformal theory \cite{DiFrancesco:1997nk}.    
Hence, before demanding conformal invariance for GFT,
one could investigate if simpler symmetries can be implemented
at this level. A way to test if a theory is scale invariant, is
to prove that its conserved EMT is traceless. 
Showing that the EMT is not locally conserved and not traceless\footnote{Of course, one of these features could have been 
a sufficient condition for claiming that the dilatation symmetry 
of the ordinary $\phi^4_4$ is explicitly broken. 
However, we have shown that the lack of traceless property of the EMT
can be improved hence only the first
condition should be considered as the actual reason for 
the breaking of scale symmetry. },
we anticipate the fact that GFT as given by the action (\ref{noncolaction}) will be not scale invariant and hence the ordinary dilatation (and so the larger conformal) symmetry of $\phi^4_4$ theory will be explicitly broken in GFTs. This is again due to the nonlocality of these theories.

\subsection{Dilatations and current tensor}
\label{sect3}

\noindent{\bf Boulatov-Ooguri model -}
We first study GFTs without dynamics in order to make the following
developments more comprehensible. Consider a $3D$ GFT model equipped  
with a quadratic part 
$
\mathcal{L}^0_{\text{kin}}[\phi]:=
\bar\phi_{1,2,3}\phi_{1,2,3}
$
and an interaction part as given by (\ref{denslagint}).
The total Lagrange density is $\mathcal{L}^{0}$ 
and we put the mass to $m=1$. 

Demanding scale invariance of the action implies that the fields transform as
\bea
\phi(g_1,g_2,g_3) \mapsto \widetilde{\phi}(g_{a\;1},g_{a\;2},g_{a\;3}) = 
[\prod_{s=1}^3 \mu(a_s,\theta_s)^{c_s}] \phi(g_{1},g_{2},g_{3})\;.
\eea
Given (\ref{deter}), the scale invariance of the interaction
(quartic in fields) and trivial kinetic term is achieved by $3 + 2c_s =0$.
Thus the scaling dimension for fields is characterized by $c =-3/2$. 
Furthermore, as it was the case for translation symmetry, 
we use complex fields and require
that group elements defined by the couples $(1,5)$, $(2,4)$ and $(3,6)$ are
all submitted to the same dilatations.

Setting $a_s=1+\epsilon_s$, where $\epsilon_s$ is an infinitesimal parameter,
angle and field infinitesimal variations (Appendix \ref{app:dilat:inftsym} 
gives useful details pertaining to the following identities) are of the form:
\bea
&& \delta_{\epsilon_s}\theta_s =   \epsilon_s \sin \theta_s \;,\crcr
&& 
\delta_{\epsilon} \phi_{1,2,3} 
:=\delta_{\epsilon_1,\epsilon_2,\epsilon_3} \phi
= \sum_{s=1}^3 \delta_{\epsilon_s} \phi_{1,2,3}
 = -\sum_{s=1}^3 
\epsilon_s\bigl( \,-c\cos \theta_s + \sin \theta_s \partial_{(s)\;\theta} \,
\bigr)\phi_{1,2,3} \;.
\eea
We introduce the functional differential operator
\bea
W(\epsilon) (\cdot)= \int [\prod_{\ell=1}^3 d\theta_\ell d\varphi_\ell^1d\varphi_\ell^2] \;
\;\; \Big{[} \delta_{\epsilon} \phi_{1,2,3} \, \frac{\delta }{\delta \phi_{1,2,3}}(\cdot) +\delta_{\epsilon} \bar\phi_{1,2,3} \, \frac{\delta }{\delta \bar\phi_{1,2,3}}(\cdot)  \Big{]} 
\label{wiopdilat}
\eea
and evaluate for the action $S^0=\int [\prod_\ell dg_\ell] \mathcal{L}^0$
\bea
&&
\frac{\partial }{\partial \epsilon_i}
W(\epsilon) S^0 = \frac{\partial }{\partial \epsilon_i}
\int [ \prod_{\ell=1}^6 d\theta_\ell d\varphi_\ell^1d\varphi_\ell^2]
 \prod_{s=1}^6  \sqrt{|\det \mathbf{g}_s|} \Big{\{ }\crcr
&& \left[ - \sum_{s=1}^3 
\epsilon_s\bigl[ \,-c\cos \theta_s  + \sin \theta_s \partial_{(s)\;\theta} \,
\bigr]\phi_{1,2,3}  \right] \left[\bar\phi_{1,2,3}  + 
\lambda \bar\phi_{5,4,3}\phi_{5,2,6} \bar\phi_{1,4,6} \right] 
+ \fifib\Big{\} }\;.
\eea
In the last above line, $\fifib$ is the symmetric of the previous
expression  under complex conjugation. 
The expansion of the last line yields 
\bea
&&
\frac{\partial }{\partial \epsilon_i}
W(\epsilon) S^0 = 
-\int [ \prod_{\ell=1}^6 d\theta_\ell d\varphi_\ell^1d\varphi_\ell^2] \crcr
&& 
\left\{   \partial_{(i)\;\theta} \left[  
[\prod_{s=1}^6 \sqrt{|\det \mathbf{g}_s|}] \sin \theta_i \mathcal{L}^0 \right] 
+  
\partial_{(i+[\alpha_i])\;\theta} \left[  
[\prod_{s=1}^6 \sqrt{|\det \mathbf{g}_s|}] \sin \theta_{i+[\alpha_i]} \Lint\right]  \right\}.
\eea
The dilatation current vector becomes a ``reduced'' and 
``stranded'' quantity
with two components  
\bea
D_s = \sin \theta_s \mathcal{L}^{0} \,, \qquad \tilde{D}_{s} 
= \sin \theta_{s+[\alpha_s]}  \Lint\;.
\eea
This fact merely comes from the absence of a true dynamics in the model.
In this ``trivial'' situation, the EMT reduces to the 
Lagrangian itself plus the interaction again, namely $T^{0}= (T^0_1,T^0_2)=(-\mathcal{L}^{0}, -\Lint)$.
From this point, one infers a slightly generalized formula 
for dilatation current in Boulatov-Ooguri tensor theory in terms of 
\bea
D_s = -\sin\theta_{s} T^{0}_{1}, \qquad 
\tilde{D}_s = - \sin \theta_{s+[\alpha_s]}  T^{0}_{2}
\eea 
where $\sin\theta_s$ should be seen as the coordinate position. Properties of the EMT $T^{0}$ 
and the current $D_s$ are direct: they 
are not locally conserved.

\noindent{\bf Dynamical GFT -}
Incorporating nontrivial dynamical part, we have to define
a correct scaling of derivative on fields. Consider a kinetic
part of the form
\bea
\Skinsca[\phi]&=& \int [\prod_{\ell=1}^{3}dg_\ell]\ \Lkinsca \crcr
\Lkinsca &:=&
 \sum_{s=1}^3 (\sin\theta_s)^{\gamma_s}\mathbf{g}_s^{ij} 
(\nabla_{(s)\;i} \;(\sin\theta_s)^{\beta_s}\bar\phi_{1,2,3})
(\nabla_{(s)\;j}\; (\sin\theta_s)^{\beta_s}\phi_{1,2,3}) \;,
\label{dilaction} 
\eea 
where the degrees $\gamma_s,\beta_s$ have to be chosen in order to satisfy
the scale invariance. It can easily inferred that $\gamma_s=\gamma = -1$ 
and $\beta_s = \beta=3/2$. The interaction part remains
as $\Sint$ and $\Lint$ so that $\Strisca = \Skinsca + \Sint$
and $\Ltrisca = \Lkinsca + \Lint$.

A direct evaluation (see Appendix \ref{app:dilat:curr}) shows that the 
dilatation current is a tensor defined by 
\bea
&&
D_{(s,s');j} = \crcr
&&
\sin \theta_{s} \Big\{
 \sin\theta_{s'}^{\frac12} \Big{[} \partial_{(s)\,\theta}\phi_{1,2,3}  \,
 \partial_{(s')\,j} (\sin\theta_{s'}^{\beta}\bar\phi_{1,2,3} )
+  \partial_{(s)\,\theta}\bar\phi_{1,2,3} \, 
 \partial_{(s')\,j} (\sin\theta_{s'}^{\beta}\phi_{1,2,3} ) \Big{]}\\
&&
- \delta_{s,s'}\mathbf{g}_{s'\, j\theta} \Ltrisca
\Big\}  
- \delta_{s+[\alpha_{s}],s'}\mathbf{g}_{s'\; j\theta} \,  \sin\theta_{s+[\alpha_{s}]} \Lint
+\beta  \cos \theta_{s}  \partial_{(s')\;j}    \left( 
  (\sin\theta_{s'})^2\bar\phi_{1,2,3}\phi_{1,2,3} \right).
\nonumber
\eea
This tensor is not covariantly conserved and its breaking
involves both the nonlocal interaction and the fact
that the Lagrangian contains explicit coordinate dependence.

Note that there exist other types of GFT interactions 
which are scale invariant under (\ref{scale}).
For instance, the following interaction
\bea
\tilde{S}_{\text{int}}[\phi]:=\frac{\lambda}{4} \int [\prod_{\ell=1}^{4}dg_\ell]\ \phi_{1,2,3}\,
\phi_{3,2,4}\,\phi_{3,4,1}\,\phi_{4,2,1}\;,
\label{inter}
\eea
assigns each group variables $g_i$ (appearing three times in the
interaction) to a vertex in the tetrahedron.
This vertex is indeed shared by three triangles, each triangle
being represented by a field. 
Hence, this model should be equivalent to a colored GFT model.
A straightforward inspection using (\ref{inter}) proves
that the scaling dimension of the fields is such that $c=-\beta_s=-1$ so that a kinetic
term of the form (\ref{dilaction}) with $\gamma_s=-1$ would be
scale invariant. Remark that the problem of non locally conserved 
quantities will be not necessarily solved by considering these interactions. 

The pattern followed by the field arguments in the 
interaction (\ref{inter}) and the pattern of the vertex of Carrozza and Oriti \cite{Carrozza:2011jn} are exactly the same. In the latter, 
the field colors are explicitly given. 
In fact, it can be shown that the assignment of 
a group variable to each vertex of a tetrahedron
and a particular pattern should be enough 
to ensure the equivalence between the said vertex and a colored vertex. 
The proof of this statement can be given
as follows: let us consider the following interaction (we change
once again the vertex in order to recover a typical colored 
theory)
\bea
\tilde{S}^{\text{col}}_{\text{int}}[\phi]:=\frac{\lambda}{4} \int [\prod_{\ell=1}^{4}dg_\ell]\ \phi_{2,3,4}\,
\phi_{1,4,3}\,\phi_{2,1,4}\,\phi_{3,2,1}\;.
\label{inter2}
\eea 
To each field $\phi_{abc}$, $a,b,c \in \{1,2,3,4\}$, 
we assign another redundant index $d \in \{1,2,3,4\}\setminus
\{a,b,c\}$. After the procedure, the same vertex can be recast as
\bea
\tilde{S}^{\text{col}}_{\text{int}}[\phi]:=\frac{\lambda}{4} \int [\prod_{\ell=1}^{4}dg_\ell]\ \phi^1_{2,3,4}\,
\phi^2_{1,4,3}\,\phi^3_{2,1,4}\,\phi^4_{3,2,1}\;.
\eea
Assign, once again, to each field variable
the redundant index of the same field. 
This procedure  yields a colored vertex 
in the sense of Gurau \cite{Gurau:2009tw}. At the
quantum level, the gluing rules between these new colored
fields can be imposed as one may require
(for instance, only fields with the same redundant index
have to be glued). Conversely, given a colored  model, 
in order to obtain, the model described by (\ref{inter2}),
affect each color to a vertex of a tetrahedron, then
remove the index from the vertex (the said index is indeed
 redundant). The model
obtained by this process coincides with (\ref{inter2}).

\section{Translations and dilatations: Colored GFT}
\label{sect:coloredsym}

\subsection{Translations and EMT}
\label{sect:coloredtranslation}

We consider now a colored GFT with Lagrangian of the form (\ref{colorlag}). 
Due the freedom of having colored fields, 
a right translation for only the fields $\phi^1$ and $\phi^4$ 
can be defined such that
\bea
\phi^1(g_1, g_2,g_3) \mapsto  \phi^1(g_1h, g_2,g_3)\;, \crcr
\phi^4(g_6, g_4,g_1)  \mapsto  \phi^4(g_6, g_4,g_1h)\;, 
\label{colortrans}
\eea 
whereas the fields of color $2,3$ remain non modified.
At the infinitesimal level, (\ref{colortrans}) gives
\bea
\delta_{X}\phi^{a=1,4} = (X^i \cdot  \partial_{(1)\;i})\, \phi^{a=1,4}\;,
\eea
where $\partial_{(1)}$ refers to only to derivative with respect to 
the strand $1$ involving the group element $g_1$. 

 The coloring of GFT defines the ``minimal'' symmetry for the GFT action in the sense that we can  transform only 
one group argument. This feature both simplifies somehow the 
derivations and, in fact, radically differs from the above translations for 
non colored GFTs as we will see. In order to recover the full symmetry of the action,
one simply has to identify  pairs of field arguments
which can be transformed independently. Thus, besides of (\ref{colortrans}), other possible field transformations are
\bea
&&
\phi^1(g_1, g_2,g_3) \mapsto  \phi^1(g_1, g_2h,g_3)\; \qquad
\text{and} \qquad
\phi^3(g_5, g_2,g_6)  \mapsto  \phi^3(g_5, g_2h,g_6)\;, \crcr
&&
\phi^1(g_1, g_2,g_3) \mapsto  \phi^1(g_1, g_2,g_3h)\;\qquad
\text{and} \qquad
\phi^2(g_3, g_4,g_5)  \mapsto  \phi^2(g_3h, g_4,g_5)\;, \crcr
&&
\phi^2(g_3, g_4,g_5) \mapsto  \phi^2(g_3, g_4h,g_5)\;  \qquad
\text{and} \qquad
\phi^4(g_6, g_4,g_1)  \mapsto  \phi^4(g_6, g_4h,g_1) \;, \crcr
&&
\phi^2(g_3, g_4,g_5) \mapsto  \phi^2(g_3, g_4,g_5h)\;  \qquad
\text{and} \qquad
\phi^3(g_5, g_2,g_6)  \mapsto \phi^3(g_5h, g_2,g_6) \;, \crcr
&&
\phi^3(g_5, g_2,g_6)  \mapsto \phi^3(g_5, g_2,g_6h)\;\qquad
\text{and} \qquad
\phi^4(g_6, g_4,g_1) \mapsto  \phi^4(g_6h, g_4,g_1)\;.
\eea
For such a $D$ dimensional GFT, there will be $D(D+1)/2$ of such 
basic transformations, one for each pair of group 
arguments in the interaction. 

We write the equations of motion of the colors $1$ and $4$:
\bea
&&0=\frac{\delta \Scol}{\delta \phi^{1}_{1,2,3}}
= -\sum_{s=1}^3\Delta_{(s)} \bar \phi_{1,2,3} +  m^2 \bar\phi^1_{1,2,3} +
\lambda \int [\prod_{\ell=4}^6 dg_\ell]\;
\phi^2_{3,4,5}\phi^3_{5,2,6} \phi^4_{6,4,1}\;, \crcr
&&0=
\frac{\delta \Scol}{\delta \phi^{4}_{6,4,1}}
= -\sum_{s=1,4,6}\Delta_{(s)} \bar\phi^4_{6,4,1} +  m^2 \bar\phi^4_{6,4,1} + \lambda \int  [\prod_{\ell\neq 1,4,6} dg_\ell]\;
\phi^1_{1,2,3}\phi^2_{3,4,5}\phi^3_{5,2,6} \;.
\label{eqmotioncol}
\eea
For the present purpose, the following functional operator
will be used to compute the EMT: 
\bea
W(X)(\cdot)&=&
\int [\prod_{\ell=1}^3 d\theta_\ell d\varphi^1_\ell d\varphi^2_\ell]  \Big\{\delta\mathbf{g}_{1}^{ij} \frac{\delta(\cdot)}{\delta \mathbf{g}_1^{ij}} 
+ \delta \phi^{1}_{1,2,3} \frac{\delta(\cdot)}{\delta \phi^{1}_{1,2,3}}
+ \delta \bar\phi^{1}_{1,2,3} \frac{\delta(\cdot)}{\delta \bar\phi^{1}_{1,2,3}}\Big\} \crcr
&+&
\int [\prod_{\ell=1,4,6} d\theta_\ell d\varphi^1_\ell d\varphi^2_\ell]\; \delta \phi^{4}_{6,4,1} \frac{\delta(\cdot)}{\delta \phi^{4}_{6,4,1}} 
+ \delta \bar\phi^{4}_{6,4,1} \frac{\delta(\cdot)}{\delta \bar\phi^{4}_{6,4,1}} 
\Big\}.
\label{colorwitrans0}
\eea
Varying the action up to a surface term, the following two-component
EMT has been identified (see Appendix \ref{app:transcolor})
\bea
T^{(1)}_{(1,s);(i,j)} &=& 
\partial_{(1)\;i}  \phi^1_{1,2,3} \;
\partial_{(s)\;j} \bar\phi^1_{1,2,3} 
+
\partial_{(1)\;i}  \bar\phi^1_{1,2,3} \;
\partial_{(s)\;j} \phi^1_{1,2,3} 
 - \delta_{1,s}\mathbf{g}_{s\; ij} \Lcol \;,
\crcr
T^{(4)}_{(1,s);(i,j)} &=& 
\partial_{(1)\;i}  \phi^4_{6,4,1} \;
\partial_{(s)\;j} \bar\phi^4_{6,4,1} 
+  
\partial_{(1)\;i}  \bar\phi^4_{6,4,1} \;
\partial_{(s)\;j} \phi^4_{6,4,1}  \;.
\eea
More generally, for any pair of colors $(a,b)$ sharing
a common group argument labelled by $g_s$, the
EMT for a translation in $g_s$ will be of the form: 
\bea
T^{(a)}_{(s,s');(i,j)} &=& 
\partial_{(s)\;i}  \phi^a_{1,2,3} \;
\partial_{(s')\;j} \bar\phi^a_{1,2,3} 
+ 
\partial_{(s)\;i}  \bar\phi^a_{1,2,3} \;
\partial_{(s')\;j} \phi^a_{1,2,3} 
- \delta_{s,s'}\mathbf{g}_{s'\; ij}\Lcol\;,
\crcr
 T^{(b)}_{(s,s');(i,j)} &=& 
\partial_{(s)\;i}  \phi^b_{1',2',3'} \;
\partial_{(s')\;j} \bar\phi^b_{1',2',3'} 
+
\partial_{(s)\;i}  \bar\phi^b_{1',2',3'} \;
\partial_{(s')\;j} \phi^b_{1',2',3'}  \;.
\eea 
Moreover, the components $T^{(1)}$ and $T^{(4)}$
satisfy the relation (see Appendix \ref{app:transcolor})
\bea
\nabla^j_{(1)}  \int [\prod_{\ell=2}^6 dg_\ell] \Big[
T^{(1)}_{(1,1);(i,j)}  + T^{(4)}_{(1,1);(i,j)}  \Big] =0
\eea
and this means that
\bea
\int [\prod_{\ell=2}^6 dg_\ell] \Big[
T^{(1)}_{(1,1);(i,j)}  + T^{(4)}_{(1,1);(i,j)}  \Big], 
\eea
being still function of $g_1$, is a conserved current.

The existence of such a conserved quantity for GFT models 
can be easily understood. This mainly comes from the definition of translations: they involve an unique field argument whereas all remaining field variables becomes integrated.
We can call these local translations. 
On the symmetry point of view, 
the colored theory with its minimal\footnote{Minimal in a sense that we previously gave, namely, a single translation symmetry on one group argument and keeping all the remaining
variables fixed and afterwards integrated. 
But, at the end, collecting all these minimal
symmetries, the colored theory will certainly fall into a category of theories with the maximal number of
symmetries. For instance, the non colored case has $D$ 
independent field translations/dilatations whereas
the colored theory has $D(D+1)/2$ of such transformations.} symmetry
acts as a kind of local
theory in $\phi^1(g_1,-)$ and $\phi^4(-,g_1)$. 
Hence, the fact the EMT is covariantly conserved here can be 
translated into a local integral in the sense of (\ref{loc}). 
Another kind analysis supporting this idea will be definitely interesting and useful. For instance, one should check that, in the noncommutative 
dual space, colored GFTs endowed with Laplacian dynamics which 
are invariant under translations will have a conserved EMT.

\subsection{Dilatations and current tensor}
\label{sect:coloredilation}

Let us assume once again that only $\phi^1$ 
and $\phi^4$ are subjected to the dilatation
$g_1 \to g_{a\;1}$: 
\bea
\phi^1(g_1, g_2,g_3) \mapsto  \mu(a,\theta_1)^{c}\phi^1(g_1, g_2,g_3)\;, \crcr
\phi^4(g_6, g_4,g_1)  \mapsto  \mu(a,\theta_1)^{c}\phi^4(g_6, g_4,g_1)\;.
\label{colortransdilat}
\eea
The action invariant under these dilatations is defined
by the Lagrangian density
\bea
&& 
\Lcolsca = \crcr
&&
 (\sin\theta_1)^{-1}\mathbf{g}^{ij}_{1}
\partial_{(1) \;i}[(\sin\theta_1)^{-c} \bar\phi^1_{1,2,3} ]
\partial_{(1) \;j}[(\sin\theta_1)^{-c}\phi^1_{1,2,3}] 
+\sum_{s=2}^3 \mathbf{g}^{ij}_{s}
\partial_{(s) \;i} \bar\phi^1_{1,2,3}  \partial_{(s) \;j}\phi^1_{1,2,3} \crcr
&+& 
 (\sin\theta_1)^{-1}\mathbf{g}^{ij}_{1}
\partial_{(1) \;i}[(\sin\theta_1)^{-c} \bar\phi^4_{6,4,1} ]
\partial_{(1) \;j}[(\sin\theta_1)^{-c}\phi^4_{6,4,1}] 
+
\sum_{s=4,6} \mathbf{g}^{ij}_{s}
\partial_{(s) \;i} \bar\phi^4_{6,4,1} 
\partial_{(s) \;j}\phi^4_{6,4,1} \crcr
& +& 
\sum_{s=3,4,5}\mathbf{g}^{ij}_{s}\partial_{(s) \;i} \bar\phi^2_{3,4,5} 
\partial_{(s) \;j}\phi^2_{3,4,5} 
+
 \sum_{s=5,2,6}\mathbf{g}^{ij}_{s}\partial_{(s) \;i} \bar\phi^3_{5,2,6} 
\partial_{(s) \;j}\phi^3_{5,2,6}  \crcr
&+&
 \lambda \;\phi^{1}_{1,2,3} \phi^2_{3,4,5} \phi^3_{5,2,6}\phi^4_{6,4,1}
+ \bar\lambda \;\bar\phi^{1}_{1,2,3} \bar\phi^2_{3,4,5} \bar\phi^3_{5,2,6}\bar\phi^4_{6,4,1}\;,
\label{eq:lagcoldilat}
\eea
where we omit to write mass terms
even though they can be  also included. Indeed, 
they possess the same scaling behaviour
as the interaction itself. 

Requiring an invariant action implies that $c=-3/2$. The associated current
can be derived using
\bea
W(\epsilon) (\cdot)&=& \int [\prod_{\ell=1}^3 d\theta_\ell d\varphi_\ell^1d\varphi_\ell^2] \; 
\;
 \Big{[}\delta_{\epsilon} \phi^1_{1,2,3} \, \frac{\delta }{\delta \phi^1_{1,2,3}}(\cdot)  
+\delta_{\epsilon} \bar\phi^1_{1,2,3} \, \frac{\delta }{\delta \bar\phi^1_{1,2,3}}(\cdot)   \Big{]}
\crcr
&+&
\int [\prod_{\ell=1,4,6} d\theta_\ell d\varphi_\ell^1d\varphi_\ell^2] \; 
\;
 \Big{[}\delta_{\epsilon} \phi^4_{1,2,3} \, \frac{\delta }{\delta \phi^2_{1,2,3}}(\cdot)  
+\delta_{\epsilon} \bar\phi^4_{1,2,3} \, \frac{\delta }{\delta \bar\phi^2_{1,2,3}}(\cdot)   \Big{]}.
\label{widilatcol}
\eea
 The current tensor for this symmetry possesses distinct
components (derivations are given in Appendix \ref{app:dilatcolor})
\bea
D^{(1)}_{(1);\;j} &=& 
\Big[(\sin\theta_{1})^2[ \,\beta\cos \theta_{1}  + \sin \theta_{1}\, \partial_{(1)\;\theta} \,
]\phi^1_{1,2,3}\; \partial_{(1)\;j} \bar\phi^1_{1,2,3} 
\crcr
&&
+ \beta  \mathbf{g}_{1\; j\theta}\cos\theta_{1}\sin\theta_{1} [ \,\beta\cos \theta_{1}  + \sin \theta_{1}\, \partial_{(1)\;\theta}]\phi^1_{1,2,3}  \; \bar\phi^1_{1,2,3} + (\phi \leftrightarrow \bar\phi)\Big] \crcr
&& - \mathbf{g}_{1\; j\theta} \,
 \sin\theta_{1} {\Lcolsca}^{\;(1,4)} \;, 
\crcr
D^{(1)}_{(s);\;j} &=&
[ \,\beta\cos \theta_{1}  + \sin \theta_{1}\, \partial_{(1)\;\theta} \,
]\phi^1_{1,2,3}\; \partial_{(s)\;j} \bar\phi^1_{1,2,3}\; +\; (\phi \leftrightarrow \bar\phi)\;, \quad s=2,3,
\eea
while the components $D^{(4)}_{(s);\;j}$ can be obtained
from $D^{(1)}_{(1);\;j} $ and $D^{(1)}_{(s=2,3);\;j} $
by taking the symmetry $(\phi^1_{1,2,3} \leftrightarrow \phi^4_{6,4,1})$
without the Lagrangian term ${\Lcolsca}^{\;(1,4)}$. The latter
is defined from $\Lcolsca$ by  considering only terms 
containing the fields of colors $1$ and $4$.
The dilatation current component $D^{(1)}_{(1);\;j}$ can be written also as
\bea
D^{(1)}_{(1);\;j}
 &=& 
\partial_{(1)\;\theta}[ (\sin\theta_1)^\beta \bar\phi^1_{1,2,3}]
\partial_{(1)\;j}[ (\sin\theta_1)^\beta \phi^1_{1,2,3}] + 
\partial_{(1)\;\theta}[ (\sin\theta_1)^\beta \bar\phi^1_{1,2,3}]
\partial_{(1)\;j}[ (\sin\theta_1)^\beta \phi^1_{1,2,3}]  \crcr
&-& \sin\theta_1{\Lcolsca}^{\;(1,4)} \;.
\eea
 Finally, we can note that the dilatation current is not covariantly conserved due to 
the explicit coordinate dependence but not
 because of the nonlocal interaction:
\bea
\int [\prod_{\ell=2}^6 dg_\ell] \left[ \sum_{s=1,2,3} \nabla^{j}_{(s)} 
D^{(1)}_{(s)\;j} +
\sum_{s=1,2,3} \nabla^{j}_{(s)} 
D^{(4)}_{(s)\;j} \right] \neq 0\;.
\eea
The explicit expression of the breaking is
given in Appendix \ref{app:dilatcolor}.

\section{Summary and outlooks}
\label{concl}

The classical formalism, i.e. the extension of Klein-Gordon field 
equation and the group symmetry study, for dynamical 
GFT over tensor copies of $SU(2)$ has been investigated in this paper.  
We find that the GFTs exhibit peculiarities 
that one could expect 
when dealing with nonlocal models. 
For translation symmetry, the EMT for the general 
GFT (without colors) proves to be symmetric but 
not locally conserved for any dimension, save $D=1$.
As a matter of fact, for $D=1$
locality is recovered and, from that, the EMT  
is covariantly conserved. In contrast and astonishingly, 
the genuine colored GFTs which are also  nonlocal
possess a covariantly conserved quantity 
obtained by integrating and summing some EMT components. 
This is another feature advocating 
in favor of these colored theories. 
In all situations (with and without color),
the EMT  possesses a nonvanishing trace, 
even though the latter property could be improved 
but the meaning of the resulting tensor remains unclear. 
We also discuss a specific 
way that group dilatations could be implemented
at the GFT level. 
Indeed, dilatation symmetry can been consistently settled in GFT
with cost to put an explicit dependence of the class angle
$\theta$ coordinate of the $SU(2)$ group variable $g(\theta,\vec n)$ in the Lagrangian. 
In any dimension $D\geq 1$ colored or not, this dissipative term 
explicitly 
breaks the conservation of the dilatation current. 
 Remark that there should be an alternative way to
implement dilatation on the sphere according to the work
by Okuyama \cite{Okuyama:2002zn} or by using Weyl transformations
that we did not consider here. 
Nevertheless, the issue of nonlocality and the current 
local conservation breaking will be
in all cases difficult to circumvent.

It can be desirable to find some improvement procedures 
adapted for the non-colored GFTs in order to render the EMTs  and
other currents covariantly conserved for $D\geq 1$. In the case of noncommutative field theory 
defined with a Moyal $\star$-product with its induced nonlocality, 
improvement procedures have been highlighted to treat the
breaking term of the EMT local conservation 
\cite{AbouZeid:2001up}-\cite{Hounkonnou:2009qt}. 
Nevertheless, these methods were successful due to the specificity of the Moyal field algebra. Here, clearly the issue is different and deserves
a better understanding. Another important
field symmetry, that we did not discuss 
and could be viewed as rotation in the context, 
would be the one associated with group adjoint transformation 
$g \to h g h^{-1}$.  
There are more infinitesimal vector field generators associated
with such transformations and so more involved
becomes the computation of the  ``angular momentum''
tensor. The fact that the EMTs found
here are symmetric is encouraging for the 
local conservation of the angular momentum tensor
but only in the color case.  
Finally, new vertices leading to orientable graphs for GFTs have been highlighted recently \cite{tanasa}. It could be interesting to see if the
techniques used in this paper could be also implemented on
these models and could lead to other or more regular properties. 

The present analysis has revealed that one of the 
simplest notions of symmetry, such as translations, for
dynamical non colored GFTs are more difficult to implement 
and so remain to be understood. In contrast, dynamical 
colored GFTs have quite regular properties with 
a conserved quantity for translation symmetry. 
A key point revealed by this analysis is that 
dynamics for colored GFTs is not incompatible 
with the notion of symmetries. 
A natural question is the implications
of such symmetries at the quantum level. 
We should comment that this study could be useful for
the larger program aiming at renormalizing GFTs \cite{vincentrev}. 
In particular, among the most interesting candidates for 
that program are colored tensor theories. Here we have proved that 
colored theories endowed with dynamics have a well defined notion of translation symmetry.
As a result, they will have more constrained Ward-Takahashi identities 
in relation with that particular symmetry.
These will be useful for both perturbative renormalization and also nonperturbative features. 
Indeed, Ward-Takahashi identities (which can be called 
quantum versions of conservation laws) provide useful relations between correlation functions which should hold even after renormalization. 
These identities have been successfully used in quantum electrodynamics renormalization (giving a link between the renormalized vertex three-point function and the renormalized wave function). 
More recently, they prove to be one of the main ingredients for the proof of asymptotic safety at all order of perturbation theory for particular matrix models \cite{Disertori:2006nq}\cite{Geloun:2008zk}. 
Note that Ward-Takahashi identities have been already discussed for non dynamical colored GFTs under unitary symmetries \cite{BenGeloun:2011xu}. 
In the present work, dynamical colored tensored theories have been investigated and should provide other kinds of Ward-Takahashi identities
with respect to translations.

Last, GFT symmetries can be inquired in 
another way by establishing a bridge between our formulation and the metric representation as described by Baratin and co-workers \cite{Baratin:2011tg}. For simplicity, we will restrict to the $1D$ GFT formalism. 
Consider a function $\phi: G \to \mathbb{C}$ 
of one group variable and use the group Fourier transform \cite{Freidel:2005me} in order
to obtain its Lie algebra representation:
\begin{equation}
\widehat\phi(x) = \int_{G} \phi(g) e_{g}(x)dg\;, \quad
e_{g}: \mathfrak{su}(2)\sim  \mathbb{R}^3 \to U(1)\;,\quad 
e_{g}(x) = e^{i\text{Tr}(xg)}\;, \quad x = \vec x \cdot \vec \tau \;,
\end{equation}
where $\vec x \in \mathbb{R}^3$, and $\vec\tau=(\tau_1,\tau_2,\tau_3)$
is the vector of anti-Hermitian $\mathfrak{su}(2)$ generators. 
 Let $g \to gh$ be a group translation such that 
under this transformation fields get modified according  to
$ \phi(g) \to \phi'(g'=gh) $, then the transformed 
dual field becomes
\bea
\widehat{(\phi')}(x) = \left(\widehat\phi \star e_{h} \right)(x)\;,
\eea
where the explicit nature of the $\star$-product can be found
in \cite{Freidel:2005me}.
Thus, the group translation invariance for fields can be translated
as a $U_\star(1)$ noncommutative gauge invariance
in the Lie algebra formulation. Note that the latter result 
is in contrast with translations 
$x\to x+a $ directly stated in the Lie algebra formulation
which yield a field transformation corresponding to a plane wave multiplication: 
$[\widehat \phi(x) \to \widehat\phi'(x+a) ] 
\;
\leftrightarrow \; [\phi(g) \to \phi'(g)= e_{g}(a) \phi(g)]$. 
Group dilatations will certainly have a noncommutative
analogue but this deserves to be investigated.
Remarkably, there is an analogue to our Noether theorem 
in this noncommutative setting. 
 An infinitesimal translation corresponding to $x\to x + a$, where $a \in \mathbb{R}^3$, is totally similar to an ordinary infinitesimal translation is
the common spacetime: 
$
\delta_{\epsilon} \,\phi  = 
\sum_{i=1}^3 \epsilon^i \partial_{i}\,\phi \;. 
$
A main difference between the group and the Lie algebra
formalisms lies in the type of field algebra under consideration. While in the
former theory, the field algebra is commutative, in the latter, 
the same algebra becomes noncommutative. However, it has been
defined a Noether procedure for noncommutative field algebras equipped with $\star$-product (\cite{Gerhold:2000ik}-\cite{Hounkonnou:2009qt} and for more general discussion on Noether's theorem in noncommutative geometry see \cite{AmelinoCamelia:2007wk}). In the present context, one needs to introduce the  operator
\bea
W_\star(\epsilon) (\cdot)= \int  d\theta d \varphi^1 d \varphi^2\;
\left(  \delta_{\epsilon} \phi \star \frac{\delta }{\delta \phi}(\cdot) \right)(x)\;.
\label{wiopinitdual}
\eea
Applying
the operator $\frac{\delta }{\delta \phi}(\cdot)$
on the action should produce the noncommutative version
of the equations of motion for fields. Note that, due to the noncommutativity theory, it is still possible to introduce other types of operators $W_\star(\epsilon)$ by considering for instance
$[\frac{\delta }{\delta \phi}\star \delta_{\epsilon} \phi ](\cdot) ](x)$ or by symmetrizing both the latter and (\ref{wiopinitdual}).
Obviously dealing with colored fields, the operator 
(\ref{wiopinitdual}) should be extended for that purpose.
An interesting exercise would be to compute and the characterize all noncommutative Noether quantities
issued from the operator (\ref{wiopinitdual}) and the similar for other
field transformations. We expect that currents will share  
the similar (broken or not) features as given in this work.

\section*{Acknowledgements}

The author warmly thanks Etera Livine for illuminating discussions 
at various stage of this work.
Helpful discussions with Tim Koslowski, Valentin Bonzom and Razvan Gurau are also deeply acknowledged.  
Research at Perimeter Institute is supported by the Government of Canada through Industry 
Canada and by the Province of Ontario through the Ministry of Research and Innovation.

\section*{Appendix}
\appendix

\renewcommand{\theequation}{\Alph{section}.\arabic{equation}}
\setcounter{equation}{0}

\section{Gauge invariant fields} 
\label{app:gaug}

Consider a gauge invariant field, namely a field satisfying
\bea
\phi(g_1,g_2,g_3) = \int dh \;\phi(g_1h,g_2h,g_3h) \;.
\label{gauginvapp}
\eea
This condition translates in Fourier modes via Peter-Weyl theorem as
\begin{equation}
\sum_{j_a,m_a,n_a} \phi^{j_a}_{m_a,n_a} \prod_{a} \sqrt{d_{j_a}} D^{j_a}_{m_a,n_a} (g_a) 
= \sum_{j_a,m_a,k_a,n_a} \phi^{ j_a}_{m_a,k_a} 
[\prod_{a}\sqrt{d_{j_a}} D^{j_a}_{m_a,n_a}(g_a)]
\int dh \prod_a D^{j_a}_{n_a,k_a}(h)
\end{equation}
where $a=1,2,3$, $\phi^{j_a}_{m_a,n_a}$ is a notation for $\phi^{j_1,j_2,j_3}_{m_1,n_1;m_2,n_2;m_3,n_3}$,
$j_a$ indexes half spin representation, $j_a \in \frac12\mathbb{N}$, $n_a,m_a,k_a$ are ordinary
associated magnetic momenta each constrained to be inside $[-j_a,+j_a]$,
$d_{j_a}=2j_a+1$ is the dimension of the representation space. 
$D^j_{mn}(g)$ denotes the Wigner matrix element 
of $g$ in the representation $j$.
The factor $\sqrt{d_{j_a}}$ is chosen as a normalization convention. 

Computing the last integral, one gets by simple identification:
\bea
 \phi^{j_a}_{m_a,n_a} 
= \sum_{k_a} \phi^{j_a}_{m_a,k_a} 
\left(\begin{array}{ccc}
j_1&j_2&j_3\\
k_1&k_2&k_3
\end{array}\right)
\left(\begin{array}{ccc}
j_1&j_2&j_3\\
n_1&n_2&n_3
\end{array}\right)\;,
\label{boujoum}
\eea
 with the arrays denoting the rotation invariant Wigner $3j$-symbols.
A field with coefficients as (\ref{boujoum}) indeed exists 
and satisfies (\ref{gauginvapp}) due the orthogonality relation of $3j$ symbols. A simple $\phi^{j_a}_{m_a,n_a}$ is given for instance 
by the product of two $3j$-symbols 
with coefficients following the pattern (\ref{boujoum}). 
For the general $D$ dimensional gauge invariant fields, the integral 
$\int dh \prod_{a=1}^D D^{j_a}_{n_a,k_a}(h)$
gives an invariant intertwiner which can be recoupled in 
more involved Wigner $3nj$-symbols. 

\section{Group translations}
\label{app:trans}

In this appendix, we give the main identities leading
to the formulas of EMTs in $1D$, $3D$ and colored GFTs and
to the (non)local conservation property of these tensors. The first
subsection explains also the method of computation of 
these quantities.

\subsection{Ward operator action for translations for 1D GFT  }
\label{app:trans:w1D}

In this paragraph, an explicit calculation of the EMT is given
for a $1D$ GFT by applying the Ward operator method in a curved
background like $S^3$ parametrized by $(\theta,\varphi_1,\varphi_2)$.
The ensuing tensor study could be deduced from this point. 

For an infinitesimal translation of parameter $X$,  we can set 
in a given local coordinate system
$\delta_X \phi = X^i \partial_i \phi$. Then one introduces
the operator
\bea
W(X)(\cdot) = \int d\theta d\varphi_1d\varphi_2\; 
\left(\delta_X \mathbf{g}^{ij} \frac{\delta(\cdot)}{\delta \mathbf{g}^{ij}} 
+ \delta_X \phi \frac{\delta(\cdot)}{\delta \phi} \right)
\eea
that acts on the action $\So$ (\ref{act1d}) such that
\bea
\frac{\partial}{ \partial X^\rho } W(X)\So
&=& \frac{\partial}{ \partial X^\rho }\int d\theta d\varphi_1d\varphi_2\; 
\Big\{\delta_X \mathbf{g}^{ij}  \left[ \frac{\partial \sqrt{|\det\mathbf{g}|}}{\partial  \mathbf{g}^{ij}} \Lo 
+  \sqrt{|\det\mathbf{g}|}\frac{\partial \Lo}{\partial \mathbf{g}^{ij}}     \right]
\crcr
&& 
+  \sqrt{|\det\mathbf{g}|}\left( X^{\kappa} \partial_\kappa \phi \right) \left(- \Delta \phi + m^2 \phi + 
\lambda \phi^3 \right)  \Big\}\;.
\eea
Keeping in mind that the Laplacian contains an inverse factor 
of the metric determinant, one gets
\bea
&&
\frac{\partial}{ \partial X^\rho } W(X)\So
= \delta_{\rho}^{\kappa} \int d\theta d\varphi_1d\varphi_2\; 
\left\{\partial_\kappa \mathbf{g}^{ij}  \left[ -\frac12 \sqrt{|\det\mathbf{g}|} \mathbf{g}_{ij} \Lo 
+  \sqrt{|\det\mathbf{g}|} \frac12 \partial_i \phi \partial_j \phi     \right] \right.
\crcr
&& \left. 
 - \partial_i (\partial_\kappa \phi \mathbf{g}^{ij} \sqrt{|\det\mathbf{g}|} \partial_j\phi )
+ (\partial_i\partial_\kappa\phi) \mathbf{g}^{ij} \sqrt{|\det\mathbf{g}|} \partial_j\phi
+  \sqrt{|\det\mathbf{g}|} \partial_\kappa(m^2 \frac12\phi + 
\frac{\lambda}{4} \phi^4) \right\} .\crcr
&&
\eea
It is customary in a field theory
to identify the EMT from 
the variations of the metric $\mathbf{g}^{ij}$.
The EMT already appears in the above expression up to some factor. However, in this paper, we will not use this route preferring instead to get a final surface term. 
To this end, further computations invoking both metric and field variations
are in order:
\bea
&&
\frac{\partial}{ \partial X^\rho } W(X)\So
= \delta_{\rho}^{\kappa} \int d\theta d\varphi_1d\varphi_2\; 
\left\{\partial_\kappa \mathbf{g}^{ij}  \left[ -\frac12 \sqrt{|\det\mathbf{g}|} \mathbf{g}_{ij} \Lo 
+  \sqrt{|\det\mathbf{g}|} \frac12 \partial_i \phi \partial_j \phi     \right] \right.
\crcr
&& \left.
 - \partial_i (\partial_\kappa \phi \mathbf{g}^{ij} \sqrt{|\det\mathbf{g}|} \partial_j\phi )
-  \sqrt{|\det\mathbf{g}|} (\partial_\kappa \mathbf{g}^{ij}) (\frac12 \partial_i\phi \partial_j\phi)
+  \sqrt{|\det\mathbf{g}|} \partial_\kappa \Lo \right\} \crcr
&& = -\int d\theta d\varphi_1d\varphi_2 \;
\partial_i  \{  \sqrt{|\det\mathbf{g}|} \mathbf{g}^{ij}\left( \partial_\rho \phi 
\partial_j\phi  -  \mathbf{g}_{\rho j} \Lo \right)\}\;.
\eea
Finally, this expression is of the form of a surface term and we identify the EMT
\bea
T_{\rho j} = \partial_\rho \phi 
\partial_j\phi  -   \mathbf{g}_{\rho j} \Lo \;.
\eea
We verify, in a covariant script, that
\bea
&&
\nabla^i T_{ij} = \nabla ^i (\nabla_i \phi \nabla_j \phi) 
- \nabla_j \left(\frac12 \mathbf{g}^{kl} \nabla_k \phi \nabla_l \phi + \frac{m^2}{2} \phi^2+\frac{\lambda}{4} \phi^4 \right) \crcr
&& = 
(\nabla^i \nabla_i \phi) \nabla_j \phi 
+  \nabla_i \phi\nabla^i\nabla_j \phi 
- \frac12 \mathbf{g}^{kl}\nabla_j (\nabla_k \phi\nabla_l \phi) 
-\nabla_j \phi\left( m^2 \phi  + \lambda \phi^3\right) 
=0\;,
\eea
where, we use the property of the Levi-Civita connection, and, in last resort, the field equation of motion $-\Delta \phi +m^2 \phi + \lambda \phi^3 =0 $. 

This method of deriving the EMT will be extended in the 
subsequent situations dealing with tensor models.

\subsection{EMT for GFT in 3D}
\label{app:trans:calc3D}

\noindent{\bf EMT calculation -}
We provide here the main stages of calculations leading to 
(\ref{emtrig}).  

First, we recall that 
the infinitesimal variation for a tensor field $\phi$ 
under a ``3-translation'' is given by $\delta_X \phi_{1,2,3}:=\delta_{X_{(1)},X_{(2)},X_{(3)}} \phi_{1,2,3} = \sum_{s=1}^3 X^{i}_{(s)} \partial_{(s)\;i}\phi_{1,2,3} $. 
Then, we require an operator symmetrization for the interaction part:
\bea
&&
\lambda 
\int [\prod_{\ell=1}^6 dg_\ell]\;
\left( \delta_X \phi_{1,2,3} \right) \bar\phi_{5,4,3}\phi_{5,2,6}\bar\phi_{1,4,6} 
+\left( \delta_X \bar\phi_{1,2,3} \right) \phi_{5,4,3}\bar\phi_{5,2,6}\phi_{1,4,6} =\crcr
&&
\frac{\lambda}{2} 
\int [\prod_{\ell=1}^6 dg_\ell]\;
\Big{\{} 
 \left(\delta_{X} \phi_{1,2,3}  \right)\bar\phi_{5,4,3}\phi_{5,2,6}\bar\phi_{1,4,6}
+ \left(\delta_{X} \phi_{5,2,6}  \right) \phi_{1,2,3} \bar\phi_{5,4,3}\bar \phi_{1,4,6}
\crcr
&& 
+
\left(\delta_{X} \bar\phi_{1,2,3}\right) \phi_{5,4,3}\bar\phi_{5,2,6} \phi_{1,4,6} 
+
\left( \delta_{X} \bar\phi_{5,2,6} \right) \phi_{5,4,3}\phi_{1,4,6}\bar\phi_{1,2,3} 
 \Big{\}} \crcr
&&  =
\frac{\lambda}{2} 
\int[\prod_{\ell=1}^6 dg_\ell]\;
\Big{\{} 
\sum_{s=1}^3 \left( X^i_{(s)} \partial_{(s)\;i} \phi_{1,2,3}\right)  \bar\phi_{5,4,3}\phi_{5,2,6}\bar\phi_{1,4,6} \crcr
&&
+ \sum_{s=1}^3 \left(X^i_{(s)} \partial_{(s+\alpha_s)\;i} \bar\phi_{5,4,3}\right)  \phi_{1,2,3} \phi_{5,2,6}\bar\phi_{1,4,6}  
+
\sum_{s=1}^3 \left(X^i_{(s)} \partial_{(s+\alpha_s)\;i}\phi_{5,2,6}\right)  \bar\phi_{5,4,3}\phi_{1,2,3} \bar\phi_{1,4,6} \crcr
&&
+
\sum_{s=1}^3 \left(X^i_{(s)} \partial_{(s+\alpha_s)\;i}\bar\phi_{1,4,6} \right)  \bar\phi_{5,4,3}\phi_{5,2,6}\phi_{1,2,3} 
 \Big{\}} \crcr
&& 
= \frac{\lambda}{2} 
\int[\prod_{\ell=1}^6 dg_\ell] \sum_{s=1}^3X^i_{(s)} \Big{(}\partial_{(s)\;i} 
+ \partial_{(s + [\alpha_s])\;i}  \Big{)}
\phi_{1,2,3} \bar\phi_{5,4,3}\phi_{5,2,6}\bar\phi_{1,4,6}\;.
\label{wiopsym}
\eea
where the index $\alpha_s=0,2,3,4$ has to be chosen appropriately.
In the last equality, the notation $[\alpha_s]$ means that 
we fix $([\alpha_1], [\alpha_2], [\alpha_3]) = (4,2,3)$.
 It is remarkable that, under integral
over all six variables, we can exchange  
$\bar\phi_{1,2,3}$ for $\bar\phi_{5,4,3}$ 
and $\bar\phi_{5,2,6}$ for $\bar\phi_{1,4,6}$ by just renaming the
variables (this is a set of
discrete symmetries which will be also used in the sequel).

We introduce the notations $(\bullet) = \prod_{s=1}^6 \sqrt{|\det \mathbf{g}_{s}|}$
and $\buss = \prod_{k\neq s} \sqrt{|\det \mathbf{g}_{k}|}$, so that
\bea
(\bullet) = \buss  \sqrt{|\det \mathbf{g}_{s}|}\;, 
\qquad \prod_{\ell=1}^6 [dg_\ell]  = \frac{1}{2\pi^2}
\prod_{\ell=1}^6 [d\theta_\ell d\varphi^1_\ell d\varphi^2_\ell] 
(\bullet)\;.
\eea
The factor $1/(2\pi^2)$ will be omitted in the following.
The EMT can be computed as follows
\bea
&&
W(X)\Stri= \int [\prod_{\ell=1}^6 d\theta_\ell d \varphi_\ell^1 d \varphi_\ell^2] \;
\left( \sum_{s=1}^6 \delta_{X} \mathbf{g}_s^{ij} \, \frac{\delta }{\delta \mathbf{g}_s^{ij}} \Stri
\right. \crcr
&&
+ \sum_{s'=1}^3 X^k_{(s')}\partial_{(s')\;k}  \phi_{1,2,3} \left[ - 
 \sum_{s=1}^3 \partial_{(s)\;j } (  \buss \sqrt{|\det \mathbf{g}_s|} 
\mathbf{g}_s^{jl}\partial_{(s)\;l} \bar\phi_{1,2,3} )\right] \crcr
&&\left. 
+  (\bullet) 
\sum_{s'=1}^3 X^k_{(s')}\partial_{(s')\;k}\phi_{1,2,3} \left[ m^2 \bar\phi_{1,2,3} 
+\lambda\;\bar\phi_{5,4,3}\phi_{5,2,6} \bar\phi_{1,4,6} )\right] 
+\fifib \right),
\label{intermenco}
\eea
where the part including  the variation $\delta_X \bar \phi$ is not
explicitly displayed but appears symbolically as $\fifib$.
The corresponding terms can be carried out in a symmetric manner. 
Adding all contributions, combining the mass terms and the
interaction using (\ref{wiopsym}), it can be seen that 
\bea
&&W(X)\Stri =\int [\prod_{\ell=1}^6 d\theta_\ell d \varphi_\ell^1 d \varphi_\ell^2] \;
\left\{ \sum_s \delta_{X} \mathbf{g}_s^{ij} \, \frac{\delta }{\delta \mathbf{g}_s^{ij}} \Stri 
\right. \crcr
&&
+ \Big\{ - \sum_{s',s=1}^3 X^k_{(s')} \partial_{(s)\;j } \left[  
\partial_{(s')\;k}  \phi_{1,2,3} \;
(\bullet)\mathbf{g}_s^{jl}  
\partial_{(s)\;l} \bar\phi_{1,2,3} \right] \crcr
&&+ \sum_{s',s=1}^3 X^k_{(s')} \partial_{(s)\;j } \left[  
\partial_{(s')\;k}  \phi_{1,2,3} \right] \;
(\bullet) 
\mathbf{g}_s^{jl}\partial_{(s)\;l } \bar \phi_{1,2,3}\; + \fifib\Big\} 
\\
&&\left. 
+(\bullet)  \sum_{s'=1}^3 X^k_{(s')} \left[ 
 m^2 \partial_{(s')\;k} (\bar\phi_{1,2,3} \phi_{1,2,3}) 
+\frac{\lambda}{2} (\partial_{(s')\;k} + 
\partial_{(s'+[\alpha_{s'}])\;k} )\phi_{1,2,3}\bar\phi_{5,4,3}\phi_{5,2,6} \bar\phi_{1,4,6} )   \right] 
\right\}.\nonumber 
\eea
Focusing now on the dynamical part and metric variations, one has:
\bea
&&K =  \int [\prod_{\ell=1}^6 d\theta_\ell d \varphi_\ell^1 d \varphi_\ell^2] \;
\left\{
\int [\prod_{\ell'=1}^6 d\theta_{\ell'} d \varphi_{\ell'}^1 d \varphi_{\ell'}^2]  \sum_{s=1}^6 \delta_{X} \mathbf{g}_s^{ij} \, 
\frac{\delta }{\delta \mathbf{g}_s^{ij}}(\buss \sqrt{|\det \mathbf{g}_s|} ) \Ltri
\right. \crcr
&&
+\int [\prod_{\ell'=1}^6 d\theta_{\ell'} 
d \varphi_{\ell'}^1 d \varphi_{\ell'}^2] \sum_{s=1}^6(\bullet)
\delta_{X} \mathbf{g}_s^{ij} \frac{\delta \Lkin}{\delta \mathbf{g}_s^{ij}} \crcr
&&
+ \Big\{ - \sum_{s',s=1}^3 X^k_{(s')} \partial_{(s)\;j } \left[  
\partial_{(s')\;k}  \phi_{1,2,3} \;
 (\bullet)\mathbf{g}_s^{jl}
\partial_{(s)\;l} \bar\phi_{1,2,3} \right] + \fifib \Big\} \crcr
&&+ (\bullet) \sum_{s',s=1}^3 X^k_{(s')} 
\partial_{(s')\;k}  \left[ \mathbf{g}_s^{jl} 
\partial_{(s)\;j } \phi_{1,2,3}\;
\partial_{(s)\;l }  \bar\phi_{1,2,3} \right]  \crcr
&& \left. - (\bullet)\sum_{s',s=1}^3 X^k_{(s')} 
\partial_{(s')\;k}  \left[ \mathbf{g}_s^{jl} \right]  
 \partial_{(s)\;j } \phi_{1,2,3}\;
\partial_{(s)\;l }  \bar\phi_{1,2,3} \right\}\;.
\label{interm2} 
\eea
Canceling the variation
$\delta \Lkin/\delta \mathbf{g}_{s}^{ij}$ with 
its partner coming from the field variations and recomposing the Lagrangian, by injecting expression $K$ (\ref{interm2}) into
(\ref{intermenco}), we get
\bea
&& W(X)\Stri=\crcr
&&\int [\prod_{\ell=1}^6 d\theta_\ell d \varphi_\ell^1 d \varphi_\ell^2] 
\left[
\Big\{ - \sum_{s',s=1}^3 X^k_{(s')} \partial_{(s)\;j } \left[ (\bullet) 
\partial_{(s')\;k}  \phi_{1,2,3} 
\mathbf{g}_s^{jl} 
\partial_{(s)\;l } \bar\phi_{1,2,3} \right] + \fifib \Big\} \right.\crcr
&& \left.  
+\sum_{s'=1}^3 X^k_{(s')} 
\Big{ \{}\; \partial_{(s')\;k} [(\bullet) \Ltri ]
+ 
\partial_{(s'+[\alpha_{s'}])\;k}  [(\bullet)\Lint] \Big{\}}
\right]\;.
\eea
We are now in position to provide the EMT for group translations 
of the GFT by just deriving the above expression by some
infinitesimal parameter: $X^k_{(s)}$, $s,k=1,2,3$, 
\bea
&& \frac{\partial}{\partial X_{(s)}^k} W(X)\Stri= 
-\int [\prod_{\ell=1}^6 d\theta_\ell d \varphi_\ell^1 d \varphi_\ell^2]
 \sum_{s'=1}^6  \partial_{(s')\;j } (\bullet) \mathbf{g}_{s'}^{jl}\Big{\{}  \crcr
&&
\partial_{(s)\;k}  \phi_{1,2,3} \;
\partial_{(s')\;l} \bar\phi_{1,2,3} + \partial_{(s)\;k}  \bar\phi_{1,2,3} \;\partial_{(s')\;l}\phi_{1,2,3} 
-\delta_{s,s'}\mathbf{g}_{s'\;lk}   \Ltri \;
-  \delta_{s+\alpha_{s},s'}\mathbf{g}_{s'\;lk}   \Lint \Big{\}} .
\crcr
&&
\eea
Hence the EMT is given by
\begin{equation}
T_{(s,s');(i,j)} = \partial_{(s)\;i}  \phi_{1,2,3} \,
\partial_{(s')\;j} \bar\phi_{1,2,3} +
\partial_{(s)\;i}  \bar\phi_{1,2,3} \,
\partial_{(s')\;j} \phi_{1,2,3}  
-\delta_{s,s'}\mathbf{g}_{s'\;ij}    \Ltri  
- \delta_{s+[\alpha_{s}],s'}\mathbf{g}_{s'\;ij} \Lint\;,
\end{equation}
for $s=1,2,3$, $s'=1,2,\dots,6$, and $i,j=1,2,3$.

\noindent{\bf Covariant conservation -}
The fact that the EMT is not covariantly conserved is proved here.
We have in covariant notations:
\bea
&&
\sum_{s'=1}^6 \nabla_{(s')}^j T_{(s,s');(ij)}
 =\crcr
&&
\Big\{ 
\sum_{s'=1}^3 \left( \nabla_{(s')}^j  \nabla_{(s)\;i}  \phi_{1,2,3} \;
\nabla_{(s')\;j} \bar\phi_{1,2,3} + \nabla_{(s)\;i}  \phi_{1,2,3} \;
\nabla_{(s')}^j \nabla_{(s')\;j} \bar\phi_{1,2,3} \right) + \fifib
\Big\} 
\crcr
&& 
- \nabla_{(s)\;i} \left( \sum_{s'=1}^3\mathbf{g}_{s'}^{kl} \nabla_{(s')\;k}\phi_{1,2,3}\nabla_{(s')\;l}\bar\phi_{1,2,3} 
+  m^2\bar\phi_{1,2,3}\phi_{1,2,3} + \frac{\lambda}{2} \phi_{1,2,3} \bar\phi_{5,4,3}\phi_{5,2,6}\bar\phi_{1,4,6}\right)   \crcr
&& 
- \nabla_{(s+[\alpha_{s}])\;i}   ( \frac{\lambda}{2} \phi_{1,2,3} \bar\phi_{5,4,3}\phi_{5,2,6}\bar\phi_{1,4,6}) \crcr
&& = 
\sum_{s'=1}^3 \left(    \nabla_{(s)\;i}  \phi_{1,2,3} \;
\nabla_{(s')}^j \nabla_{(s')\;j} \bar\phi_{1,2,3} \right) 
- m^2 \bar\phi_{1,2,3} (\nabla_{(s)\;i} \phi_{1,2,3}) \crcr
&&+ \sum_{s'=1}^3 \left(    \nabla_{(s)\;i}  \bar\phi_{1,2,3} \;
\nabla_{(s')}^j \nabla_{(s')\;j} \phi_{1,2,3} \right) 
- m^2\phi_{1,2,3} (\nabla_{(s)\;i}  \bar\phi_{1,2,3}) \crcr
&&  - \frac{\lambda}{2} (\nabla_{(s)\;i} \phi_{1,2,3}) \bar\phi_{5,4,3}\phi_{5,2,6}\bar\phi_{1,4,6}  
-\frac{\lambda}{2} \phi_{1,2,3}  (\nabla_{(s)\;i}\bar\phi_{5,4,3})\phi_{5,2,6}\bar\phi_{1,4,6}  \crcr
&&
- \frac{\lambda}{2}\phi_{1,2,3} \bar\phi_{5,4,3} (\nabla_{(s)\;i} \phi_{5,2,6})\bar\phi_{1,4,6}  
-\frac{\lambda}{2} \phi_{1,2,3}  \bar\phi_{5,4,3}\phi_{5,2,6}(\nabla_{(s)\;i}\bar\phi_{1,4,6} )\crcr
&& 
- \frac{\lambda}{2}   \phi_{1,2,3} 
( \nabla_{(s+[\alpha_{s}])\;i} \bar\phi_{5,4,3})\phi_{5,2,6}\bar\phi_{1,4,6} 
- \frac{\lambda}{2} \phi_{1,2,3} \bar\phi_{5,4,3}  (\nabla_{(s+[\alpha_{s}])\;i}  \phi_{5,2,6})\bar\phi_{1,4,6}
\crcr
&&
- \frac{\lambda}{2}   \phi_{1,2,3} 
 \bar\phi_{5,4,3}\phi_{5,2,6}( \nabla_{(s+[\alpha_{s}])\;i}\bar\phi_{1,4,6})\;.
\label{loco}
\eea
Using the equation of motion of $\phi_{1,2,3}$ and $\bar\phi_{1,2,3}$ after integrating by $g_4,g_5$ and $g_6$,  (\ref{loco}) becomes
\bea
&&  \int [\prod_{\ell=4}^6 dg_\ell] \Big{\{} 
  + \frac{\lambda}{2} (\nabla_{(s)\;i} \phi_{1,2,3}) \bar\phi_{5,4,3}\phi_{5,2,6}\bar\phi_{1,4,6}  
-\frac{\lambda}{2} \phi_{1,2,3}  (\nabla_{(s)\;i}\bar\phi_{5,4,3})\phi_{5,2,6}\bar\phi_{1,4,6}  \crcr
&&
- \frac{\lambda}{2}\phi_{1,2,3} \bar\phi_{5,4,3} (\nabla_{(s)\;i} \phi_{5,2,6})\bar\phi_{1,4,6}  
-\frac{\lambda}{2} \phi_{1,2,3}  \bar\phi_{5,4,3}\phi_{5,2,6}(\nabla_{(s)\;i}\bar\phi_{1,4,6} )\crcr
&& 
- \frac{\lambda}{2}   \phi_{1,2,3} 
( \nabla_{(s+[\alpha_{s}])\;i} \bar\phi_{5,4,3})\phi_{5,2,6}\bar\phi_{1,4,6} 
- \frac{\lambda}{2} \phi_{1,2,3} \bar\phi_{5,4,3}  (\nabla_{(s+[\alpha_{s}])\;i}  \phi_{5,2,6})\bar\phi_{1,4,6}\crcr
&&
- \frac{\lambda}{2}   \phi_{1,2,3} 
 \bar\phi_{5,4,3}\phi_{5,2,6}( \nabla_{(s+[\alpha_{s}])\;i}\bar\phi_{1,4,6})
+ \lambda (\nabla_{(s)\;i}\bar\phi_{1,2,3}) \phi_{5,4,3}
\bar\phi_{5,2,6} \phi_{1,4,6}
 \Big{\}}  
\label{lasteq}
\eea
which is not a vanishing quantity. Hence the EMT is not
covariantly conserved and the breaking term is clearly
a factor of the coupling constant $\lambda$.
As a proof that actually the EMT is globally
(covariantly) conserved, we have to check that, for $s=1,2,3$, 
the above remainder can only vanish under integration over the 
full six group copies. Due the particular symmetric form
of the strands, it is sufficient to check that 
claim for $s=1$. We have:
\bea
&&  \int [\prod_{\ell=4}^6 dg_\ell] \Big{\{} 
   \frac{\lambda}{2} (\nabla_{(1)\;i} \phi_{1,2,3}) \bar\phi_{5,4,3}\phi_{5,2,6}\bar\phi_{1,4,6}  
-\frac{\lambda}{2} \phi_{1,2,3}  \bar\phi_{5,4,3}\phi_{5,2,6}(\nabla_{(1)\;i}\bar\phi_{1,4,6} )\crcr
&& 
- \frac{\lambda}{2}   \phi_{1,2,3} 
( \nabla_{(5)\;i} \bar\phi_{5,4,3})\phi_{5,2,6}\bar\phi_{1,4,6} 
- \frac{\lambda}{2} \phi_{1,2,3} \bar\phi_{5,4,3}  (\nabla_{(5)\;i}  \phi_{5,2,6})\bar\phi_{1,4,6} \crcr
&&
+ \lambda (\nabla_{(1)\;i}\bar\phi_{1,2,3}) \phi_{5,4,3}
\bar\phi_{5,2,6} \phi_{1,4,6}
 \Big{\}}  \;.
\label{lasteq2}
\eea
We use two further integrations in $g_1$ and $g_3$, in order
to put the last expression, in the form
\bea
\int [\prod_{\ell=1,3,4,5,6} dg_\ell] \Big{\{} 
  -\lambda\phi_{1,2,3}  \bar\phi_{5,4,3}\phi_{5,2,6}(\nabla_{(1)\;i}\bar\phi_{1,4,6} )
+ \lambda (\nabla_{(1)\;i}\bar\phi_{1,2,3}) \phi_{5,4,3}
\bar\phi_{5,2,6} \phi_{1,4,6}
 \Big{\}}  \;,
\eea
and, finally, in order to cancel this term, we need the last integration
in $g_2$.

\subsection{EMT for the colored model}
\label{app:transcolor}

{\bf EMT calculation -} Let us start by 
considering the functional operator (\ref{colorwitrans0})
where the variations of the fields are $\delta_X \phi^a_{1,2,3} = X^i \partial_i \phi^a_{1,2,3}$ and equations of motion (\ref{eqmotioncol}). 
We introduce further notations $(\bullet)_{a,b,c} = \prod_{s=a,b,c} \sqrt{|\det \mathbf{g}_s|}$. Using the field variations and equations of motion,
the interaction has to be reconstructed as follows
\bea
&&
\lambda \int [\prod_{\ell=1}^6 dg_\ell]\;  
\Big[
\delta_X \phi^1_{1,2,3} \phi^2_{3,4,5}\phi^3_{5,2,6} \phi^4_{6,4,1} 
+ \delta_X \phi^4_{6,4,1}\phi^1_{1,2,3}\phi^2_{3,4,5}\phi^3_{5,2,6}
 + \fifib \Big]\crcr
&& =
\lambda \int [\prod_{\ell=1}^6 dg_\ell]\;  
\Big[
X^i \partial_{(1)\;i} (\phi^1_{1,2,3} \phi^2_{3,4,5}\phi^3_{5,2,6} \phi^4_{6,4,1} ) 
 + \fifib\Big]\;.
\label{wicolorsym}
\eea
The EMT can be computed from a similar routine as previously performed. We sum up the main steps:
\bea
&&
W(X)\Scol= \int [\prod_{\ell=1}^3 d\theta_\ell d \varphi_\ell^1 d \varphi_\ell^2] \;
\Big\{  \delta_{X} \mathbf{g}_1^{ij} \, \frac{\delta }{\delta \mathbf{g}_1^{ij}} \Scol
\crcr
&&
+ X^k \partial_{(1)\;k}  \phi^1_{1,2,3} \left[
 -  \sum_{s=1}^3 \partial_{(s)\;j } (  \buss \sqrt{|\det \mathbf{g}_s|} 
\mathbf{g}_s^{jl}\partial_{(s)\;l} \bar\phi^1_{1,2,3} )\right] \crcr
&&
+  (\bullet)_{1,2,3} 
 X^k \partial_{(1)\;k}\phi^1_{1,2,3} \left[ m^2 \bar\phi^1_{1,2,3} 
+\lambda \int [\prod_{\ell=4}^6 dg_\ell]\;\phi^2_{3,4,5}\phi^3_{5,2,6} \phi^4_{6,4,1} \right] +\fifib
\Big\}
\crcr
&&
+ \int [\prod_{\ell=1,4,6} d\theta_\ell d \varphi_\ell^1 d \varphi_\ell^2] 
\Big\{ \; 
X^k \partial_{(1)\;k}  \phi^4_{6,4,1} \left[
 -  \sum_{s=1}^3 \partial_{(s)\;j } ( \buss \sqrt{|\det \mathbf{g}_s|} 
\mathbf{g}_s^{jl}\partial_{(s)\;l} \bar\phi^4_{6,4,1} )\right]\crcr
&& 
+  (\bullet)_{1,4,6} 
X^k \partial_{(1)\;k}\phi^4_{6,4,1} \left[ m^2 \bar\phi^4_{6,4,1} 
+\lambda \int [\prod_{\ell\neq 1,4,6} dg_\ell]\;\phi^1_{1,2,3}\phi^2_{3,4,5} \phi^3_{5,2,6}  \right] +\fifib
 \Big\} .
\eea
Adding the contributions to the mass term and using (\ref{wicolorsym}), we get 
\bea
&&W(X)\Scol =\int [\prod_{\ell} d\theta_\ell d \varphi_\ell^1 d \varphi_\ell^2] \;
\Bigl\{  \delta_{X} \mathbf{g}_1^{ij} \, \left[  
(\bullet)_{\check{1}}\frac{\delta \sqrt{|\det \mathbf{g}|_1}}{\delta \mathbf{g}_1^{ij}}  \Lcol + (\bullet) \frac{\delta \Lkincol }{\delta \mathbf{g}_1^{ij}}\right] \crcr  
&&
+\Big\{ 
- X^k\sum_{s=1}^3  \partial_{(s)\;j } \left[  
\partial_{(1)\;k}  \phi^1_{1,2,3} \;
(\bullet) \mathbf{g}_s^{jl}  
\partial_{(s)\;l} \bar\phi^1_{1,2,3} \right] \crcr
&&
- X^k \sum_{s=1,4,6}  \partial_{(s)\;j } \left[  
\partial_{(1)\;k}  \phi^4_{6,4,1} \;
(\bullet)\mathbf{g}_s^{jl}  
\partial_{(s)\;l} \bar\phi^4_{6,4,1} \right] \crcr
&&+ X^k\sum_{s=1}^3  
(\bullet)
\mathbf{g}_s^{jl}
\partial_{(1)\;k}
\left[   
\partial_{(s)\;j }\phi^1_{1,2,3}\;
 \partial_{(s)\;l }   \bar \phi^1_{1,2,3}\right]
\label{intermclor1}\\
&&+ X^k\sum_{s=6,4,1} (\bullet) \mathbf{g}_s^{jl}
\partial_{(1)\;k}\left[  
\partial_{(s)\;j }   \phi^4_{6,4,1} \;
\partial_{(s)\;l } \bar \phi^4_{6,4,1} \right] + \fifib\Big\}
\label{intermclor2}\\
&&
+(\bullet)  X^k \Big[ 
 m^2 \partial_{(1)\;k} (\bar\phi^1_{1,2,3} \phi^1_{1,2,3}+ \bar\phi^4_{6,4,1} \phi^4_{6,4,1}) \crcr
&&
+\Big( \lambda \partial_{(1)\;k}(\phi^1_{1,2,3}\phi^2_{3,4,5}\phi^3_{5,2,6} \phi^4_{6,4,1} ) + \fifib \Big) 
\Big] 
\Bigr\} \;, 
\nonumber
\eea
where we used the symmetric part in $\bar\phi$ for completing the partial derivative.  Expanding the metric variations, one has
\bea
&&
 \delta_{X} \mathbf{g}_1^{ij} \, \left[  
(\bullet)_{\check{1}}\frac{\delta \sqrt{|\det \mathbf{g}|_1}}{\delta \mathbf{g}_1^{ij}}  \Lcol
+ (\bullet) \frac{\delta \Lcol}{\delta \mathbf{g}_1^{ij}}  \right]= \crcr
&& X^k \partial_{(1)\;k}\mathbf{g}_1^{ij} 
\Big[ 
 -\frac12 (\bullet)_{\check{1}} \sqrt{|\det \mathbf{g}|_1} \mathbf{g}_{1\; ij} \Lcol  ]
\label{term1}\\
&&
+ (\bullet) [\partial_{(1)\;i} \bar\phi^1_{1,2,3} \partial_{(1)\;l} \phi^1_{1,2,3}
+ \partial_{(1)\;i} \bar\phi^4_{6,4,1} \partial_{(1)\;l} \phi^4_{6,4,1}
\Big].
\label{term2}
\eea
The term (\ref{term1}) completes the derivative of $(\bullet) \Lcol$ while the second (\ref{term2}) cancels exactly the term 
with $-\partial_{(1)\;k} \mathbf{g}_s^{ij} $ obtained after integrating
by parts (\ref{intermclor1}) and (\ref{intermclor2}).
We obtain
\bea
&&
\frac{\partial}{\partial X^\rho}W(X)S =
 - \delta^{\rho k}\int [\prod_{\ell} d\theta_\ell d \varphi_\ell^1 d \varphi_\ell^2] \;
\Bigl\{  \crcr
&&
\Big[ \sum_{s=1}^3  \partial_{(s)\;j } \left[  
\partial_{(1)\;k}  \phi^1_{1,2,3} \;
(\bullet) \mathbf{g}_s^{jl}  
\partial_{(s)\;l} \bar\phi^1_{1,2,3} \right] \crcr
&&
+ \sum_{s=1,4,6}  \partial_{(s)\;j } \left[  
\partial_{(1)\;k}  \phi^4_{6,4,1} \;
(\bullet)\mathbf{g}_s^{jl}  
\partial_{(s)\;l} \bar\phi^4_{6,4,1} \right]  + \fifib \Big] \crcr
&& - \partial_{(1)\; k}\;[(\bullet)\mathcal{L}^{ (1,4)}] 
 -[\partial_{(1)\;k}\mathbf{g}_1^{ij} ][-\frac12 (\bullet)_{\check{1}} \sqrt{|\det \mathbf{g}|_1} \mathbf{g}_{1\; ij} \mathcal{L}^{ (\check{1},\check{4})} ]
\Bigr\} , 
\label{intervaria}
\eea
where, by definition,
\bea
&&
\mathcal{L}^{(1,4)}:=
 \sum_{s=1}^3\mathbf{g}^{ij}_{s}\partial_{(s) \;i} \bar\phi^1_{1,2,3} 
\partial_{(s) \;j}\phi^1_{1,2,3} 
 + 
\sum_{s=1,4,6}\mathbf{g}^{ij}_{s}\partial_{(s) \;i} \bar\phi^4_{6,4,1} 
\partial_{(s) \;j}\phi^4_{6,4,1} \crcr
&&
+ m^2\left[  \bar\phi^1_{1,2,3} \phi^1_{1,2,3}  +
 \bar\phi^4_{6,4,1} \phi^4_{6,4,1}\right] +
 \lambda \;\phi^{1}_{1,2,3} \phi^2_{3,4,5} \phi^3_{5,2,6}\phi^4_{6,4,1}
+ \bar\lambda \;\bar\phi^{1}_{1,2,3} \bar\phi^2_{3,4,5} \bar\phi^3_{5,2,6}\bar\phi^4_{6,4,1}\;,
\cr\cr
&&
\mathcal{L}^{ (\check{1},\check{4})} :=
\sum_{s=3,4,5}\mathbf{g}^{ij}_{s}\partial_{(s) \;i} \bar\phi^2_{3,4,5} 
\partial_{(s) \;j}\phi^2_{3,4,5} 
+
 \sum_{s=5,2,6}\mathbf{g}^{ij}_{s}\partial_{(s) \;i} \bar\phi^3_{5,2,6} 
\partial_{(s) \;j}\phi^3_{5,2,6} \crcr
 && 
+
m^2\left[
\bar\phi^2_{3,4,5} \bar\phi^2_{3,4,5}  + 
\bar\phi^3_{5,2,6} \phi^3_{5,2,6} \right]  .
\eea
Since $\mathcal{L}^{ (\check{1},\check{4})}$ does not contain the variable $g_1$,  the last term in (\ref{intervaria}) computes to a surface
term $\partial_{(1)\;k}[ (\bullet)\mathcal{L}^{ (\check{1},\check{4})}]$. Thence, the variations (\ref{intervaria}) can be 
written
\bea
&&
\frac{\partial}{\partial X^\rho}W(X)S =\crcr
&&
 - \int [\prod_{\ell=1}^6 d\theta_\ell d \varphi_\ell^1 d \varphi_\ell^2] \;
\Bigl\{ 
\sum_{s=1}^3  \partial_{(s)\;j }(\bullet) \mathbf{g}_s^{jl} \Big[  
\Big( \partial_{(1)\;\rho}  \phi^1_{1,2,3} \;
\partial_{(s)\;l} \bar\phi^1_{1,2,3} + \fifib \Big)
 - \delta_{s,1}\mathbf{g}_{s\; l\rho}\Lcol \Big]
\crcr
&&
+\sum_{s=1,4,6}  \partial_{(s)\;j } (\bullet)\mathbf{g}_s^{jl}  \left[  
 \partial_{(1)\;\rho}  \phi^4_{6,4,1} \;
\partial_{(s)\;l} \bar\phi^4_{6,4,1}+ \fifib\right]  
\Bigr\} .
\eea
From these last lines,  the EMT can be readily identified as a 
two-component tensor
\bea
T^{(1)}_{(1,s);(i,j)} &=& 
\partial_{(1)\;i}  \phi^1_{1,2,3} \;
\partial_{(s)\;j} \bar\phi^1_{1,2,3} 
+
\partial_{(1)\;i}  \bar\phi^1_{1,2,3} \;
\partial_{(s)\;j} \phi^1_{1,2,3} 
 - \delta_{1,s}\mathbf{g}_{s\; ij}\Lcol\;,
\crcr
T^{(4)}_{(1,s);(i,j)} &=& 
\partial_{(1)\;i}  \phi^4_{6,4,1} \;
\partial_{(s)\;j} \bar\phi^4_{6,4,1} 
+  
\partial_{(1)\;i}  \bar\phi^4_{6,4,1} \;
\partial_{(s)\;j} \phi^4_{6,4,1}  \;.
\eea
Of course it is a matter of choice to put the Lagrangian term
in one or the other component.

\noindent{\bf Covariant conservation -}
The conservation property of the EMT should be checked. 
We first evaluate:
\bea
&&
\sum_{s=1}^3\nabla^j_{(s)} 
T^{(1)}_{(1,s);(i,j)}  +\sum_{s=1,4,6}\nabla^j_{(s)} 
T^{(4)}_{(1,s);(i,j)} = \crcr
&&
\sum_{s=1}^3 \Big[ 
\nabla_{(1)\;i}  \phi^1_{1,2,3} \;
\nabla^j_{(s)}\nabla_{(s)\;j} \bar\phi^1_{1,2,3} 
+
 \nabla_{(1)\;i}  \bar\phi^1_{1,2,3} \;
\nabla^j_{(s)}\nabla_{(s)\;j} \phi^1_{1,2,3}  \Big] \crcr
&&
+  
\sum_{s=1,4,6}^3 \Big[ 
\nabla_{(1)\;i}  \phi^4_{6,4,1}\nabla^j_{(s)}\nabla_{(s)\;j}  \bar\phi^4_{6,4,1} + 
\nabla_{(1)\;i} \bar \phi^4_{6,4,1}\nabla^j_{(s)}\nabla_{(s)\;j}  \phi^4_{6,4,1} \Big] \crcr
&&
-  \Big[
  m^2\left[  \nabla_{(1)\;i} \bar\phi^1_{1,2,3} \phi^1_{1,2,3}+ 
 \nabla_{(1)\;i} \bar\phi^4_{6,4,1} \phi^4_{6,4,1} \right] \crcr
&&+ 
\lambda \;[\nabla_{(1)\;i} \phi^{1}_{1,2,3}] \phi^2_{3,4,5} \phi^3_{5,2,6}\phi^4_{6,4,1} +
  \lambda \; \phi^{1}_{1,2,3}\phi^2_{3,4,5} \phi^3_{5,2,6}
[\nabla_{(1)\;i}\phi^4_{6,4,1} ]
+ \fifib
\Big].
\eea 
Integrating first by $g_{4},g_5$ and $g_6$, and using equations 
of motion of $\phi^1$ and $\bar\phi_1$, we obtain
\bea
&&
\int [\prod_{\ell=4}^6 dg_\ell] 
\Big[\sum_{s=1}^3\nabla^j_{(s)} 
T^{(1)}_{(1,s);(i,j)}  +\sum_{s=1,4,6}\nabla^j_{(s)} 
T^{(4)}_{(1,s);(i,j)} \Big]  = \\
&&  \int  [\prod_{\ell=4}^6 dg_\ell] 
\Big\{ \nabla_{(1)\;i}  \phi^4_{6,4,1}
\Big[
\sum_{s=1,4,6}\nabla^j_{(s)}\nabla_{(s)\;j}  \bar\phi^4_{6,4,1}  
-
 m^2 \bar\phi^4_{6,4,1}
-
  \lambda \int dg_5\, \phi^{1}_{1,2,3}\phi^2_{3,4,5} \phi^3_{5,2,6} 
\Big] \crcr
&&+ \fifib\Big\}\;.
\nonumber
\eea
Performing a second integration with respect to $g_2$ and $g_{3}$,
using this time equations of motion of $\phi^4$ and $\bar\phi^4$,
one gets
\bea
&&
\int [\prod_{\ell=2}^6 dg_\ell] 
\Big[\sum_{s=1}^3\nabla^j_{(s)} 
T^{(1)}_{(1,s);(i,j)}  +\sum_{s=1,4,6}\nabla^j_{(s)} 
T^{(4)}_{(1,s);(i,j)} \Big]  = 0\;.
\label{cocons}
\eea
The following quantity
\bea
\int[\prod_{\ell=2}^6 dg_\ell] 
[T^{(1)}_{(1,1);(i,j)} + T^{(4)}_{(1,1);(i,j)} ]
\label{courantcolor}
\eea
 is therefore covariantly conserved. Indeed, starting from (\ref{cocons}), 
a  calculation yields
\bea
&& 0=\nabla^j_{(1)}  \int [\prod_{\ell=2}^6 dg_\ell]  \Big[
T^{(1)}_{(1,1);(i,j)}  + T^{(4)}_{(1,1);(i,j)}  \Big] 
+\crcr
&&  \int [\prod_{\ell=2}^6 dg_\ell]
\Big[\sum_{s=2,3}\nabla^j_{(s)} 
T^{(1)}_{(1,s);(i,j)}  +\sum_{s=4,6}\nabla^j_{(s)} 
T^{(4)}_{(1,s);(i,j)} \Big]  \crcr
&& 
0=\nabla^j_{(1)}  \int[\prod_{\ell=2}^6 dg_\ell] \Big[
T^{(1)}_{(1,1);(i,j)}  + T^{(4)}_{(1,1);(i,j)}  \Big] \crcr
&&
+  \sum_{s=2,3}\int \prod_{\ell=2,3}[d\theta_\ell d\varphi^1_\ell d \varphi^2_\ell]
\partial_{(s)\; k} [(\bullet) \mathbf{g}^{kj} T^{(1)}_{(1,s);(i,j)}] 
\crcr
&& 
+
\sum_{s=4,6}
\int \prod_{\ell=4,6}[d\theta_\ell d\varphi^1_\ell d \varphi^2_\ell] \partial_{(s)\; k} [(\bullet) \mathbf{g}^{kj} 
T^{(4)}_{(1,s);(i,j)}] \;,
\eea
where we used the fact that $\nabla_{(s)\;j} $ and $\nabla_{(s')\;i}$
commute for $s\neq s'$, and some integrations by parts 
for trading covariant derivatives for partial derivatives. 
Thus (\ref{courantcolor}) is a conserved current.

\section{Group dilatations}
\label{app:dilat}

\subsection{Dilatations on the sphere $S^D$}
\label{app:dilat:sph}

Consider the sphere $S^D$ with spherical local
coordinates $(\theta, \phi_1, \phi_2,\dots, \phi_{D-1})$,
and the transformation $d_a:\theta \mapsto \theta_a$
such that 
\bea
\tan \frac{\theta_a}{2} = a \tan \frac{\theta }{2}\;.
\eea
Note that this transformation is invertible $(d_a)^{-1} =d_{\frac1a}$.
We define the mapping on $S^D$
\bea
&&
 (\theta,\phi_1,\phi_2,\dots,\phi_{D-1}) 
\mapsto (y^0= \theta_a,y^1=\phi_1,y^2=\phi_2,\dots,y^{D-1}=\phi_{D-1}) 
\label{dilappp}\\
&&\theta_a = 2 \arctan \{ a \tan \frac{\theta}{2}\}
\eea
The mapping defines a conformal transformation of the
sphere with metric tensor $\mathbf{g}$ if the  metric induced by (\ref{dilappp})
satisfies, $\forall p\in S^D$,
\bea
\mathbf{g}_{\mu\nu}|_{(\theta_a,\phi_i)}
\frac{\partial y^\mu}{\partial x^\alpha}
\frac{\partial y^\nu}{\partial x^\beta}
  = \mu^{2} (\theta,\phi_i) \; \mathbf{g}_{\alpha\beta} |_{(\theta,\phi_i)} 
\; .
\label{conform}
\eea
The first component of the induced metric can be computed as, for $t =\theta/2$,
\bea
&&
\mathbf{g}_{\theta\theta} |_{(\theta_a,\phi_i)} 
\frac{\partial \theta_a}{\partial \theta}
\frac{\partial \theta_a }{\partial \theta}
= 1 \cdot \left(\frac{\partial  \theta_a}{\partial \theta}\right)^2 
= \mu^{2}(a,\theta) \\
&&\frac12(1+ \tan^2 t_a)d\theta_a
=  \frac{a}{2}(1+ \tan^2 t)d\theta \;,\quad
\frac{d  \theta_a }{d \theta}
=\frac{ 2a}{(1-a^2)\cos \theta+ 1+ a^2}  = \mu(a,\theta)\;.
\nonumber
\eea
The other metric components are of the
form $ \mathbf{g}_{\phi_i,\phi_i} |_{(\theta_a,\phi_i)} \cdot 1$,
$i=1,\dots,D-1$, such that
one can prove that the metric tensor is conformally invariant.
Indeed, the central points for that are: (1) $\sin^2\theta_a$ is a factor 
shared by all these components and (2) $\sin\theta_a/\sin\theta$ 
scales as $d\theta_a/ d\theta$. 
Hence the relation (\ref{conform}) is verified.

\subsection{Infinitesimal dilatations}
\label{app:dilat:inftsym}

Under an infinitesimal dilatation with parameter such that 
$a = 1+\epsilon$,  $\delta_\epsilon \theta = \epsilon \sin \theta$,
a field with scaling factor $c$ defined on a single copy of $G\simeq S^3$ transforms as
\bea
&&
\delta_\epsilon\phi (g)= \tilde\phi(d_a d_{\frac1a}(g)) - \phi(g) = 
\textstyle{\left[\frac{2(1+\epsilon)}{(1-(1+2\epsilon))\cos [\theta -\epsilon\sin\theta] + 1+(1+2\epsilon)}\right] ^{c}}
\phi(\theta - \epsilon \sin \theta)  - \phi(\theta) \crcr
&&= -\epsilon\bigl( \,- c\cos \theta  + \sin \theta \partial_\theta \,
\bigr)\phi(g)\;.
\label{inftfield1d}
\eea
Thus $\mathcal D (\cdot):=  \,[- c\cos \theta  + \sin \theta \partial_\theta](\cdot)$ is the generator of this dilatation.
For tensor fields, it can be shown similarly that the corresponding
operator becomes 
\bea
&&
\delta_{\epsilon} \phi_{1,2,3} 
= \sum_{s=1}^3 \delta_{\epsilon_s} \phi_{1,2,3}
 = \sum_{s=1}^3 
-\epsilon_s\bigl( \,-c\cos \theta_s + \sin \theta_s \,\partial_{(s)\;\theta} \,
\bigr)\phi_{1,2,3} \;,
\label{dilatinftensor}\\
&& \mathcal{D}_{(s)} := -c\cos \theta_s + \sin \theta_s \,\partial_{(s)\;\theta}\;.
\eea

\subsection{Dilatation current for 1D GFT}
\label{app:dilat:1dgft}

\noindent{\bf Current calculation -} Consider the operator (\ref{wiopdilat1d}),
we evaluate the variation of the action under this operator
using the equation of motion (\ref{eqscale1d}):
\bea
&&
\frac{\partial }{\partial \epsilon}
W(\epsilon) \Sosca = \frac{\partial }{\partial \epsilon}
\int  d\theta d\varphi^1d\varphi^2
\Big{\{ } \left(  -\epsilon  \mathcal{D} \phi  \right) \times \crcr
&& \Big{[} 
(\bullet)\frac{ (\cos\theta)^2}{\sin\theta} \phi
+ (\bullet)
 \cos\theta \partial_{\theta} \phi 
-  \partial_{\theta} [  (\bullet)\cos\theta \phi ] 
 - \widetilde{\Delta} \phi 
+ (\bullet)\lambda \sin\theta \phi^3 \Big{]}\Big{\}} .
\eea
First, we recombine the interaction in a surface term
\bea
A_0=\int  d\theta d\varphi^1d\varphi^2
\Big{\{ } \left(  -\epsilon  \mathcal{D} \phi  \right) 
 \Big{[} (\bullet)\lambda \sin\theta\phi^3 \Big{]}\Big{\}} 
= -\epsilon
\int  d\theta d\varphi^1d\varphi^2\;
\partial_{\theta}  
\left[(\bullet) \frac{\lambda}{4}  (\sin\theta)^2 \phi^4 \right],
\label{interma0}
\eea
and then reduce the following terms
\bea
&&
B_0= \int  d\theta d\varphi^1d\varphi^2
\Big{\{ } \left(  -\epsilon  \mathcal{D} \phi  \right) 
 \Big{[} 
(\bullet)\frac{ (\cos\theta)^2}{\sin\theta} \phi
+ (\bullet)
 \cos\theta \partial_{\theta} \phi 
-  \partial_{\theta} [  (\bullet)\cos\theta \phi ] 
\Big{]}\Big{\}} \crcr
&& = -\epsilon 
\int  d\theta d\varphi^1d\varphi^2
\Big{\{ } -  \partial_{\theta} \Bigl{[} (\bullet)
\left( \cos\theta +\sin\theta \partial_\theta   \right) \phi 
\cos\theta  \phi \Bigr{]} \crcr
&&
+ (\bullet)\frac{ (\cos\theta)^3}{\sin\theta} \phi^2
+  (\bullet)  (\cos\theta)^2 \partial_{\theta}\frac12\phi^2 \crcr
&&
+ (\bullet)  \Bigl{[} 
\{ -\cos\theta\sin\theta \phi^2 +3 (\cos\theta)^2  \phi\partial_{\theta}  \phi  + \cos\theta\sin\theta\partial_\theta  [\phi \partial_\theta \phi] \Big{]}\Big{\} } 
\crcr
&& =  \epsilon 
\int  d\theta d\varphi^1d\varphi^2
\Big{\{ } 
  \partial_{\theta}(\bullet) \Bigl{[} 
\cos\theta \{\left( \cos\theta +\sin\theta \partial_\theta   \right) \phi\} 
 \phi 
-
\cos\theta(\cos\theta +\sin\theta \partial_\theta) \frac12 \phi^2 \Bigl{]} 
\crcr
&& \qquad - (\bullet) (\sin\theta)^2 \phi \partial_\theta \phi
\Big{\} }  \;.
\label{intermb0}
\eea
Evaluating the Laplacian term, one finds:
\bea
&&
C_0 = 
\epsilon \int d\theta d\varphi^1 d \varphi^2\;
 \mathcal{D}\phi\; \widetilde{\Delta} \phi
\crcr
&& = 
\epsilon\int d\theta d\varphi^1d\varphi^2\;
\Big{\{} 
 \partial_{k}\left\{ \bigl[ \cos \theta  + \sin \theta\, \partial_{\theta} \,
\bigr]\phi\; (\bullet) \sin\theta\,\mathbf{g}^{kl} \partial_{l} \phi\right\} \crcr
&& - \partial_{k}\left\{ \bigl[ \cos \theta  + \sin \theta\, \partial_{\theta} \,
\bigr]\phi\right\}\; (\bullet) \sin\theta\,\mathbf{g}^{kl} \partial_{l} \phi
\;\Big{\}}\crcr
&&=\epsilon \int d\theta d\varphi^1 d \varphi^2\;\Big{\{} 
 \partial_{k}\left\{ \bigl[ \cos \theta  + \sin \theta\, \partial_{\theta} 
\bigr]\phi\; (\bullet) \sin\theta\,\mathbf{g}^{kl} \partial_{l} \phi\right\} \crcr
&& + (\bullet)\sin\theta \phi\, \partial_{\theta} \phi 
-\partial_{\theta}  \Bigl{[}(\bullet)\;
\frac12 (\sin\theta)^2\, \,\mathbf{g}^{kl}\partial_{k}\phi  \partial_{l} \phi \Bigr{]}
 \Big{\}}\;,
\label{intermlap}
\eea
where we use some integrations by parts and the fact that 
$\partial_\theta[\sin^2\theta \mathbf{g}^{kl}]=
 2\delta_{k,\theta}\delta_{l,\theta} \cos\theta\sin\theta$.

One notices that in the last expression, the term which is not of the form of a surface term 
cancels exactly the similar expression in (\ref{intermb0}).
Adding  the three contributions $A_0$ (\ref{interma0}),
$B_0$ (\ref {intermb0}) and $C_0$ (\ref{intermlap}), we get
\bea
&&
\frac{\partial}{\partial \epsilon} W(\epsilon) \Sosca 
=  \int d\theta d\varphi^1 d\varphi^2
\Big{\{} 
\partial_{k}\left\{ (\bullet) \sin\theta\,\mathbf{g}^{kl}\bigl[ \cos \theta  + \sin \theta\, \partial_{\theta} 
\bigr]\phi\;  \partial_{l} \phi\right\} \crcr
&& 
-\partial_{\theta}\; (\bullet) \Big{[} \sin\theta \Bigl{(}\;
\frac12 \sin\theta\, \,\mathbf{g}^{kl}\partial_{k}\phi  \partial_{l} \phi
+
\frac{(\cos\theta)^2}{\sin\theta} \frac12 \phi^2 +\cos\theta \phi    \partial_\theta \phi  + 
\frac{\lambda}{4} \sin\theta\phi^4 \Big{)} \crcr
&& -
\cos\theta \{\left( \cos\theta +\sin\theta \partial_\theta   \right) \phi\} 
 \phi 
 \Big{]}\;\;
\Bigr{\}} \crcr
&& 
=  \int d\theta d\varphi^1 d\varphi^2
\Big{\{} 
\partial_{k}\left\{ (\bullet) \sin\theta\,\mathbf{g}^{kl}\bigl[ \cos \theta  + \sin \theta\, \partial_{\theta} 
\bigr]\phi\;  \partial_{l} \phi\right\} \crcr
&& 
-\partial_{\theta} \;(\bullet)\Big{[} \, \sin\theta \Losca
-
\cos\theta \{\left( \cos\theta +\sin\theta \partial_\theta   \right) \phi\} 
 \phi 
 \Big{]}\;\;
\Bigr{\}} \crcr
&& 
=  \int d\theta d\varphi^1 d\varphi^2\;\;
\partial_{k}  \Big{\{}  \;(\bullet) \mathbf{g}^{kl}
\Bigl{[}
 \sin\theta\,\bigl[ \cos \theta  + \sin \theta\, \partial_{\theta} 
\bigr]  \phi\;  \partial_{l} \phi      \crcr
&& 
-\mathbf{g}_{l\theta} 
\sin\theta \Losca
+\mathbf{g}_{l\theta} \cos\theta \phi \left( \cos\theta +\sin\theta \partial_\theta   \right) \phi
 \Bigr{]}\Big{\}}  \;.
\eea
The dilatation current can be written 
\bea
D_j &=& \sin\theta\,\bigl[ \cos \theta  + \sin \theta\, \partial_{\theta} 
\bigr]  \phi\;  \partial_{j} \phi     
+\mathbf{g}_{j\theta} \cos\theta \phi \left( \cos\theta +\sin\theta \partial_\theta   \right) \phi
-\mathbf{g}_{j\theta} 
\sin\theta \Losca \;,\crcr
&=& \partial_{\theta} (\sin\theta \phi)\partial_{j} (\sin\theta\phi)     
-\mathbf{g}_{j\theta} 
\sin\theta \Losca\;.
\eea

\noindent{\bf Covariant conservation -} 
The equation of motion for this model can be calculated further as:
\bea
0 = -  (\bullet)\frac{\cos^2\theta - \sin^2\theta}{\sin\theta} \phi
 - \widetilde{\Delta} \phi 
+ (\bullet)\lambda \sin\theta \phi^3\;.
\eea
We compute still using the Levi-Civita connection:
\bea
&&
\nabla^j D_j  = 
\bigl[ \cos \theta  +\sin \theta\, \nabla_{\theta} 
\bigr]  \phi\; \frac{1}{(\bullet)}\widetilde{\Delta} \phi  \crcr
&&
+
 \,\bigl[ -\sin \theta  + \cos \theta\, \nabla_{\theta} 
\bigr]  \phi\;  (\sin\theta \nabla_{\theta} \phi  )+
 \,(\bigl[ \cos \theta  + \sin \theta\, \nabla_{\theta} 
\bigr]   \nabla^j   \phi)\; (\sin\theta \nabla_{j} \phi )   \cr\cr
&&-\sin\theta  \phi \left[ \cos\theta +\sin\theta \nabla_\theta   \right] \phi
+\cos\theta(\nabla_{\theta} \phi) \left[\cos\theta +\sin\theta \nabla_\theta   \right] \phi \crcr
&&
+ \cos\theta \phi \left[ -\sin\theta +\cos\theta \nabla_\theta   \right] \phi 
+ \cos\theta \phi\left[ \cos\theta +\sin\theta \nabla_\theta   \right] 
 \nabla_{\theta}\phi   \cr\cr 
&&
- \cos\theta 
\Big[  \frac12\frac{(\cos\theta)^2}{\sin\theta} \phi^2
+ \cos\theta\phi \partial_{\theta}\phi
 + \frac12 \sin\theta \mathbf{g}^{kl} \nabla_{k}\phi \nabla_{l}\phi    +\frac{\lambda}{4} \sin\theta \phi^4 \Big] \crcr
&& 
-  \sin\theta \Big{[}  \; 
   \frac12 [(-2)- \cot^2\theta]\cos\theta\phi^2 
+  \frac{(\cos\theta)^2}{\sin\theta}\phi \nabla_{\theta} \phi\crcr
&& 
+  (\nabla_{\theta}\phi) [ \cos\theta \nabla_{\theta}  \phi]
+    \phi   [-\sin\theta\nabla_{\theta} \phi + \cos\theta\nabla_{\theta}\nabla_{\theta}\phi] \crcr
&&
 + \frac12 \nabla_{\theta}\{ \sin\theta \mathbf{g}^{kl} \nabla_{k}\phi \nabla_{l}\phi\}    
+\frac{\lambda}{4} [\cos\theta \phi^4 +
4(\sin\theta \phi^3)\nabla_{\theta}\phi] 
 \Big{]}\;. \eea
Canceling the equation of motion,
substituting the remaining term in $\widetilde{\Delta}$ making use of
the equation of motion and performing 
some direct simplifications yields
\bea
\nabla^j D_j  = 
 \cos \theta    \sin\theta \;
\Big[ -(\cot\theta)^2 \phi^2  
 + 
 \nabla_{\theta} \phi\;  \nabla_{\theta} \phi 
+\frac{\lambda}{2} \phi^4 
\Big]\;,
\label{breakdilat1d}
\eea
which is not vanishing without further assumptions. 
Hence the dilatation current is not conserved as expected
from a system with an explicit coordinate dependence
in the Lagrangian unless the field satisfies both the 
equation of motion and (\ref{breakdilat1d}) equated to zero. The latter statement is far
to be an obvious issue or to even have nontrivial solutions
for fields on $SU(2)$. In order to have a taste of
that problem and for simplicity, by considering only
class angle fields $\phi = \phi(\theta)$, the system to be solved
is of the form
\bea
\left\{
\begin{array}{c} 
-(\cos\theta)^2 \phi^2  
 + 
 (\sin\theta)^2 ( \phi' )^2 
+\frac{\lambda}{2}  (\sin\theta)^2\phi^4  =0 \\
\\
   (\cos^2\theta - \sin^2\theta) \phi
 + 3\cos\theta\sin\theta \phi'
+ \sin^2\theta  \phi''
- \lambda \sin^2\theta \phi^3  =0
\end{array}\right. 
\eea

\subsection{Dilatation current for GFT in 3D}
\label{app:dilat:curr}

In this section, we compute the dilatation current
for a dynamical GFT in $3D$.

We need again to symmetrize the variation operator on the complex interaction, recalling that $\delta_\epsilon \phi_{1,2,3}$ 
assumes the form (\ref{dilatinftensor}):
\bea
&&
\lambda 
\int [\prod_{\ell=1}^6 dg_\ell]\;
\left( \delta_{\epsilon} \phi_{1,2,3} \bar\phi_{5,4,3}\phi_{5,2,6}\bar\phi_{1,4,6} 
+ \delta_{\epsilon} \bar\phi_{1,2,3}  
\phi_{5,4,3}\bar \phi_{5,2,6} \phi_{1,4,6} 
\right)
=\crcr
&&
\frac{\lambda}{2} 
\int [\prod_{\ell=1}^6 dg_\ell]\;
\Big{\{} 
 \left(\delta_{\epsilon} \phi_{1,2,3}  \right)\bar\phi_{5,4,3} \phi_{5,2,6} \bar\phi_{1,4,6}
+ \left(\delta_{\epsilon} \phi_{5,2,6}  \right) \phi_{1,2,3} \bar\phi_{5,4,3} \bar \phi_{1,4,6}
\crcr
&& 
+
\left(\delta_{\epsilon} \bar\phi_{5,4,3}\right) \phi_{1,2,3}\phi_{5,2,6}\bar \phi_{1,4,6} 
+
\left( \delta_{\epsilon} \bar\phi_{1,4,6} \right) \phi_{1,2,3}\bar\phi_{5,4,3}\phi_{5,2,6}
 \Big{\}} \crcr
&&  =
-\frac{\lambda}{2} 
\int[\prod_{\ell=1}^6 dg_\ell]\;
\Big{\{} 
\sum_{s=1}^3 \left( \epsilon_{(s)} \mathcal{D}_{(s)} \phi_{1,2,3}\right)  \bar\phi_{5,4,3}\phi_{5,2,6}\bar\phi_{1,4,6} \crcr
&&
+ \sum_{s=1}^3 \left(\epsilon_{(s)} \mathcal{D}_{(s+\alpha_s)} \bar\phi_{5,4,3}\right)  \phi_{1,2,3} \phi_{5,2,6}\bar\phi_{1,4,6}  
+
\sum_{s=1}^3 \left(\epsilon_{(s)} \mathcal{D}_{(s+\alpha_s)}\phi_{5,2,6}\right)  \bar\phi_{5,4,3}\phi_{1,2,3} \bar\phi_{1,4,6} \crcr
&&
+
\sum_{s=1}^3 \left(\epsilon_{(s)} \mathcal{D}_{(s+\alpha_s)}\bar\phi_{1,4,6} \right)  \bar\phi_{5,4,3}\phi_{5,2,6}\phi_{1,2,3} 
 \Big{\}} \crcr
&& 
= -\frac{\lambda}{2} 
\int[\prod_{\ell=1}^6 dg_\ell] \sum_{s=1}^3\epsilon_{(s)}  \Big{(} \mathcal{D}^{(2)}_{(s)} 
+ \mathcal{D}^{(2)}_{(s + [\alpha_s])}  \Big{)}
\phi_{1,2,3} \bar\phi_{5,4,3}\phi_{5,2,6}\bar\phi_{1,4,6}\;, \label{wiopdilasym}\\
&& \mathcal{D}^{(2)}_{(s)}  :=  -2c\cos \theta_s + \sin \theta_s \,\partial_{(s)\;\theta}\;,
\eea
where the definition of indices $\alpha_s$ and $[\alpha_s]$ remains the same as in the section dealing with translations.

Let us consider the action $\Skinsca[\phi]$ (\ref{dilaction}) without mass,\footnote{In fact, 
a mass term can be included but for simplicity purpose, we do not consider a massive field.} with $\gamma=-1$, $\beta=3/2$. The equation of motion for the field $\phi_{1,2,3}$ can be 
inferred as:
\bea
&&
\frac{\delta \Strisca}{\delta \phi_{1,2,3}} 
= \sum_{s=1}^3  \Big{\{} 
(\bullet)_{1,2,3}\beta^2 (\cos\theta_s)^2 \bar\phi_{1,2,3} 
+ (\bullet)_{1,2,3}\beta
 \cos\theta_s\sin\theta_s \partial_{(s)\;\theta} \bar\phi_{1,2,3} \\
&&- \beta \partial_{(s)\;\theta} \left[  (\bullet)_{1,2,3}
 \cos\theta_s\sin\theta_s\bar\phi_{1,2,3} \right] 
 - \widetilde{\Delta}_{(s)}\bar\phi_{1,2,3} \Big{\}} + (\bullet)_{1,2,3}\int [\prod_{\ell=4}^{6}dg_\ell]\lambda \bar\phi_{5,4,3}\phi_{5,2,6} \bar\phi_{1,4,6}\;,\crcr
&&(\bullet)_{1,2,3} := \prod_{s=1}^3 \sqrt{|\det \mathbf{g}_s|} \;,\quad
\widetilde{\Delta}_{(s)}\phi_{1,2,3}:= \partial_{(s)\;k}\left\{(\bullet)_{1,2,3}    
(\sin\theta_s)^{2} \mathbf{g}_s^{kl}(\partial_{(s)\;l}\phi_{1,2,3})\right\}  ,
\nonumber
\eea
where $\widetilde{\Delta}_{(s)}$ is again a modified Laplacian due the presence of
the sine function.  The functional operator for dilatations given by
(\ref{wiopdilat}) allows us to compute 
the variations of the action up to a surface term:
\bea
&&
\frac{\partial }{\partial \epsilon_i}
W(\epsilon) \Strisca = \frac{\partial }{\partial \epsilon_i}
\int [ \prod_{\ell=1}^6 d\theta_\ell d\varphi_\ell^1d\varphi_\ell^2]
\Big{\{ } \left(  -\sum_{s=1}^3 
\epsilon_s  \mathcal{D}_s \phi_{1,2,3}  \right) \times \crcr
&& \Big{[}\sum_{s=1}^3  \Big{\{} 
(\bullet)\beta^2 (\cos\theta_s)^2 \bar\phi_{1,2,3} 
+ (\bullet)\beta
 \cos\theta_s\sin\theta_s \partial_{(s)\;\theta}  \bar\phi_{1,2,3} \crcr
&&- \beta \partial_{(s)\;\theta} \left[  (\bullet)
 \cos\theta_s\sin\theta_s \bar\phi_{1,2,3} \right] 
 - \widetilde{\Delta}_{(s)}  \bar\phi_{1,2,3} \Big{\}}
 +(\bullet)\lambda  \bar\phi_{5,4,3}\phi_{5,2,6}  \bar\phi_{1,4,6} \Big{]} \; + \fifib \Big{\} }.
\nonumber
\eea
By first recombining the variations of the interaction, we get an
expression like (\ref{wiopdilasym}):
\bea
&&
A =\int [ \prod_{\ell=1}^6 d\theta_\ell d\varphi_\ell^1d\varphi_\ell^2]
\Big{\{ }  
  -\sum_{s=1}^3 
\epsilon_s[\mathcal{D}_s \phi_{1,2,3}]   \Big{[}(\bullet)\lambda \bar\phi_{5,4,3}\phi_{5,2,6} \bar\phi_{1,4,6} \Big{]}
+ \fifib \Big{\} }\crcr
&& = 
-\int [ \prod_{\ell=1}^6 d\theta_\ell d\varphi_\ell^1d\varphi_\ell^2]
\sum_{s=1}^3 \epsilon_s (\bullet)\Big{\{} \mathcal{D}^{(2)}_{s} 
+\mathcal{D}^{(2)}_{s+ [\alpha_s]}  \Big{\}}
\frac\lambda2  \phi_{1,2,3}\bar\phi_{5,4,3} \phi_{5,2,6} \bar\phi_{1,4,6}\crcr
&& = -\sum_{s=1}^3 \epsilon_s \int [ \prod_{\ell=1}^6 d\theta_\ell d\varphi_\ell^1d\varphi_\ell^2]
 \Big{\{} \partial_{(s)\;\theta} 
+\partial_{(s+ [\alpha_s])\;\theta}  \Big{\}}[(\bullet)
\frac\lambda2  \phi_{1,2,3}\bar\phi_{5,4,3} \phi_{5,2,6} \bar\phi_{1,4,6}]\;.
\label{eq:alin}
\eea
Second, we treat the terms with no or an unique derivative in the kinetic part:
\bea
&&
B = -\sum_{s'=1}^3 
\epsilon_{s'}\int [ \prod_{\ell=1}^6 d\theta_\ell d\varphi_\ell^1d\varphi_\ell^2]
\sum_{s=1}^3 \Big{\{ }\crcr
&&  \mathcal{D}_{s'}\phi_{1,2,3} \Big{[} 
(\bullet)\beta^2 (\cos\theta_s)^2 \bar\phi_{1,2,3} 
- \beta \partial_{(s)\;\theta}[(\bullet)\cos\theta_s\sin\theta_s]  \bar \phi_{1,2,3}  \Big{]}  + \fifib\Big{\} }\crcr
&&
=-\sum_{s'=1}^3 
\epsilon_{s'}\int [ \prod_{\ell=1}^6 d\theta_\ell d\varphi_\ell^1d\varphi_\ell^2]
\sum_{s=1}^3 \Big{\{ }\crcr
&&   
+ \Big[- \beta\partial_{(s)\;\theta}[(\bullet) \bigl( \,\beta\cos \theta_{s'}  + \sin \theta_{s'}\, \partial_{(s')\;\theta} \,
\bigr)\phi_{1,2,3} \cos\theta_s\sin\theta_s  \bar\phi_{1,2,3} ]  
+\fifib  \Big]\cr\cr
&&
+(\bullet) 3\beta^2 (\cos\theta_s)^2 \cos \theta_{s'} \bar\phi_{1,2,3}\phi_{1,2,3}
+ 
 (\bullet)\beta^2 (\cos\theta_s)^2\sin \theta_{s'} \partial_{(s')\;\theta}[\bar\phi_{1,2,3}\phi_{1,2,3}] 
\crcr
&& +  (\bullet)\delta_{s,s'} \beta \cos\theta_s\sin\theta_s     
\bigl( \,-2\beta \sin \theta_{s'}\bar\phi_{1,2,3}\phi_{1,2,3}
 \bigr)  \;  \label{line1}\\\cr
&& +  (\bullet) \delta_{s,s'}\beta \cos\theta_s\sin\theta_s     
\cos\theta_{s'} ( \bar\phi_{1,2,3} \partial_{(s')\;\theta}\phi_{1,2,3} + \phi_{1,2,3} \partial_{(s')\;\theta}\bar\phi_{1,2,3} ) \;  \crcr
&&
+ (\bullet)2\beta^2 \cos\theta_s \sin\theta_s \cos\theta_{s'}  
(\bar\phi_{1,2,3}\partial_{(s)\;\theta} \phi_{1,2,3} 
+\phi_{1,2,3}\partial_{(s)\;\theta} \bar\phi_{1,2,3} ) \crcr
&& + (\bullet)\beta\sin\theta_{s'}\cos\theta_s\sin\theta_s  [\partial_{(s')\;\theta}(\bar\phi_{1,2,3}\partial_{(s)\;\theta}\phi_{1,2,3}) + 
\partial_{(s')\;\theta}(\phi_{1,2,3}\partial_{(s)\;\theta}\bar\phi_{1,2,3})]
\Big{\}}.
\label{line2}
\eea
Since
$
\partial_{(s')\;\theta}  [(\bullet)\sin\theta_{s'} \cos^2\theta_s]
=   (\bullet)\left( 3 \cos\theta_{s'}\cos^2\theta_s
- 2\delta_{s,s'}\sin^2\theta_{s'}\cos\theta_{s'} \right)$,
the intermediate line (\ref{line1}) reduces to a surface term
\bea
&&
(\bullet) 3 \cos \theta_{s'}  \sum_{s=1}^3\beta^2(\cos\theta_s)^2
\bar\phi_{1,2,3}\phi_{1,2,3} 
+ 
 (\bullet) \sin \theta_{s'}\sum_{s=1}^3\beta^2 (\cos\theta_s)^2  \partial_{(s')\;\theta}[\bar\phi_{1,2,3}\phi_{1,2,3}] 
\\
&& - 2(\bullet) \sin \theta_{s'}\sum_{s=1}^3\delta_{s,s'} \beta^2 \cos\theta_s\sin\theta_s     
\bar\phi_{1,2,3}\phi_{1,2,3} = \partial_{(s')\;\theta}\left[\sin\theta_{s'} (\bullet) 
\sum_{s=1}^3\beta^2(\cos\theta_s)^2
\bar\phi_{1,2,3}\phi_{1,2,3} \right]\nonumber
\eea
Furthermore, using 
\begin{equation}
\partial_{(s')\;\theta}  [(\bullet)\sin\theta_{s'} \cos\theta_s\sin\theta_s]
= (\bullet)\left[3\cos\theta_{s'}\cos\theta_s\sin\theta_s 
+\delta_{s,s'}\sin\theta_{s'} (-\sin^2\theta_{s'} + \cos^2\theta_{s'}) \right],
\end{equation}
 we have
\bea
&&
3\cos\theta_{s'} (\bullet) \sum_{s=1}^3 \beta \cos\theta_s \sin\theta_s
( \bar\phi_{1,2,3} \partial_{(s)\;\theta}\phi_{1,2,3} + \phi_{1,2,3} \partial_{(s)\;\theta}\bar\phi_{1,2,3} )\crcr
&&
+\sum_{s=1}^3
  (\bullet)\delta_{s,s'} \beta \cos\theta_s\sin\theta_s     
\cos\theta_{s'} ( \bar\phi_{1,2,3} \partial_{(s)\;\theta}\phi_{1,2,3} + \phi_{1,2,3} \partial_{(s)\;\theta}\bar\phi_{1,2,3} ) \;  \crcr
&& + \sin\theta_{s'}(\bullet)\sum_{s=1}^3 \beta\cos\theta_s\sin\theta_s  \partial_{(s')\;\theta}( \bar\phi_{1,2,3} \partial_{(s)\;\theta}\phi_{1,2,3} + \phi_{1,2,3} \partial_{(s)\;\theta}\bar\phi_{1,2,3} )\crcr
&& = \partial_{(s')\; \theta}
\left[\sin\theta_{s'}\sum_{s=1}^3 (\bullet) \beta \cos\theta_s \sin\theta_s( \bar\phi_{1,2,3} \partial_{(s)\;\theta}\phi_{1,2,3} + \phi_{1,2,3} \partial_{(s)\;\theta}\bar\phi_{1,2,3} )\right] \crcr
&&
+  \beta\sin^3\theta_{s'}( \bar\phi_{1,2,3} \partial_{(s')\;\theta}\phi_{1,2,3} + \phi_{1,2,3} \partial_{(s')\;\theta}\bar\phi_{1,2,3} )\;.
\label{line3}
\eea
Hence, the quantity $B$ can be rewritten as
\bea
&&
B =  -\sum_{s'=1}^3 
\epsilon_{s'}\int [ \prod_{\ell=1}^6 d\theta_\ell d\varphi_\ell^1d\varphi_\ell^2]
 \Big{\{ }\label{eq:blin}\\
&&\Big[  -\sum_{s=1}^3 
\beta\partial_{(s)\;\theta}[(\bullet) \bigl( \,\beta\cos \theta_{s'}  + \sin \theta_{s'}\, \partial_{(s')\;\theta} \bigr)\phi_{1,2,3} \cos\theta_s\sin\theta_s  \bar\phi_{1,2,3} ]
+ \fifib\Big] \crcr
&&+ \partial_{(s')\; \theta} \Big{[}\sin\theta_{s'} (\bullet) 
\sum_{s=1}^3\beta^2(\cos\theta_s)^2\bar\phi_{1,2,3}\phi_{1,2,3} \Big{]} \crcr
&&
 +  \partial_{(s')\; \theta}
\Big{[}\sin\theta_{s'}(\bullet)\sum_{s=1}^3  \beta \cos\theta_s \sin\theta_s  \partial_{(s)\;\theta}(\bar\phi_{1,2,3}\phi_{1,2,3}) \Big{]}   +  \beta\sin^3\theta_{s'} \partial_{(s')\;\theta}(\bar\phi_{1,2,3}\phi_{1,2,3}) \Big{\} }.
\nonumber
\eea
Last, the Laplacian terms have to be calculated as follows:
\bea
&&
C = \sum_{s'=1}^3 
\epsilon_{s'}\int [ \prod_{\ell=1}^3 d\theta_\ell d\varphi_\ell^1d\varphi_\ell^2]
 \sum_{s=1}^3 \Big[
 \mathcal{D}_{s'}\phi_{1,2,3}\; \widetilde{\Delta}_{(s)} \bar\phi_{1,2,3} \; + \fifib \big]
\crcr
&&
=\sum_{s'=1}^3 
\epsilon_{s'}\int [ \prod_{\ell=1}^3 d\theta_\ell d\varphi_\ell^1d\varphi_\ell^2]
 \sum_{s=1}^3\Big{\{} \crcr
&& 
\Big[ \partial_{(s)\;k}\left\{ \bigl[ \,\beta\cos \theta_{s'}  + \sin \theta_{s'}\, \partial_{(s')\;\theta} \,
\bigr]\phi_{1,2,3}\; (\bullet) (\sin\theta_{s})^2\,\mathbf{g}^{kl}_{s} \partial_{(s)\;l} \bar\phi_{1,2,3}\right\} \crcr
&& + (\bullet) \delta_{s,s'}\delta_{k,\theta} \beta (\sin\theta_{s'})^3 \phi_{1,2,3} \,\mathbf{g}^{kl}_{s} \partial_{(s)\;l} \bar\phi_{1,2,3} \crcr
&& 
- (\bullet) \beta \cos \theta_{s'} (\sin\theta_s)^2 \partial_{(s)\;k}\phi_{1,2,3}
\,\mathbf{g}^{kl}_{s} \partial_{(s)\;l} \bar\phi_{1,2,3}\crcr
&&  
- (\bullet) \delta_{s,s'}\delta_{k,\theta} \cos\theta_{s'} (\sin \theta_{s})^2 \partial_{(s)\;\theta}
\phi_{1,2,3}\,\mathbf{g}^{kl}_{s} \partial_{(s)\;l} \bar\phi_{1,2,3} + \fifib\Big]\crcr
&& 
- (\bullet) \sin \theta_{s'}
\partial_{(s')\;\theta} 
[\sin^2\theta_s\,\mathbf{g}^{kl}_{s}\partial_{(s)\;k}\phi_{1,2,3}  \partial_{(s)\;l} \bar\phi_{1,2,3} ]\crcr
&& 
+ (\bullet) \sin \theta_{s'}
\partial_{(s')\;\theta} 
[\sin^2\theta_s\,\mathbf{g}^{kl}_{s}]\partial_{(s)\;k}\phi_{1,2,3}  \partial_{(s)\;l} \bar\phi_{1,2,3} \; \Big{\}}.
\label{interm10}
\eea
The  identity 
$\partial_{(s')\;\theta}  (\sin^2\theta_s\mathbf{g}^{kl}_{s})  =
\delta_{s,s'} \delta_{k,\theta}\delta_{l,\theta} 2 \cos\theta_{s'}\sin\theta_{s'}$,
allows one to rewrite (\ref{interm10}) as
\bea
&&C = \sum_{s'=1}^3 
\epsilon_{s'}\int [ \prod_{\ell=1}^3 d\theta_\ell d\varphi_\ell^1d\varphi_\ell^2]
 \Big{\{} \label{line4}\\
&& 
\Big[  \sum_{s=1}^3\partial_{(s)\;k}\left\{ \bigl[ \,\beta\cos \theta_{s'}  + \sin \theta_{s'}\, \partial_{(s')\;\theta} \,
\bigr]\phi_{1,2,3}\; (\bullet) (\sin\theta_{s})^2\mathbf{g}^{kl}_{s} \partial_{(s)\;l} \bar\phi_{1,2,3}\right\} + \fifib\Big]
\crcr
&& 
- \partial_{(s')\;\theta} \left((\bullet) \sin \theta_{s'}
 \sum_{s=1}^3\sin^2\theta_s
\mathbf{g}^{kl}_{s}\partial_{(s)\;k}\phi_{1,2,3}  \partial_{(s)\;l} \bar\phi_{1,2,3} \right) 
+ (\bullet)\beta (\sin\theta_{s'})^3 
 \partial_{(s')\;\theta} (\bar\phi_{1,2,3} \phi_{1,2,3})\; .
\nonumber
\eea
The non-like surface term appearing in (\ref{line4})
cancels the extra term appearing in (\ref{line3}). 
Summing all contributions, $A$ (\ref{eq:alin}), $B$ (\ref{eq:blin}) and $C$ (\ref{line4}) affords
\bea
&&
\frac{\partial}{\partial \epsilon_q} W(\epsilon) \Strisca = \int [\prod_{\ell=1}^6 d\theta_\ell d\varphi^1_{\ell}d\varphi^2_{\ell}]
\Big{\{}
\crcr
&&
\Big[ \sum_{s=1}^3\partial_{(s)\;k}\left\{ (\bullet) (\sin\theta_{s})^2[ \,\beta\cos \theta_{q}  + \sin \theta_{q}\, \partial_{(q)\;\theta} \,
]\phi_{1,2,3}\; \mathbf{g}^{kl}_{s} \partial_{(s)\;l} \bar\phi_{1,2,3}\right\}\crcr
&&
+\sum_{s=1}^3 
\beta\partial_{(s)\;\theta}[(\bullet)\cos\theta_s\sin\theta_s \bigl( \,\beta\cos \theta_{q}  + \sin \theta_{q}\, \partial_{(q)\;\theta} \bigr)\phi_{1,2,3}  \; \bar\phi_{1,2,3} ]+ \fifib\Big] \crcr
&& - \partial_{q\;\theta}[(\bullet) \sin\theta_{q} \Ltrisca] 
 - \partial_{q+[\alpha_{q}]\;\theta}[(\bullet) \sin\theta_{q+[\alpha_{q}]} \Lint] \Big{\}} \crcr
&& =  \int [\prod_{\ell=1}^6 d\theta_\ell d\varphi^1_{\ell}d\varphi^2_{\ell}]
\Big{\{}
\crcr
&&
\Big[ 
 \sum_{s=1}^3\partial_{(s)\;k} 
(\bullet)\mathbf{g}^{kl}_{s} \Big\{  (\sin\theta_{s})^2[ \,\beta\cos \theta_{q}  + \sin \theta_{q}\, \partial_{(q)\;\theta} \,
]\phi_{1,2,3}\; \partial_{(s)\;l} \bar\phi_{1,2,3} 
\crcr
&&
+ \beta  \mathbf{g}_{s\; l\theta}\cos\theta_s\sin\theta_s [ \,\beta\cos \theta_{q}  + \sin \theta_{q}\, \partial_{(q)\;\theta}]\phi_{1,2,3}  \;\bar \phi_{1,2,3} + \fifib\Big] \crcr
&& - \delta_{q,s}\mathbf{g}_{s\; l\theta} \,
 \sin\theta_{q} \Ltrisca
- \delta_{q+[\alpha_{q}],s}\mathbf{g}_{s\; l\theta} \,  \sin\theta_{q+[\alpha_{q}]} \Lint\Big{\}} \;.
\eea
 The current for this symmetry becomes a stranded tensor expressed by
\bea
&&
D_{(s,s');\;j} = \crcr
&& 
\sin \theta_{s} \Big\{
 \sin\theta_{s'}^{\frac12} \Big{[} \partial_{(s)\;\theta}\phi_{1,2,3}  
 \partial_{(s')\;j} (\sin\theta_{s'}^{\beta}\bar\phi_{1,2,3} )
+  \partial_{(s)\;\theta}\bar\phi_{1,2,3}  
 \partial_{(s')\;j} (\sin\theta_{s'}^{\beta}\phi_{1,2,3} ) \Big{]} \\
&&
- \delta_{s,s'}\mathbf{g}_{s'\; j\theta} \Ltrisca
\Big\}  
- \delta_{s+[\alpha_{s}],s'}\mathbf{g}_{s'\; j\theta} \,  \sin\theta_{s+[\alpha_{s}]} \Lint
+\beta  \cos \theta_{s}  \partial_{(s')\;j}    \left( 
  (\sin\theta_{s'})^2\bar\phi_{1,2,3}\phi_{1,2,3} \right) .
\nonumber
\eea
Again due to both the presence of the nonlocal interaction and
the explicit coordinate appearance in the Lagrangian, 
the dilatation current is not covariantly conserved.

\subsection{Dilatation current for the colored model}
\label{app:dilatcolor}

{\bf Current calculation -} 
We start by giving the equations of motion for the fields $\phi^{1}$
 and $\phi^4$, using $\Lcolsca= {\Lcolsca}^{(1,4)}+ {\Lcolsca}^{(\check{1},\check{4})} $ in the form (\ref{eq:lagcoldilat}), with $-c =\beta=3/2$, with
\bea
&&
{\Lcolsca}^{(1,4)}=\crcr
&&
 (\sin\theta_1)^{-1}\mathbf{g}^{ij}_{1}
\partial_{(1) \;i}[(\sin\theta_1)^{\beta} \bar\phi^1_{1,2,3} ]
\partial_{(1) \;j}[(\sin\theta_1)^{\beta}\phi^1_{1,2,3}] 
+\sum_{s=2}^3 \mathbf{g}^{ij}_{s}
\partial_{(s) \;i} \bar\phi^1_{1,2,3}  \partial_{(s) \;j}\phi^1_{1,2,3} \crcr
&&+ 
 (\sin\theta_1)^{-1}\mathbf{g}^{ij}_{1}
\partial_{(1) \;i}[(\sin\theta_1)^{\beta} \bar\phi^4_{6,4,1} ]
\partial_{(1) \;j}[(\sin\theta_1)^{\beta}\phi^4_{6,4,1}] 
+
\sum_{s=4,6} \mathbf{g}^{ij}_{s}
\partial_{(s) \;i} \bar\phi^4_{6,4,1} 
\partial_{(s) \;j}\phi^4_{6,4,1} \crcr
&&+
 \lambda \;\phi^{1}_{1,2,3} \phi^2_{3,4,5} \phi^3_{5,2,6}\phi^4_{6,4,1}
+ \bar\lambda \;\bar\phi^{1}_{1,2,3} \bar\phi^2_{3,4,5} \bar\phi^3_{5,2,6}\bar\phi^4_{6,4,1}\;, \cr\cr
&&
{\Lcolsca}^{(\check{1},\check{4})}= 
\sum_{s=3,4,5}\mathbf{g}^{ij}_{s}\partial_{(s) \;i} \bar\phi^2_{3,4,5} 
\partial_{(s) \;j}\phi^2_{3,4,5} 
+
 \sum_{s=5,2,6}\mathbf{g}^{ij}_{s}\partial_{(s) \;i} \bar\phi^3_{5,2,6} 
\partial_{(s) \;j}\phi^3_{5,2,6}  \;.
\label{eq:lagcoldilatapp}
\eea
The equation of motion obtained for $\phi^1_{1,2,3}$ is
\bea
&&
\frac{\delta \Solsca}{\delta \phi^1_{1,2,3}} 
=
(\bullet)_{1,2,3}\beta^2 (\cos\theta_1)^2 \bar\phi^1_{1,2,3} 
+ (\bullet)_{1,2,3}\beta
 \cos\theta_1\sin\theta_1 \partial_{(1)\;\theta} \bar\phi^1_{1,2,3} \crcr
&&\qquad 
\qquad- \beta \partial_{(1)\;\theta} \left[  (\bullet)_{1,2,3}
 \cos\theta_1 \sin\theta_1\bar\phi^1_{1,2,3} \right] 
 - \widetilde{\Delta}_{(1)}\bar\phi^1_{1,2,3} 
-(\bullet)_{1,2,3}\sum_{s=2,3} \Delta_{(s)} \bar\phi^1_{1,2,3}   \crcr
&& 
 \qquad\qquad +\lambda (\bullet)_{1,2,3}\int [\prod_{\ell=4}^{6}dg_\ell] \phi^2_{3,4,5}\phi^3_{5,2,6} \phi^4_{6,4,1}\;,
\label{eqmotiii}\\ 
&&(\bullet)_{a,b,c} := \prod_{s=a,b,c} \sqrt{|\det \mathbf{g}_s|} \;,\quad
\widetilde{\Delta}_{(1)}\phi_{1,2,3}:= \partial_{(1)\;k}\left\{(\bullet)_{1,2,3}    
(\sin\theta_1)^{2} \mathbf{g}_1^{kl}(\partial_{(1)\;l}\phi_{1,2,3})\right\}.
\nonumber
\eea
The equation of motion of $\phi^4$ and complex conjugate
fields are therefore obvious from (\ref{eqmotiii}). 
The functional operator for dilatations is given by
(\ref{widilatcol})
where the infinitesimal field variations possess an unique
parameter $\epsilon$. 
Let us evaluate the variations of the action up to the point we obtain a
surface term:
\bea
&&
\frac{\partial }{\partial \epsilon}
W(\epsilon) \Solsca= \frac{\partial }{\partial \epsilon}
( -\epsilon)\int [ \prod_{\ell=1}^6 d\theta_\ell d\varphi_\ell^1d\varphi_\ell^2]
\Big{\{ } \crcr
&&  \mathcal{D}_{(1)} \phi^1_{1,2,3}  
 \Big{[}(\bullet)\beta^2 (\cos\theta_1)^2 \bar\phi^1_{1,2,3} 
+ (\bullet)\beta
 \cos\theta_1\sin\theta_1 \partial_{(1)\;\theta} \bar\phi^1_{1,2,3} \crcr
&&- \beta \partial_{(1)\;\theta} \left[  (\bullet)
 \cos\theta_1 \sin\theta_1\bar\phi^1_{1,2,3} \right] 
 - \widetilde{\Delta}_{(1)}\bar\phi^1_{1,2,3} 
-(\bullet)\sum_{s=2,3} \Delta_{(s)} \bar\phi^1_{1,2,3}   +(\bullet)\lambda \phi^2_{3,4,5} \phi^3_{5,2,6} \phi^4_{6,4,1}  \Big{]} \crcr
&& +
 \mathcal{D}_{(1)} \phi^4_{6,4,1} 
 \Big{[}(\bullet)\beta^2 (\cos\theta_1)^2 \bar\phi^4_{6,4,1} 
+ (\bullet)\beta
 \cos\theta_1\sin\theta_1 \partial_{(1)\;\theta} \bar\phi^4_{6,4,1} \crcr
&&- \beta \partial_{(1)\;\theta} \left[  (\bullet)
 \cos\theta_1 \sin\theta_1\bar\phi^4_{6,4,1} \right] 
 - \widetilde{\Delta}_{(1)}\bar\phi^4_{6,4,1} 
-(\bullet)\sum_{s=4,6} \Delta_{(s)} \bar\phi^4_{6,4,1}  +(\bullet)\lambda  \phi^1_{1,2,3}\phi^2_{3,4,5} \phi^3_{5,2,6}\Big{]}
\crcr
&& + \fifib \Big{\} }.
\eea
Following the same steps as in Appendix \ref{app:dilat:curr},
the variations of the interaction can be recombined as
\bea
A & = &
-\epsilon\int [ \prod_{\ell=1}^6 d\theta_\ell d\varphi_\ell^1d\varphi_\ell^2]
 (\bullet)\Big[ \mathcal{D}^{(2)}_{(1)} 
[\lambda  \phi^1_{1,2,3}\phi^2_{3,4,5}\phi^3_{5,2,6} \phi^4_{6,4,1}]
+ \fifib\Big] \label{eq:alincol}\\
&=& - \epsilon \int [ \prod_{\ell=1}^6 d\theta_\ell d\varphi_\ell^1d\varphi_\ell^2]
 \Big{\{} \partial_{(1)\;\theta}\{ 
(\bullet) \sin\theta_1 \;
[\lambda \phi^1_{1,2,3}\phi^2_{3,4,5}\phi^3_{5,2,6} \phi^4_{6,4,1}]\} + \fifib \Big{\} } .
\nonumber
\eea
Second, we treat the terms with no or a single derivative in the kinetic part:
\bea
&&
B = -\epsilon\int [ \prod_{\ell=1}^6 d\theta_\ell d\varphi_\ell^1d\varphi_\ell^2]
 \Big{\{ }\crcr
&&  \Big[\Big(\mathcal{D}_{(1)}\phi_{1,2,3} \Big{[} 
(\bullet)\beta^2 (\cos\theta_1)^2 \bar\phi^1_{1,2,3} 
- \beta \partial_{(1)\;\theta}[(\bullet)\cos\theta_1\sin\theta_1]  \bar \phi^1_{1,2,3}  \Big{]} \crcr
&&
  \; +
(\phi^1_{1,2,3} \leftrightarrow \phi^4_{6,4,1}) \Big) \;+\; (\phi\leftrightarrow \bar\phi) \Big] \Big{\}}\crcr
&&
=  -
\epsilon\int [ \prod_{\ell=1}^6 d\theta_\ell d\varphi_\ell^1d\varphi_\ell^2]
 \Big{\{ }\crcr
&&
\Big[ 
 -
\beta\partial_{(1)\;\theta}[(\bullet) \bigl( \,\beta\cos \theta_{1}  + \sin \theta_{1}\, \partial_{(1)\;\theta} \bigr)\phi^1_{1,2,3} \cos\theta_1\sin\theta_1  \bar\phi^1_{1,2,3} ]
+\fifib \Big] \crcr
&&+ \partial_{(1)\; \theta} \Big{[}\beta^2(\bullet)\sin\theta_{1}  
(\cos\theta_1)^2\bar\phi^1_{1,2,3}\phi^1_{1,2,3} \Big{]} \crcr
&&
 +  \partial_{(1)\; \theta}
\Big{[}  \beta(\bullet)\cos\theta_1 (\sin\theta_1)^2  \partial_{(1)\;\theta}(\bar\phi_{1,2,3}\phi^1_{1,2,3}) \Big{]}   +  \beta(\bullet)\sin^3\theta_{1} \partial_{(1)\;\theta}(\bar\phi^1_{1,2,3}\phi^1_{1,2,3}) \crcr
 &&  \; +
(\phi^1_{1,2,3} \leftrightarrow \phi^4_{6,4,1}) 
\Big{\} }.
\label{eq:blincol}
\eea
Last, the Laplacian terms have to be calculated following the same steps
as done for the case without color. We find:
\bea
&&
C = 
\epsilon \int [ \prod_{\ell=1}^3 d\theta_\ell d\varphi_\ell^1d\varphi_\ell^2] \Big\{ \crcr
&&
\Big[\mathcal{D}_{1}\phi^1_{1,2,3}\; \widetilde{\Delta}_{(1)} \bar\phi^1_{1,2,3} +(\bullet)_{1,2,3}\sum_{s=2,3}\mathcal{D}_{1}\phi^1_{1,2,3}\; \Delta_{(s)} \bar\phi^1_{1,2,3} + \fifib\Big]  \; +
(\phi^1_{1,2,3} \leftrightarrow \phi^4_{6,4,1}) 
\Big{\} }
\crcr
&&  = \epsilon\int [ \prod_{\ell=1}^3 d\theta_\ell d\varphi_\ell^1d\varphi_\ell^2]
\Big{\{} \crcr
&& 
 \Big[\partial_{(1)\;k}\left\{ \bigl[ \,\beta\cos \theta_{1}  + \sin \theta_{1}\, \partial_{(1)\;\theta} \,
\bigr]\phi^1_{1,2,3}\; (\bullet) (\sin\theta_{1})^2\,\mathbf{g}^{kl}_{1} \partial_{(1)\;l} \bar\phi^1_{1,2,3}\right\}  \crcr
&&
 +
\sum_{s=2,3} \partial_{(s)\; k}[(\beta\cos\theta_1 + \sin\theta_1 \partial_{(1)\;\theta} )\phi^1_{1,2,3}\; (\bullet)\mathbf{g}^{kl}_{s} \partial_{(s)\;l}\bar\phi^1_{1,2,3} ] + \fifib\Big]   \crcr
&& + (\bullet) \beta (\sin\theta_{1})^3\partial_{(1)\;\theta}
( \phi^1_{1,2,3} \,  \bar\phi^1_{1,2,3} )\crcr
&&-   \partial_{(1)\;\theta} \Big[\sum_{s=2,3}\; (\bullet)\sin\theta_1\mathbf{g}^{kl}_{s} 
\partial_{(s)\; k} \phi^1_{1,2,3}\; \partial_{(s)\; l} \bar\phi^1_{1,2,3} \Big]
\crcr
&& 
- 
\partial_{(1)\;\theta}  \Big[(\bullet) \sin \theta_{1}
[(\sin\theta_1)^2\,\mathbf{g}^{kl}_{1}\partial_{(1)\;k}\phi^1_{1,2,3}  \partial_{(1)\;l} \bar\phi^1_{1,2,3}] \Big]
+ 
(\phi^1_{1,2,3} \leftrightarrow \phi^4_{6,4,1}) 
\; \Big{\}}\;.
\label{line4col}
\eea
and again the non-like surface term in (\ref{line4col})
cancels the extra term in (\ref{eq:blincol}). 
By adding all contributions, $A$ (\ref{eq:alincol}), $B$ (\ref{eq:blincol}) and $C$ (\ref{line4col}), one writes
\bea
&&
\frac{\partial}{\partial \epsilon} W(\epsilon) \Solsca= \int [\prod_{\ell=1}^6 d\theta_\ell d\varphi^1_{\ell}d\varphi^2_{\ell}]
\Big{\{}
\crcr
&&
\partial_{(1)\;k}(\bullet)\mathbf{g}^{kl}_{1}\Big\{ 
\Big[ \Big( (\sin\theta_{1})^2[ \,\beta\cos \theta_{1}  + \sin \theta_{1}\, \partial_{(1)\;\theta} \,
]\phi^1_{1,2,3}\;  \partial_{(1)\;l} \bar\phi^1_{1,2,3} 
\crcr
&&
+ \beta\mathbf{g}_{1\;l\theta}
\cos\theta_1\sin\theta_1  [ \,\beta\cos \theta_{1}  + \sin \theta_{1}\, \partial_{(1)\;\theta} ]\phi^1_{1,2,3}  \bar\phi^1_{1,2,3} 
+ \fifib \Big)\crcr
&&
+ (\phi^1_{1,2,3} \leftrightarrow \phi^4_{6,4,1}) \Big]  - \mathbf{g}_{1\;l\theta} \sin\theta_{1} {\Lcolsca}^{(1,4)}  \Big\}\crcr 
&&
 +
\sum_{s=2,3} \partial_{(s)\; k}(\bullet)  \mathbf{g}^{kl}_{s}
\Big[\Big(
[\beta\cos\theta_1 + \sin\theta_1 \partial_{(1)\;\theta} ]\phi^1_{1,2,3}\; \partial_{(s)\;l}\bar\phi^1_{1,2,3} ] +\fifib \Big) \crcr
&&
+ 
(\phi^1_{1,2,3} \leftrightarrow \phi^4_{6,4,1})  \Big]
  \Big{\}}.
\eea
 The current tensor for this symmetry possesses the distinct
components:
\bea
D^{(1)}_{(1);\;j} &=& 
\Big[(\sin\theta_{1})^2[ \,\beta\cos \theta_{1}  + \sin \theta_{1}\, \partial_{(1)\;\theta} \,
]\phi^1_{1,2,3}\; \partial_{(1)\;j} \bar\phi^1_{1,2,3} 
\crcr
&&
+ \beta  \mathbf{g}_{1\; j\theta}\cos\theta_{1}\sin\theta_{1} [ \,\beta\cos \theta_{1}  + \sin \theta_{1}\, \partial_{(1)\;\theta}]\phi^1_{1,2,3}  \; \bar\phi^1_{1,2,3} + (\phi \leftrightarrow \bar\phi)\Big] \crcr
&& - \mathbf{g}_{1\; j\theta} \,
 \sin\theta_{1}  {\Lcolsca}^{(1,4)} \;, 
\crcr
D^{(1)}_{(s);\;j} &=&
[ \,\beta\cos \theta_{1}  + \sin \theta_{1}\, \partial_{(1)\;\theta} \,
]\phi^1_{1,2,3}\; \partial_{(s)\;j} \bar\phi^1_{1,2,3}\; +\; (\phi \leftrightarrow \bar\phi)\;, \quad s=2,3,
\eea
and the other components $D^{(4)}_{(s);\;j}$ can be obtained
from $D^{(1)}_{(1);\;j} $ and $D^{(1)}_{(s=2,3);\;j} $
by taking the symmetry $(\phi^1_{1,2,3} \leftrightarrow \phi^4_{6,4,1})$
and  omitting  the Lagrangian part.
We can rewrite the dilatation tensor component $D^{(1)}_{(1);\;j}$ in the more compact form:
\bea
D^{(1)}_{(1);\;j}&=& 
\partial_{(1)\;\theta}[ (\sin\theta_1)^\beta \bar\phi^1_{1,2,3}]\,
\partial_{(1)\;j}[ (\sin\theta_1)^\beta \phi^1_{1,2,3}] + 
\partial_{(1)\;\theta}[ (\sin\theta_1)^\beta \phi^1_{1,2,3}]\,
\partial_{(1)\;j}[ (\sin\theta_1)^\beta \bar\phi^1_{1,2,3}]  \crcr
&-&  \mathbf{g}_{1\; j\theta}\sin\theta_1 {\Lcolsca}^{(1,4)}\;.
\eea
\noindent{\bf Covariant conservation -} We write the equation of motion
for the color $1$ field as
\bea
&&
  0=   \beta\left[- \beta\cos\theta^2_1  + \sin^2\theta \right] \bar\phi^1_{1,2,3} 
 - \frac{1}{ (\bullet)_{1,2,3}} \widetilde{\Delta}_{(1)}\bar\phi^1_{1,2,3} 
-\sum_{s=2,3} \Delta_{(s)} \bar\phi^1_{1,2,3}   \crcr
&& 
 \qquad\qquad +\lambda \int [\prod_{\ell=4}^{6}dg_\ell] \phi^2_{3,4,5}\phi^3_{5,2,6} \phi^4_{6,4,1}\;.
\nonumber
\eea
In a covariant form, we evaluate 
\bea
 &&
\sum_{s=1,2,3} \nabla^j_{(s)} D^{(1)}_{(s)\; j} + 
\sum_{s=1,4,6} \nabla^j_{(s)} D^{(4)}_{(s)\; j} = \crcr
&&
 \nabla^j_{(1)} \Big[
[\beta \cos\theta_1 +\sin\theta_1\nabla_{(1)\;\theta}]\bar\phi^1_{1,2,3}\; [(\sin\theta_1)^{2}\nabla_{(1)\;j}\phi^1_{1,2,3}]
\crcr
&&
+\beta \delta_{\theta j}\sin\theta_1\cos\theta_1\phi^1_{1,2,3} 
  [ \beta \cos\theta_1 +
\sin\theta_1\nabla_{(1)\;\theta}] \bar\phi^1_{1,2,3} 
 + (\phi \leftrightarrow \bar\phi) 
\Big] \crcr
&&
+ \sum_{s=2,3}  \Big[
[ \,\beta\cos \theta_{1}  + \sin \theta_{1}\, \nabla_{(1)\;\theta} \,
]\nabla^j_{(s)}\phi^1_{1,2,3}\; \nabla_{(s)\;j} \bar\phi^1_{1,2,3} \crcr
&&
+ [ \,\beta\cos \theta_{1}  + \sin \theta_{1}\, \nabla_{(1)\;\theta} \,
]\phi^1_{1,2,3}\; \nabla^j_{(s)}\nabla_{(s)\;j} \bar\phi^1_{1,2,3}
\; +\; (\phi \leftrightarrow \bar\phi) \Big] \crcr
&&
 +\nabla^j_{(1)} \Big[
[\beta \cos\theta_1 +\sin\theta_1\nabla_{(1)\;\theta}]\bar\phi^4_{6,4,1} \; [(\sin\theta_1)^{2}\nabla_{(1)\;j}\phi^4_{6,4,1} ]
\crcr
&&
+\beta \delta_{\theta j}\sin\theta_1\cos\theta_1\phi^4_{6,4,1} 
  [ \beta \cos\theta_1 +
\sin\theta_1\nabla_{(1)\;\theta}] \bar\phi^4_{6,4,1} 
 + (\phi \leftrightarrow \bar\phi) 
\Big] \crcr
&&
+ \sum_{s=4,6}  \Big[
[ \,\beta\cos \theta_{1}  + \sin \theta_{1}\, \nabla_{(1)\;\theta} \,
]\nabla^j_{(s)}\phi^4_{6,4,1}\; \nabla_{(s)\;j} \bar\phi^4_{6,4,1} \crcr
&&
[ \,\beta\cos \theta_{1}  + \sin \theta_{1}\, \nabla_{(1)\;\theta} \,
]\phi^4_{6,4,1}\; \nabla_{(s)\;j} \nabla^j_{(s)}\bar\phi^4_{6,4,1} 
\; +\; (\phi \leftrightarrow \bar\phi) \Big] \crcr
&&-\cos\theta_1 \mathcal{L}^{(1,4)} \crcr
&&-  \sin\theta_1
\Big[ \crcr
&&
\Big( 2\cos\theta_1 \sin\theta_1
 \nabla_{(1)}^j \bar\phi^1_{1,2,3}\nabla_{(1) \;j}\phi^1_{1,2,3} \crcr
&&
+ (\sin\theta_1)^2 \mathbf{g}_{1}^{jk}
[\nabla_{(1)\;\theta}\nabla_{(1)\;j} \bar\phi^1_{1,2,3}]\nabla_{(1) \;k}\phi^1_{1,2,3}  +
 (\sin\theta_1)^2 \mathbf{g}_{1}^{jk}\nabla_{(1)\;j} \bar\phi^1_{1,2,3}
[\nabla_{(1)\;\theta}\nabla_{(1) \;k}\phi^1_{1,2,3} ]  \cr\cr
&& 
+\beta [ -(\sin\theta_1)^2  + (\cos\theta_1 )^2   ]
\bar\phi^1_{1,2,3}\nabla_{(1) \;\theta}\phi^1_{1,2,3} \crcr
&&
 + 
\beta \cos\theta_1  \sin\theta_1  
[\nabla_{(1) \;\theta}\bar\phi^1_{1,2,3}\nabla_{(1) \;\theta}\phi^1_{1,2,3}
 +\bar\phi^1_{1,2,3}\nabla_{(1) \;\theta}\nabla_{(1) \;\theta}\phi^1_{1,2,3}  ] 
\cr\cr
&&
+ \beta[-(\sin\theta_1)^2  + (\cos\theta_1 )^2  ]
\phi^1_{1,2,3}\nabla_{(1)\;\theta}\bar\phi^1_{1,2,3} \crcr
&&
+ 
\beta \cos\theta_1  \sin\theta_1  
 [\nabla_{(1)\;\theta}\phi^1_{1,2,3}\nabla_{(1)\;\theta}\bar\phi^1_{1,2,3} 
+\phi^1_{1,2,3}\nabla_{(1)\;\theta}\nabla_{(1)\;\theta}\bar\phi^1_{1,2,3} ]\cr\cr
&& 
  + \beta^2[-2 \sin\theta_1\cos\theta_1] \bar\phi^1_{1,2,3} 
\phi^1_{1,2,3} 
+\beta^2 (\cos\theta)^2 [\nabla_{(1)\;\theta}\bar\phi^1_{1,2,3} 
\phi^1_{1,2,3}  +\bar\phi^1_{1,2,3} 
\nabla_{(1)\;\theta}\phi^1_{1,2,3}  ] 
 \cr\cr
&& 
+\sum_{s=2,3} 
[\nabla_{(1)\;\theta}\nabla_{(s)}^j \bar\phi^1_{1,2,3}  \nabla_{(s) \;j}\phi^1_{1,2,3} +\nabla_{(s)}^j \bar\phi^1_{1,2,3}  \nabla_{(1)\;\theta}\nabla_{(s) \;j}\phi^1_{1,2,3} ] \crcr
&&
+(\phi^1 \leftrightarrow \phi^4) \Big)\crcr
&&
 \lambda \;[\nabla_{(1)\;\theta}\phi^{1}_{1,2,3}] \phi^2_{3,4,5} \phi^3_{5,2,6}\phi^4_{6,4,1} +
\phi^{1}_{1,2,3} \phi^2_{3,4,5} \phi^3_{5,2,6}[\nabla_{(1)\;\theta}\phi^4_{6,4,1}]\crcr
&& 
+ \bar\lambda \;[\nabla_{(1)\;\theta}\bar\phi^{1}_{1,2,3}] \bar\phi^2_{3,4,5} \bar\phi^3_{5,2,6}\bar\phi^4_{6,4,1}
+ \bar\phi^{1}_{1,2,3} \bar\phi^2_{3,4,5} \bar\phi^3_{5,2,6}
[\nabla_{(1)\;\theta}\bar\phi^4_{6,4,1}]
\Big] 
\eea
which yields after canceling equations of motion of $\phi^{1}$, $\bar\phi^1$, $\phi^4$ and $\bar\phi^4$, by integrating all variables save $g_1$
and trading the remaining modified Laplacian using
once again the equations of motion:
\bea
&&
\int [\prod_{\ell=2}^6 dg_\ell]
\Big[ \sum_{s=1,2,3} \nabla^j_{(s)} D^{(1)}_{(s)\; j} + 
\sum_{s=1,4,6} \nabla^j_{(s)} D^{(4)}_{(s)\; j} \Big] =\crcr
&&
=\int [\prod_{\ell=2}^6 dg_\ell]
\Big\{\Big[ 
\Big(
\beta \cos\theta_1 \bar\phi^1_{1,2,3}
\Big[  \frac{1}{(\bullet)_{123}}\widetilde{\Delta}_{(1)}\phi^1_{1,2,3}
+\sum_{s=2,3}\; \nabla^j_{(s)}\nabla_{(s)\;j}\phi^1_{1,2,3} \Big]  
 + (\phi \leftrightarrow \bar\phi) 
\Big) \crcr &&
+2 \cos\theta_1(\sin\theta_1)^{2}\nabla_{(1)\;\theta}\bar\phi^1_{1,2,3}\; \nabla_{(1)\;\theta}\phi^1_{1,2,3} 
+\beta^2  \cos\theta
[(\cos\theta_1)^2 - 2(\sin\theta_1)^2]\phi^1_{1,2,3}\bar\phi^1_{1,2,3} \crcr\crcr
&&
+ 2\cos \theta_{1} \sum_{s=2,3} 
\nabla^j_{(s)}\bar\phi^1_{1,2,3}\; \nabla_{(s)\;j} \phi^1_{1,2,3} 
 \; +\; (\phi^1 \leftrightarrow \phi^4) \Big]\Big\} \crcr
&&- \Big[ 
\lambda \cos\theta_1\phi^{1}_{1,2,3}\phi^2_{3,4,5} \phi^3_{5,2,6}\phi^4_{6,4,1} 
+
  \bar\lambda \cos\theta_1 \bar\phi^{1}_{1,2,3} \bar\phi^2_{3,4,5} \bar\phi^3_{5,2,6}\bar\phi^4_{6,4,1} \Big] \Big\} \crcr
&&
=\int [\prod_{\ell=2}^6 dg_\ell]\Big{\{}
\Big[
2 \cos\theta_1(\sin\theta_1)^{2}\nabla_{(1)\;\theta}\bar\phi^1_{1,2,3}\; \nabla_{(1)\;\theta}\phi^1_{1,2,3} \crcr
&&
+ 2\cos \theta_{1} \sum_{s=2,3} 
\nabla^j_{(s)}\bar\phi^1_{1,2,3}\; \nabla_{(s)\;j} \phi^1_{1,2,3}
-\frac{9}{2} (\cos\theta_1)^3
 \bar\phi^1_{1,2,3}\phi^1_{1,2,3} 
 \; +\; (\phi^1 \leftrightarrow \phi^4) \Big] \crcr
&&+\frac12 \Big[ 
\lambda \cos\theta_1\phi^{1}_{1,2,3}\phi^2_{3,4,5} \phi^3_{5,2,6}\phi^4_{6,4,1} 
+
  \bar\lambda \cos\theta_1 \bar\phi^{1}_{1,2,3} \bar\phi^2_{3,4,5} \bar\phi^3_{5,2,6}\bar\phi^4_{6,4,1} \Big] \Big\}.
\eea
One can compare the latter expression with the breaking (\ref{breakdilat1d}) for the $1D$ case and discover than
they have in fact the same structure.

\end{document}